\documentclass[%
 superscriptaddress,
 reprint,
 showpacs,preprintnumbers,
 nofootinbib,
 amsmath,amssymb,
 aps,
 prd,
 floatfix,
]{revtex4-2}

\usepackage{graphicx}

\usepackage{xcolor}
\usepackage{lineno}

\usepackage{amssymb,amsmath,amstext,amsthm,amsfonts,amsfonts}
\usepackage[utf8]{inputenc}
\usepackage{subcaption}
\usepackage[linesnumbered,ruled]{algorithm2e}
\usepackage{url}
\usepackage{soul}
\usepackage{bm}
\usepackage[normalem]{ulem}
\usepackage{multirow, makecell}

\newcommand{\V}{\mathcal{V}}
\newcommand{\E}{\mathcal{E}}
\newcommand{\G}{\mathcal{G}}

\newcommand{\mb}{\mathbf}

\newcommand{\review}[1]{\textcolor{black}{#1}}

\DeclareMathAlphabet{\mathsfit}{\encodingdefault}{\sfdefault}{m}{sl}
\SetMathAlphabet{\mathsfit}{bold}{\encodingdefault}{\sfdefault}{bx}{n}

\begin{document}

	\title{Graph neural network for 3D classification of ambiguities and optical crosstalk in scintillator-based neutrino detectors}

    \author{Sa\'ul Alonso-Monsalve}
    \email[E-mail: ]{saul.alonso.monsalve@cern.ch}
    \affiliation{CERN, The European Organization for Nuclear Research, 1211 Meyrin, Switzerland}
    \affiliation{Universidad Carlos III de Madrid, Av. de la Universidad, 30, 28911 Madrid, Spain}
    
    \author{Dana Douqa}
    \email[E-mail: ]{dana.douqa@unige.ch}
	\affiliation{University of Geneva, Section de Physique, DPNC, 1205 Genève, Switzerland}
    
    \author{C\'esar Jes\'us-Valls}
	\affiliation{IFAE, Institut de F\'isica d'Altes Energies, Carrer de Can Magrans, 08193 Barcelona, Spain}
    
   \author{Thorsten Lux}
	\affiliation{IFAE, Institut de F\'isica d'Altes Energies, Carrer de Can Magrans, 08193 Barcelona, Spain}
	
	\author{Sebastian Pina-Otey}
	\affiliation{IFAE, Institut de F\'isica d'Altes Energies, Carrer de Can Magrans, 08193 Barcelona, Spain}
	\affiliation{Aplicaciones en Inform\'atica Avanzada (AIA), 08172 Sant Cugat del Vall\`es Barcelona, Spain}
	
	\author{Federico S\'anchez}
	\affiliation{University of Geneva, Section de Physique, DPNC, 1205 Genève, Switzerland}

    \author{Davide Sgalaberna}
    \affiliation{ETH Zurich, Institute for Particle Physics and Astrophysics, CH-8093 Zurich, Switzerland}

    \author{Leigh H. Whitehead}
    \affiliation{Cavendish Laboratory, University of Cambridge, Cambridge, CB3 0HE, United Kingdom}

\begin{abstract}
\noindent 
Deep-learning tools are being used extensively in high energy physics and are becoming central in the reconstruction of neutrino interactions in particle detectors. In this work, we report on the performance of a graph neural network in assisting with particle \review{set} event reconstruction. The three-dimensional reconstruction of particle tracks produced in neutrino interactions can be subject to ambiguities due to high multiplicity signatures in the detector or leakage of signal between neighboring active detector volumes. Graph neural networks potentially have the capability of identifying all these features to boost the reconstruction performance. As an example case study, we tested a graph neural network, inspired by the GraphSAGE algorithm, on a novel 3D-granular plastic-scintillator detector, that will be used to upgrade the near detector of the T2K experiment. The developed neural network has been trained and tested on diverse neutrino interaction samples, showing very promising results: the classification of particle track voxels produced in the detector can be done with efficiencies and purities of 94-96\% per event and most of the ambiguities can be identified and rejected, while being robust against systematic effects.
\end{abstract}

\maketitle

\section{Introduction}
\label{sec:introduction}

Since 1999, a series of neutrino oscillation experiments have provided deep insight into the nature of neutrinos~\cite{oscSuperK,oscSNO,oscKamland,oscK2K,oscDayaBay,MINOSFinal3Flav,Acero:2019ksn,Abe:2019vii}. A number of these experiments are long-baseline neutrino oscillation experiments that use two detectors to characterize a beam of (anti-)neutrinos: a near detector, located a few hundred meters away from the target that measures the original beam composition, and a far detector, located several hundred kilometres away, that allows for the determination of the beam composition after neutrino flavor oscillations.

The energy of these beam neutrinos ranges from a few hundred MeV up to several GeV. Charged particles can be produced in neutrino interactions, and the energy that they deposit as they traverse the detector can be used to reconstruct the events. In general, the larger the energy transferred from the neutrino to the nucleus, the larger the number of particles and particle types produced in the final state. Modeling nuclear interactions in the target nuclei is highly complex, particularly for high energy transfers where the hadronic component of the interaction is more important. As a result, current long-baseline neutrino oscillation experiments mostly analyze interactions with low particle multiplicity. This situation, however, is expected to change in the coming years. On one hand, the statistical and systematic uncertainties of current experiments have decreased significantly over recent years such that neutrino-nucleus modeling is becoming a dominant source of uncertainty~\cite{NOvA:2018gge,Abe:2019vii}. On the other hand, future experiments like DUNE~\cite{duneTDR} will use a broad-band energy neutrino beam, expecting a significant fraction of the neutrino interactions to have a high energy transfer to the nucleus. 

As a result, in recent years, the neutrino physics community has turned its attention to measuring neutrino-nucleus interaction cross-sections for different ranges of energies and target materials \cite{Tanabashi:2018oca} as a way to constrain the oscillation uncertainties while providing new measurements to further develop the interaction models. In parallel, a new generation of neutrino detectors are under development that aim to resolve and reliably identify short particle tracks even in very complex interactions. To achieve this, two main detector technologies stand out: one is based on Liquid Argon Time-Projection-Chambers (LArTPCs)~\cite{Rubbia:1977zz} and the other is based on finely segmented plastic scintillators with three readout views~\cite{Blondel_2018} that will form part of the near detectors for T2K~\cite{Abe:2019whr} and, possibly, DUNE~\cite{3DST_ICHEP}.

For the latter, the detector response to a charged particle is read out into three orthogonal 2D projections. When reconstructing the 3D neutrino event, different types of hits are rebuilt, introducing non-physical entities that can hinder the reconstruction process. Due to the spatial disposition of such hits, an approach of utilizing Graph Neural Networks (GNNs)~\cite{SperdutiFirstGNN} is proposed to perform the classification of 3D hits to provide clean tracks for event reconstruction.

The article proceeds in the following way: Sec.~\ref{sec:motivation} describes properly the motivation behind the methodology given the details of the detector technology. Section~\ref{sec:background} introduces deep-learning techniques and explains the specific GNN algorithm used. The simulated data samples and GNN training are discussed in Sec.~\ref{sec:methodology}. Results and a study of systematic uncertainties are given in Secs.~\ref{sec:results} and ~\ref{sec:systematics}, respectively, followed by concluding remarks in Sec.~\ref{sec:conclusions}.

\section{Motivation}
\label{sec:motivation}

\begin{figure}[hbt]
    \centering
		\includegraphics[width=0.7\linewidth]{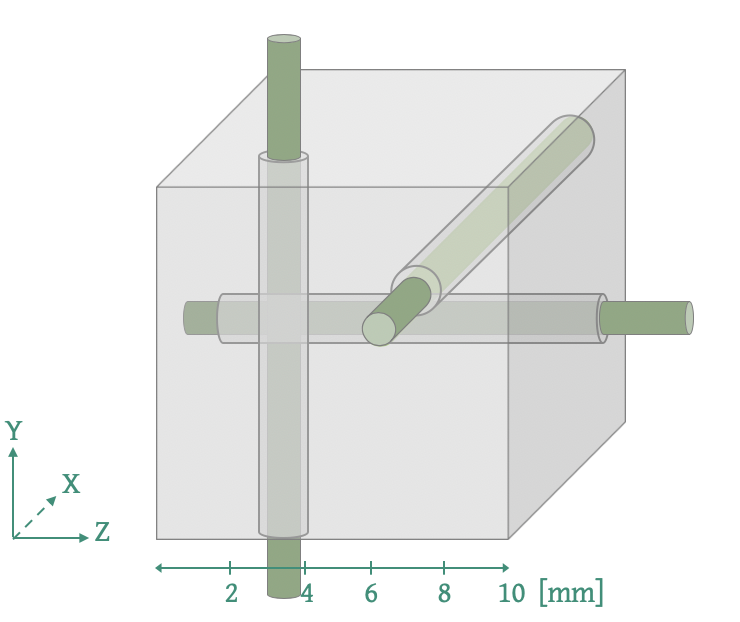}
	\caption{Geometry of a single SuperFGD element. \review{Each cube (gray) is intersected by three WLS fibers (green). The whole SuperFGD will be an array of $56 \times 184 \times 192$ of these elements ($H \times L \times W$).}}
    \label{fig:cube3d}
\end{figure}

\begin{figure}[htb]
    \centering
		\includegraphics[width=1.1\linewidth]{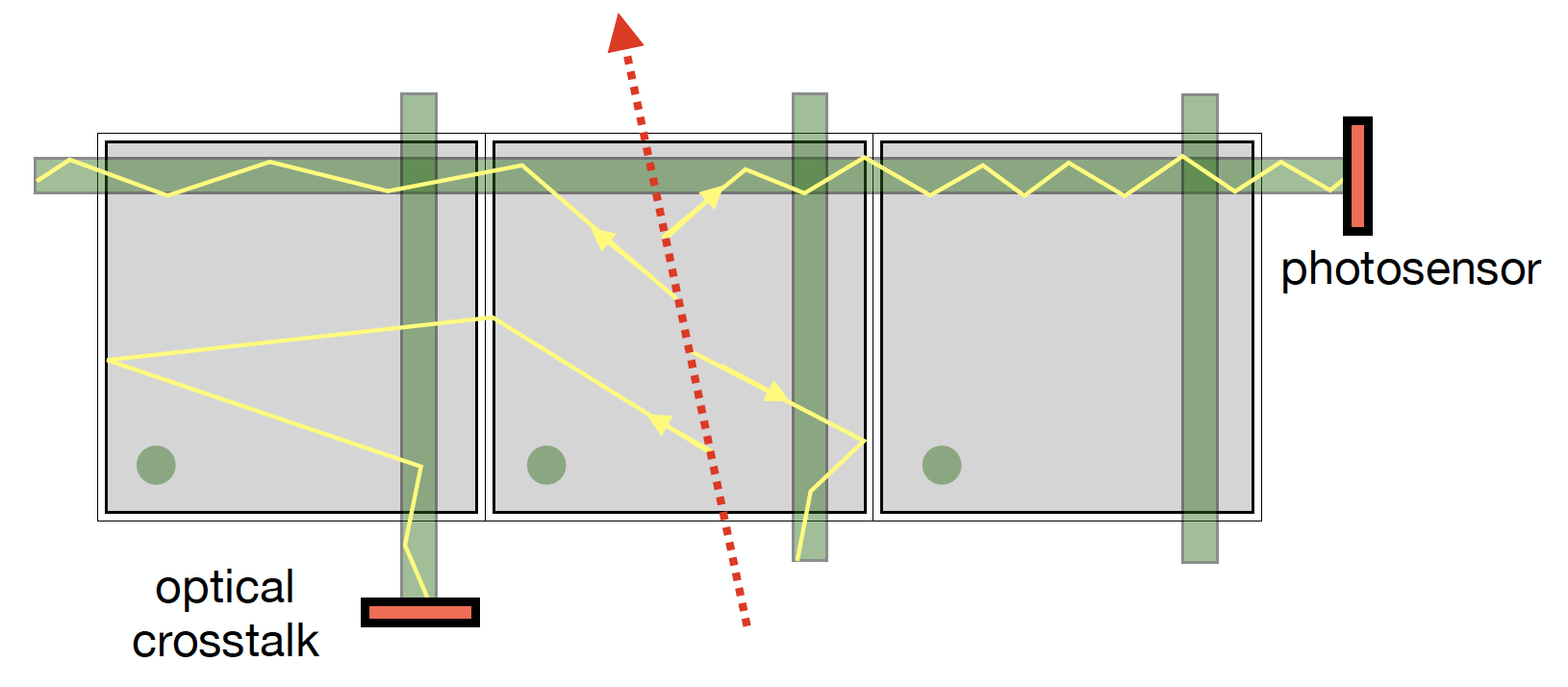}
	\caption{Sketch of the signal generation, fiber transport, and signal detection processes highlighting the production of optical crosstalk signals. \review{The cubes are depicted in gray, the WLS fibers in green, the dashed red line is a charged track and photons are illustrated in yellow.}}
    \label{fig:xtalk}
\end{figure}

A finely segmented scintillator detector consists of a 3D matrix of plastic scintillator cubes. The scintillation light produced by charged particles traversing the cubes is read out by three orthogonal wavelength-shifting (WLS) fibers that transport the scintillation light out of the detector where silicon photomultipliers (SiPMs) convert it into a certain number of photoelectrons (p.e.), as illustrated in Figs.~\ref{fig:cube3d} and~\ref{fig:xtalk}.

\begin{figure*}[ht!] 
  \subcaptionbox{Projections of the observed neutrino interaction onto the three 2D detector views (XY, XZ, and YZ). \label{fig:projections}}%
  [.495\linewidth]{\includegraphics[width=8.8cm]{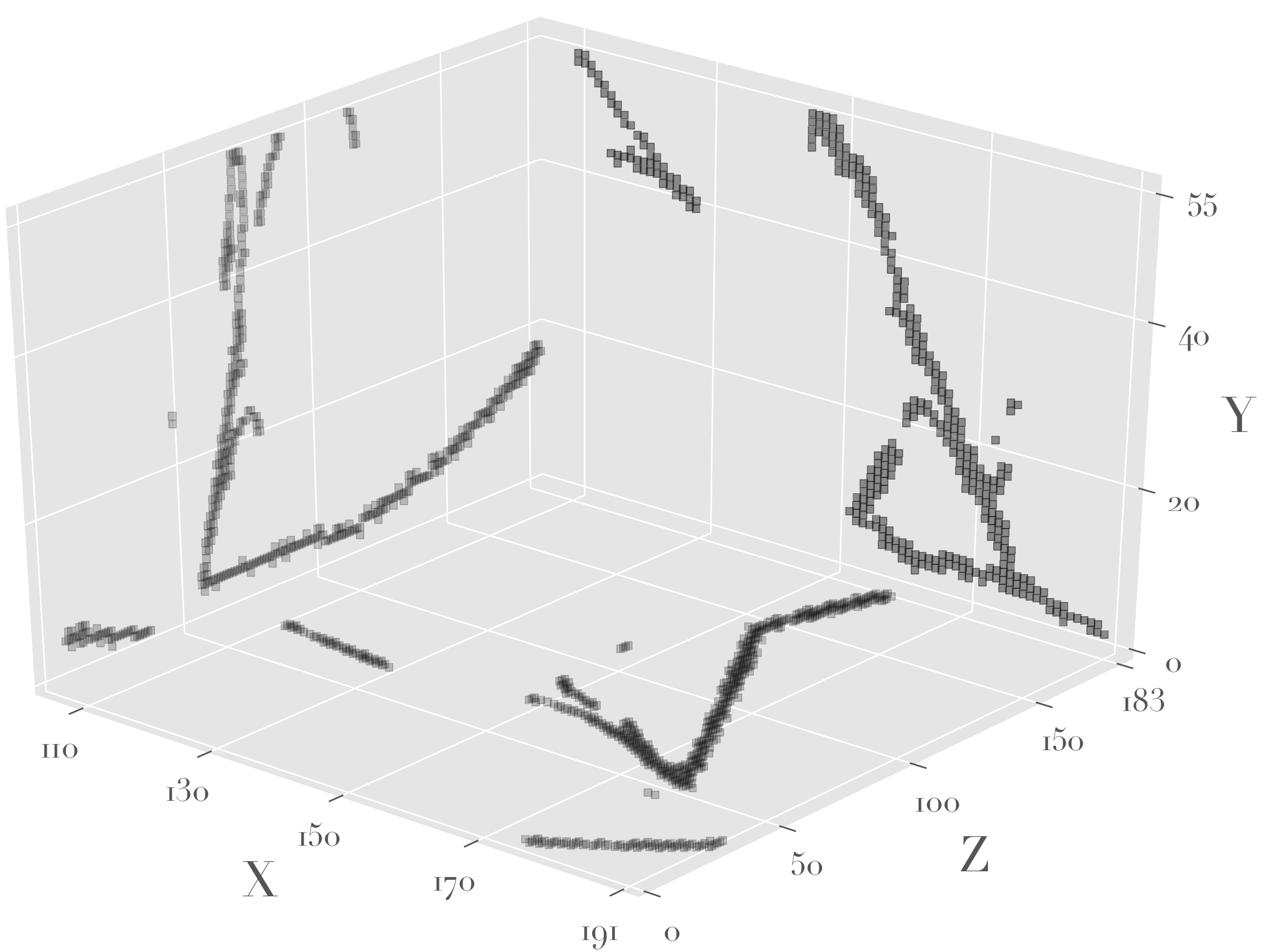}}
  \subcaptionbox{3D view of the neutrino interaction after the 3D matching of the three 2D views  \review{in Figure~\ref{fig:projections}}. The 3D voxels are shown as dark points. Projections of the observed neutrino interaction onto the three 2D detector views (XY, XZ, and YZ) are shown as shadow. 
  \label{fig:3D}}%
  [.495\linewidth]{\includegraphics[width=8.8cm]{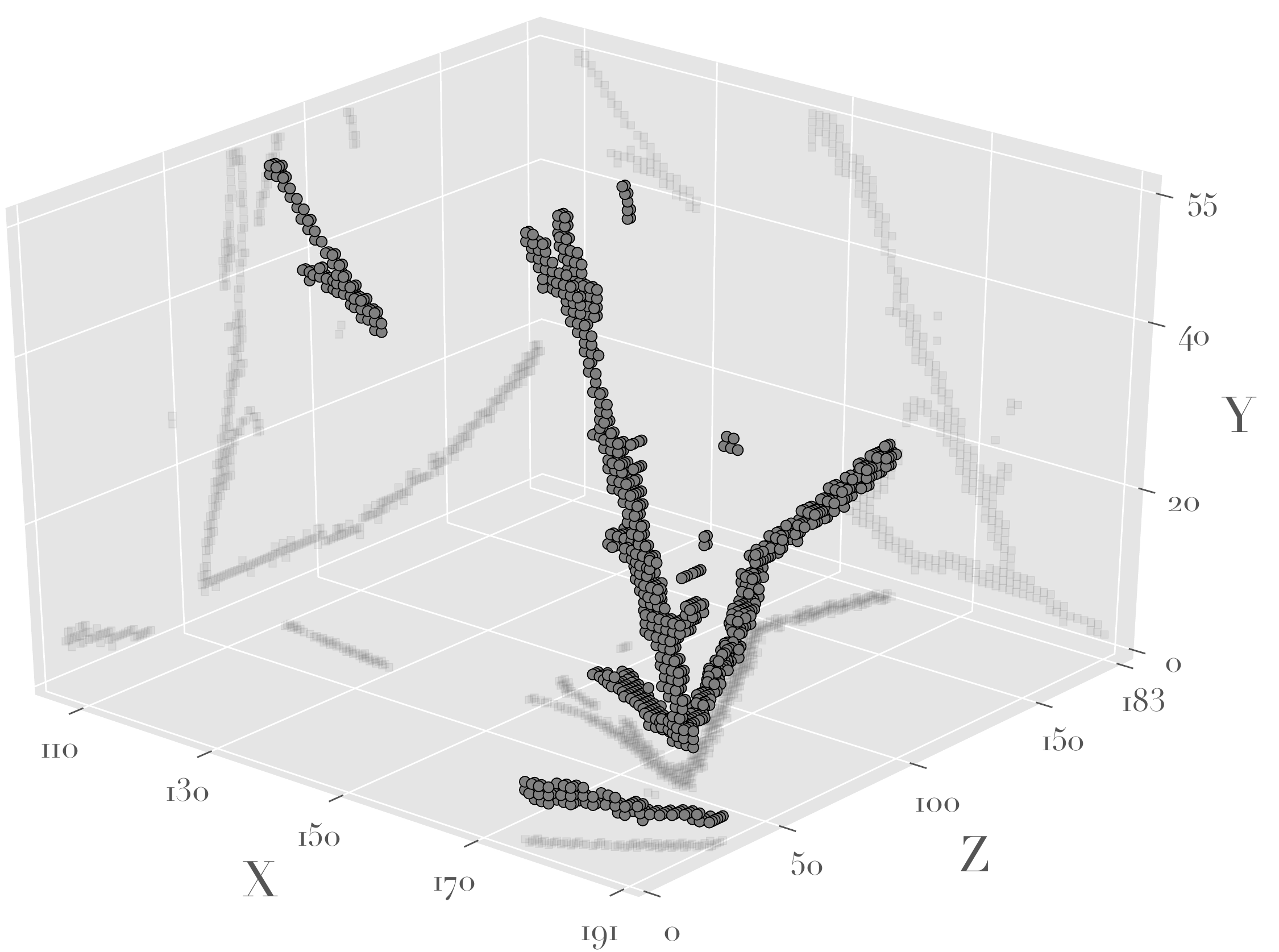}}
  \subcaptionbox{
  3D view of the neutrino interaction after the 3D matching of the three 2D views  \review{in Fig.~\ref{fig:projections}}. The 3D voxels labelled as track (red), crosstalk (blue), and ghost (yellow) according to the truth information from the simulation. Projections of the observed neutrino interaction onto the three 2D detector views (XY, XZ, and YZ) are shown as shadows. \label{fig:goal}}%
  [.75\linewidth]{\includegraphics[width=13cm]{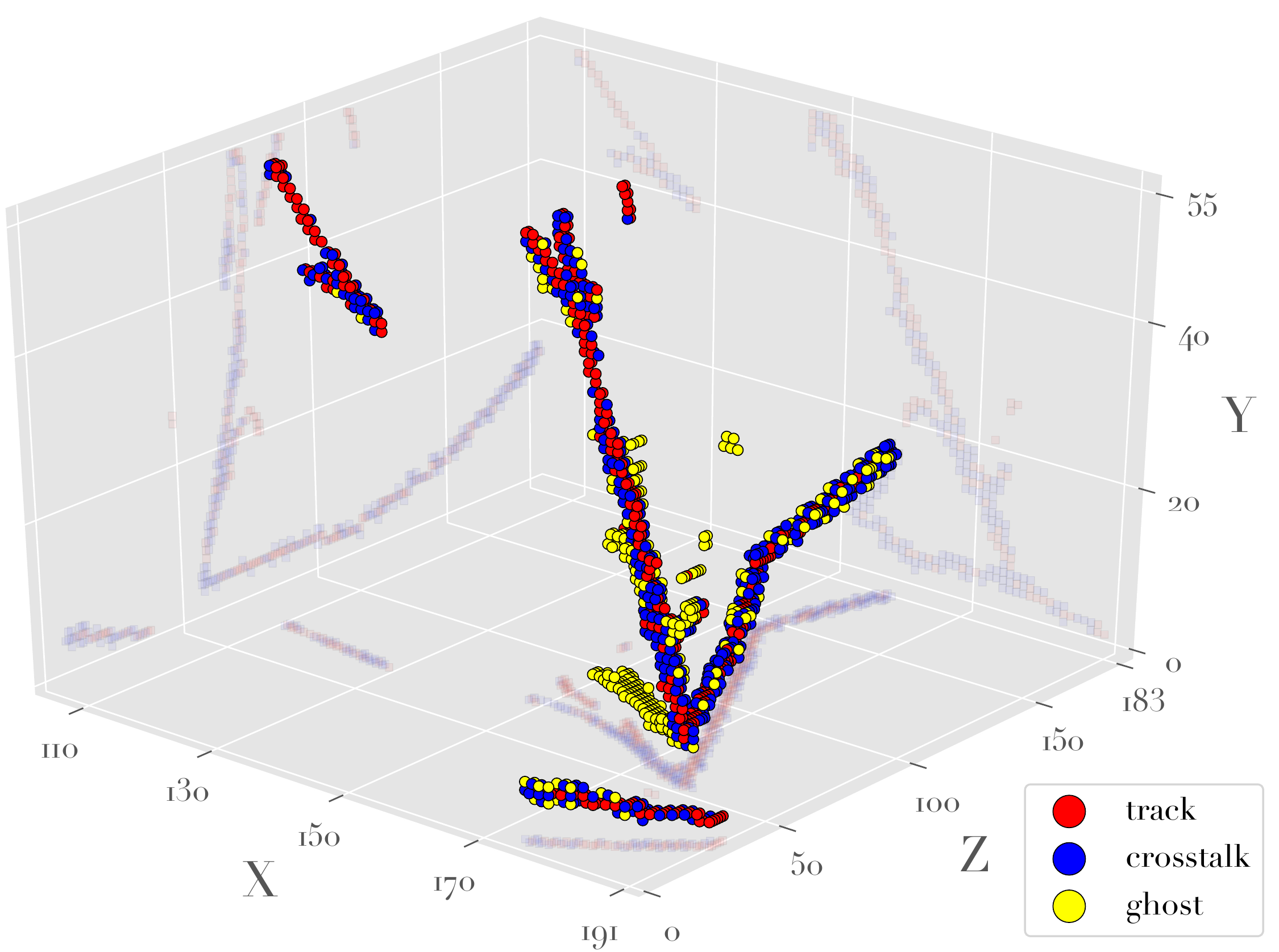}}
  \caption{Visualization of a neutrino interaction in a finely segmented 3D scintillator detector, demonstrating the relationship between the observed 2D projections onto the three orthogonal 2D views \review{(Fig.~\ref{fig:projections})}, the reconstructed 3D voxels \review{(Fig.~\ref{fig:3D})}, and the true classification of the voxels \review{(Fig.~\ref{fig:goal})}. \review{The energy of the incoming neutrino is 4.754~GeV. The axes are in cm.}
  }
  \label{fig:2Dto3D}
\end{figure*}

Here, we consider the Super Fine-Grained Detector (SuperFGD)~\cite{Abe:2019whr}\review{, which will be used in T2K,} as a specific case-study. The detector \review{will have 2 million} plastic scintillator cubes, each 1$\times$1$\times$1\,cm$^3$ in size, and provides three orthogonal 2D projections of particle tracks produced by a neutrino interaction, as depicted in Fig.~\ref{fig:projections}. 

\begin{figure}[htb]
    \centering
		\includegraphics[width=0.8\linewidth]{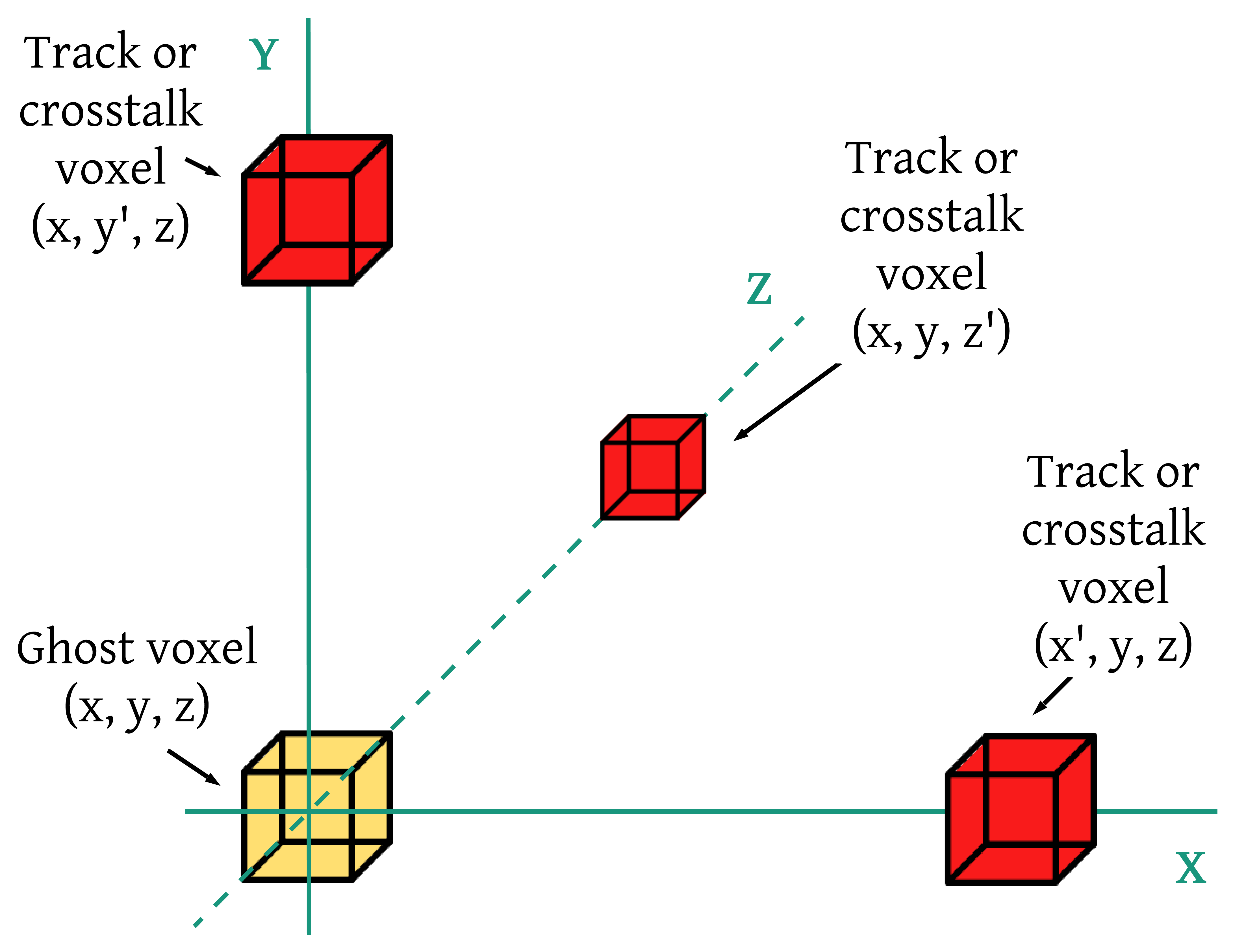}
	\caption{
	\review{Example of a ghost voxel arising from a 2D to 3D matching ambiguity. A 2D hit from each of the three track or crosstalk voxels (red) intersect generating a ghost voxel (yellow).}}
    \label{fig:ghost}
\end{figure}

To reconstruct neutrino interactions in three dimensions, the light yield measurements in the three 2D views are matched together, as shown in Fig.~\ref{fig:3D}. The 3D objects, corresponding to the cubes where the energy deposition is reconstructed, are referred to as \emph{voxels}. In addition to the cubes where a particle has passed and deposited energy, light-leakage between neighboring cubes can create additional \emph{crosstalk} signals \cite{blondel2020superfgd,mineev:2019beamtest}, as depicted in Fig.~\ref{fig:xtalk}. Moreover, ambiguities in the matching process can give rise to \emph{ghost} voxels, shown in Fig.~\ref{fig:ghost}.

To accurately reconstruct neutrino interactions in these detectors, it is crucial to be able to classify each voxel as one of the three types:
\begin{itemize}

\item  \textbf{Track}: \review{a voxel whose energy deposit comes, partially or totally, from scintillation light generated in that same cube.}
\item  \textbf{Crosstalk}: \review{ a voxel whose energy deposit comes exclusively from light-leakage from neighboring cubes.}
\item \textbf{Ghost}: \review{ a voxel with no physical energy deposit with an apparent signal arising from ambiguities when matching the three 2D views into 3D.}
\end{itemize}

Figure~\ref{fig:goal} shows the three types of voxels using truth information after 3D matching has been performed for an example neutrino interaction. Once these voxels are \review{properly labeled (by a classification algorithm)}, the ghost voxels can be removed before the full event reconstruction proceeds, while simultaneously cleaning the particle tracks of crosstalk.

In this article, we represent the voxels as nodes in a graph and classify the signals using a deep-learning technique based on a GNN. The abstract data representation provided by graphs makes this method very versatile and applicable to any experiment where the output data from the detector elements can be represented as a list of features with arbitrary dimensionality.

\review{In the case study presented here, focused on the SuperFGD detector of the T2K experiment, this method shows great potential to assist reconstruction by assigning a probability to each voxel as being track, crosstalk or ghost. Detector response simulations show that about half of the reconstructed voxels in the SuperFGD will not be of track type. For neutrino physics studies, both in terms of cross-sections and oscillation measurements, correct event topology identification and kinematic reconstruction of outgoing tracks are paramount. Ghost and crosstalk voxels, if not dealt with, will smear the track range estimations, and hence the reconstructed momentum and angle. Similarly, more densely populated events in term of voxels will merge highly colinear tracks and will make it more difficult to identify short tracks key for correct topology assignment. The method presented here is therefore expected to benefit future physics measurements in T2K and in any other experiments with similar conceptual challenges as the ones here described. 
}

\review{\section{Deep-learning methods}}
\label{sec:background}
\review{\subsection{Convolutional neural networks and data sparsity}}
\label{sec:introNN}

Deep-learning techniques are now commonly applied within the field of neutrino physics. In particular, Convolutional Neural Network (CNN)~\cite{firstCNN} algorithms that operate on two-dimensional images of the neutrino interactions have been very successful in a number of tasks, such as event classification~\cite{novacvn,Acciarri:2016ryt,duneTDR,dunecvn,Adams-2020-cye,Kronmueller-2019-jzh,Li-2019-qgs}, \review{hit}-level identification of track-like (linear) and shower-like (locally dense) energy deposits~\cite{Abi:2017aow,PhysRevD.99.092001}\review{, or energy reconstruction~\cite{Baldi-2019-improved,Seong-2019-convolutional,Huennefeld-2019-rrf}}. \review{Despite the success of CNNs in the neutrino world}, images of neutrino interactions are typically very sparse as only those readout channels with a detected signal \review{contribute} non-zero values \review{to the images}, and in the case of the detector presented in Sec.~\ref{sec:motivation} the average occupancy of the detector for a neutrino interaction is less than 0.02\%. Thus, much of the computation time is spent unnecessarily applying convolutions \review{to empty regions of the images.}

The goal of this work is to classify 3D voxels as one of three categories (track, crosstalk or ghost), which is natively a three dimensional problem. To apply a 3D CNN-based algorithm to this detector would require two million voxels to avoid any downsampling or cropping of the input data, which is computationally prohibitive.
\review{A popular approach to deal with the sparsity of neutrino interactions is the submanifold sparse convolutional network (SSCN)~\citep{Graham-2017-submanifold}. Standard ``dense'' CNNs are very inefficient when applied on images of neutrino interactions, whereas SSCNs require considerably less computation and report almost identical (or even better) results in terms of accuracy~\citep{Graham-2017-3d}. Some neutrino experiments have improved their reconstruction deep-learning algorithms by moving to SSCNs. For example, MicroBooNE recently updated the implementation of their semantic segmentation CNN~\citep{PhysRevD.99.092001} to an SCNN-based model~\citep{MicroBooNE-2020-semantic}, reporting improvements at inference by a factor of 354 and 33 in memory and wall-time, respectively. The NEXT collaboration are also exploring the idea of using an SSCN for track classification~\citep{nextcollaboration2020demonstration}.}

\review{\subsection{Graph neural networks}}

\begin{figure}[htb]
    \centering
		\includegraphics[width=1.0\linewidth]{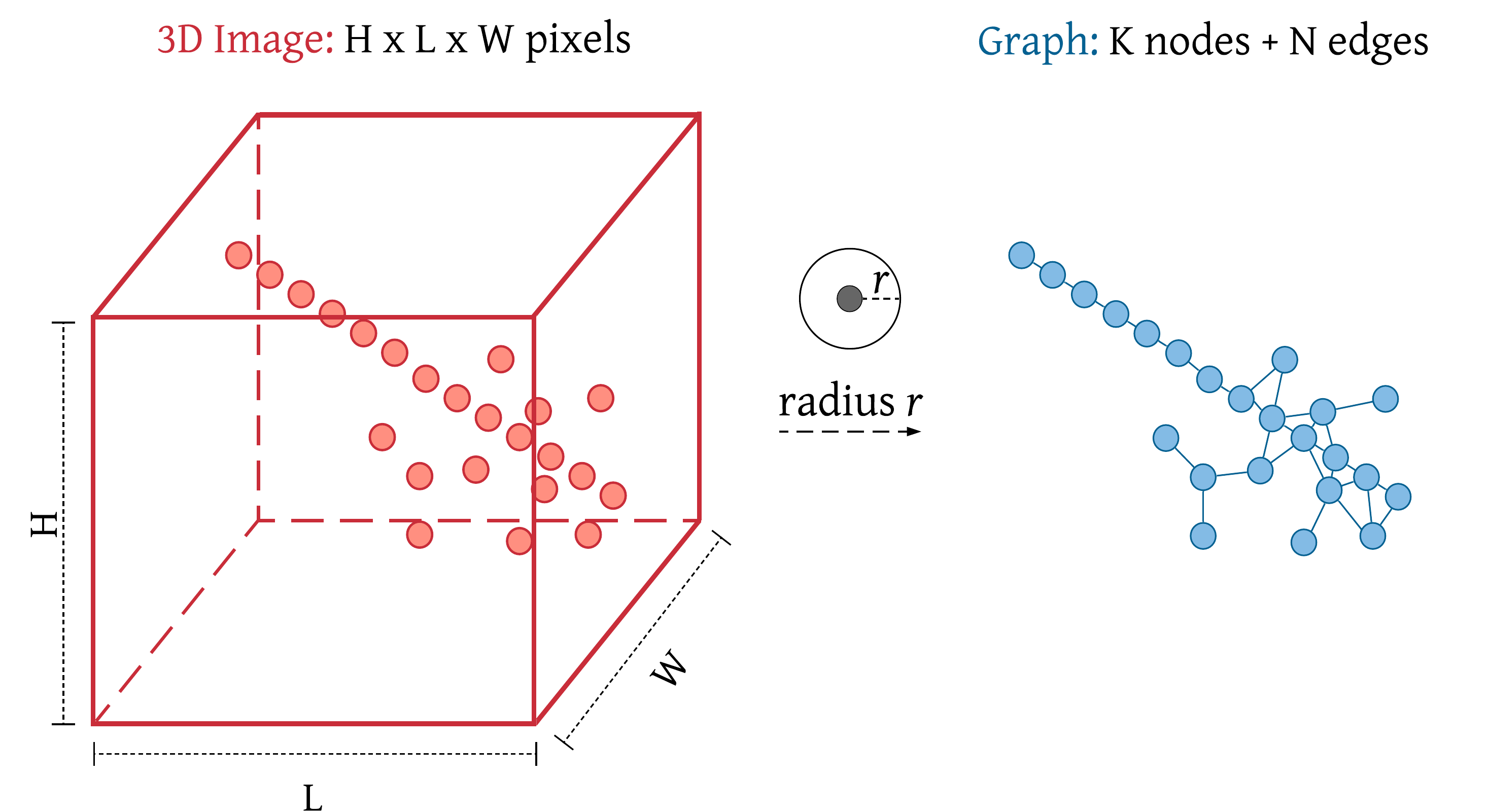}
	\caption{Data and computation size comparison between a 3D image and a graph. The size of the 3D image on the left is fixed ($H \times L \times W$) regardless of the number of hits as CNNs require fixed image sizes \review{(in most cases)}. The connected graph shown on the right is a much more efficient representation of the data. Each hit is represented as a graph node and connections, called edges, are made between neighboring hits within a sphere of radius $r$.}
    \label{fig:3dvsgraph}
\end{figure}

\review{An alternative approach for handling with sparse data is to represent hits (or voxels) as nodes in a graph.} In computer science, a graph $\G$ is a data structure that represents a mathematical concept consisting of \review{nodes}~$\V$ and edges~$\E$:
\begin{equation}
    \G = (\V,\E).
\end{equation}
A graph can be directed, where each edge has a starting and an ending \review{node} that define a direction, or undirected, where the edge simply connects two \review{nodes} without inducing a sense of direction. In our case, we use an undirected graph, since we are only interested in the spatial connections between \review{nodes}. Figure~\ref{fig:3dvsgraph} shows a comparison of the 3D CNN and graph data structures, as well as the radial search method used for defining edges between nodes.

As mentioned above, each detector \review{voxel} cube is represented as a node in a graph, and each node consists of a list of input variables called features that describe the physical properties of the detected signal (see Section~\ref{sec:methodology} and Appendix~\ref{sec:inputvariables}). The deep-learning algorithm that operates on graphs is the Graph Neural Network (GNN)~\cite{SperdutiFirstGNN,zhou2018graph}. GNNs are used in many different fields~\cite{NIPS2015_5954,Shen_2018_ECCV} and can be applied for graph classification~\cite{wang2018dynamic,NIPS2018_7729} or node classification~\cite{10.1145/2623330.2623732,kipf2016semisupervised,hamilton2017inductive}. In this article, a GNN inspired by the GraphSAGE algorithm~\cite{hamilton2017inductive} is used to classify individual voxels in SuperFGD events. The application of GNNs to data from neutrino experiments has been recently demonstrated by the IceCube experiment in order to identify entire events as atmospheric neutrino interactions, outperforming a 3D CNN~\cite{icecubeGCN}. \review{The work in Ref.~\citep{Drielsma-2020-clustering} also shows an application of GNNs for both node and edge classification for a neutrino detector, where a GNN-based reconstruction chain is used for clustering both electromagnetic showers and particle interactions.} Other GNN-based studies have been performed for particle reconstruction in high energy physics detectors~\cite{farrell2018novel,Qasim_2019,ju2020graph}. \review{The main drawback of GNNs with respect to SSCNs is that the former needs to pre-process the events to perform the neighborhood computation (defining edges) whilst no pre-processing is needed for the SSCN images. However, the advantage of GNNs in this field is that they can use a strong node representation, where a large number of features can define each node without reducing the scalability of the model.} To the best of our knowledge, the approach we present in this paper \review{is one of the first attempts of using} GNNs for node classification in neutrino experiments.

\subsection{GraphSAGE}
\label{sec:graphsage}

GraphSAGE~\cite{hamilton2017inductive} is a technique that leverages the features of graph nodes $\V$ - which can range from physical information to text attributes - to generate efficient representations on previously unseen samples by learning aggregator functions from training nodes. These aggregators can be simple functions (e.g., mean or maximum) or more complex ones, such as \review{Long short-term memory} (LSTM) cells~\cite{LSTM}, and must be functions that take an arbitrary number of inputs without any given order. The model learns not only $K$ aggregator functions that combine information from neighboring nodes but also a set of weight matrices $\mb{W}^{k}, \forall k \in \{1,...,K\}$, which are used to propagate information through the $K$ layers of the model and combines local information of the node with the aggregator information of its neighbors into an encoding vector (see Algorithm~\ref{alg:graphsage}). The number of aggregator functions is also used to define the depth of the model, meaning that a GraphSAGE model has a depth of $K$. \review{In each layer of the aggregator information, a new representation of the node $v$ is computed, denoted by $\mb{h}^k_v$ (with $\mb{h}^0_v$ being the initial node features $\mb{x}_v$).} Once trained, it can produce the embedding of a new node given its input features and neighborhood\review{, in the form of the vector of the last layer $\mb{h}^K_v$}; this embedding is then used as the input of a multilayer perceptron (MLP)~\cite{Rosenblatt1957} that is responsible for predicting the label.

\begin{algorithm}[htb]
\caption{GraphSAGE embedding generation (i.e., forward propagation) algorithm (from~\cite{hamilton2017inductive})}
\label{alg:graphsage}
        \SetKwInOut{Input}{Input}\SetKwInOut{Output}{Output}
    \Input{~Graph $\G(\V,\E)$; input features $\{\mb{x}_v, \forall v\in \V\}$; depth $K$; weight matrices $\mb{W}^{k}, \forall k \in \{1,...,K\}$; non-linearity $\sigma$; differentiable aggregator functions $\textsc{aggregate}_k, \forall k \in \{1,...,K\}$; neighborhood function $\mathcal{N} : v \rightarrow 2^{\V}$}
    \Output{~Vector representations $\mb{z}_v$ for all $v \in \V$}
    \BlankLine
    $\mb{h}^0_v \leftarrow \mb{x}_v, \forall v \in \V$ \;
    \For{$k=1...K$}{
          \For{$v \in \V$}{
          $\mb{h}^{k}_{\mathcal{N}(v)} \leftarrow \textsc{aggregate}_k(\{\mb{h}_u^{k-1}, \forall u \in \mathcal{N}(v)\})$\;
                        $\mb{h}^k_v \leftarrow \sigma\left(\mb{W}^{k}\cdot\textsc{concat}(\mb{h}_v^{k-1}, \mb{h}^{k}_{\mathcal{N}(v)})\right)$
          }
          $\mb{h}^{k}_v\leftarrow \mb{h}^{k}_v/ \|\mb{h}^{k}_v\|_2, \forall v \in \V$
        }
     $\mb{z}_v\leftarrow \mb{h}^{K}_v, \forall v \in \V$
\end{algorithm}

Since GraphSAGE learns from node features, it allows us to decide which physical information to use for each voxel. This means that the model can follow the particle \review{set}, i.e., by predicting the label for each voxel based on the physical attributes of the target voxel as well as the features of its neighbors.

\section{Methodology}
\label{sec:methodology}
\subsection{Data sample generation}\label{sec:generation}

In order to generate data sets of neutrino interactions with true labels that allow to train and benchmark the classification algorithm, the steps below are followed. For each neutrino interaction:
\begin{enumerate}
    \item Initial particle types and initial kinematics are specified for all final-state particles produced in the interaction.
    \item Initial particles are propagated through the detector geometry producing further particles and leaving signals in the form of energy deposits.
    \item Using particle energy deposits, the detector response is simulated.
    \item The information is stored as a list of voxels with a \review{unique integer} known true label\review{: track, crosstalk or ghost}.
\end{enumerate}

\noindent\textbf{Initial particle types and kinematics}\\
The initial particle types and their associated kinematics were simulated following two approaches. Firstly, GENIE datasets were created using \texttt{GENIE-G18.10b} neutrino interaction software~\cite{Andreopoulos:2009rq}. For a given neutrino flux\footnote{We used the T2K flux, which peaks at 600\,MeV/c, see Ref.~\cite{Abe:2012av}.} and target geometry specification, it generates a list of realistic neutrino event interactions both in the number and type of outgoing particles, often referred to as event topologies, and in their individual initial kinematics. Secondly,
\review{Particle bomb (P-Bomb) datasets have been constructed as a complementary group of data not affected by the specific tunings provided by a neutrino interaction generator, such as GENIE. The motivation underlying this is to show in the later sections that the reported algorithm performance is not highly dependent on the neutrino interaction modelling. Moreover, given that neutrino generators do not perfectly model real interactions, these two datasets (GENIE and P-Bomb) are also used in the following sections to discuss the reliability of training in GENIE (or other generators) and classifying real data. Hence the purpose of P-Bomb datasets is to provide events similar to those found in neutrino interactions in terms of the outgoing particles but with no realistic kinematic modeling and without considering any kinematic correlations among the outgoing particles.  
To achieve this the P-Bomb dataset is constructed adding equal numbers of events with the following particle gun combinations, each of which has random flat solid angle and momentum [10-1000 MeV/c] distributions: 1 $\mu^{-}$; 1 $\mu^{-}$ and 1 proton; 1 $\mu^{-}$ and 1 $\pi^{-}$; 1 $\mu^{-}$ and 1 $\pi^{+}$;  1 $\mu^{-}$ and 2 protons;  and 1 $\mu^{-}$, 1 $\pi^{+}$ and 3 protons. 
An illustrative comparison between GENIE and P-Bomb neutrino interaction modelling can be found in Appendix~\ref{sec:datasets_distributions}.}
A summary regarding the number of events and voxels in the two datasets, as well as of the class distribution is presented in Tab.~\ref{tab:datasets}.

\begin{table}[htb]
    \centering
    \begin{tabular}{l|lccc}
    \hline \hline
    \multirowcell{5}{\textbf{GENIE}\\\textbf{dataset}} & \rule{0pt}{12pt} & \textbf{Training} & \textbf{Validation} & \textbf{Testing}\\
    & \# Events & 6k & 2k & 11.5k\\
    & \# Voxels & 1.83M & 606.7k & 3.58M\\
    & \rule{0pt}{12pt} & \textbf{Track} & \textbf{Crosstalk} & \textbf{Ghost}\\
    & Fraction & 43\% & 37\% & 20\%\\
    \hline
    \multirowcell{5}{\textbf{P-Bomb}\\\textbf{dataset}} & \rule{0pt}{12pt} & \textbf{Training} & \textbf{Validation} & \textbf{Testing}\\
    & \# Events & 6k & 2k & 39.5k\\
    & \# Voxels & 1.84M & 618k & 12.3M\\
    & \rule{0pt}{12pt} & \textbf{Track} & \textbf{Crosstalk} & \textbf{Ghost}\\
    & Fraction & 49\% & 38\% & 13\%\\
    \hline \hline
    \end{tabular}
    \caption{Descriptions of both GENIE and P-Bomb datasets, displaying the number of events and number of voxels used for training, validating and testing the models. Additionally the fractions of the different classes of voxels are shown\review{, which are conserved through the training, validating, and testing sets}.}
    \label{tab:datasets}
\end{table}

\noindent\textbf{Particle propagation simulation in the detector}\\
The SuperFGD detector geometry was simulated as described in Ref.~\cite{Abe:2019whr}. The particle propagation and physics simulation is done by means of \texttt{GEANT v4-10.6.1}~\cite{Agostinelli:2002hh}. GEANT is a Monte Carlo based toolkit that provides realistic propagation of particles through matter. It outputs a list of energy deposits.

All energy deposits\footnote{Only signals in the first 100\,ns are considered. Further delayed signals, such as decays, can be treated as independent graphs.} occurring in the same detector cube, including the effect of Birks’ quenching~\cite{Birks_1951},  are summed to form the list of \textit{track voxels}. To simulate imperfect cube light-tightness, the 3D voxelized energy is then shared with the neighboring cubes, creating a new set of voxels that originally had no energy deposits, the \textit{crosstalk voxels} (see Figure~\ref{fig:xtalk}). \review{For the energy sharing, a fraction of the energy in the original cube is leaked into each of its six neighbors. The fraction that is shared is sampled from a Poisson distribution, with $\mu=2.7\%$. Given that the probability for the energy to leak twice is $O(\mu^2)$, only leakage to immediate neighbors is considered.}
The 3D voxelized energy of both track and crosstalk voxels is projected onto its three orthogonal planes where the detector 2D signals are simulated, converting the continuous energy deposit into discretized photons\footnote{\review{No waveform processing is simulated. A single conversion factor is used from energy deposit to number of photons in the WLS fiber, based on laboratory data \cite{blondel2020superfgd}.}}, weighted by distance-dependent attenuation factors, which are detected with 35\% probability. To mimic a minimum threshold detection sensitivity, only 2D hits with three or more detected photons are kept. \review{SuperFGD thresholds at this level are expected to remove virtually all dark rate hits \cite{blondel2020superfgd}, henceforth we have not included noise hits in our simulation.} Then, the 2D hits are matched into 3D reconstructed voxels only if the same XYZ coordinate combination can be made using two different combinations of 2D planes. In this process, due to ambiguities some extra voxels are created, the \textit{ghost voxels} (see Fig.~\ref{fig:ghost}). Finally, those track and crosstalk voxels not reconstructed after the 3D matching are discarded from the original lists. An example of the 2D to 3D reconstruction is shown in Figures~\ref{fig:projections} and~\ref{fig:3D}.
\\

\noindent\textbf{Simulation output} \\
The resulting output from the simulation is a list of voxels and their associated energy deposits in the three planes, each with one of the following three labels that we want to classify, as described in Sec.~\ref{sec:motivation}: track, crosstalk or ghost voxel. \review{Using the list of voxels of each event, further features are computed for each voxel as described in Appendix~\ref{sec:inputvariables}. The correlation matrices of the features for the GENIE and P-Bomb datasets are presented in Fig.~\ref{fig:correlationMatrices} in Appendix~\ref{sec:correlations}. The graph's adjacency matrix is built utilizing the position of each voxel, as will be detailed below. Both the new list of expanded voxel features plus the corresponding adjacency matrix of the event are fed into the GNN algorithm.}

\subsection{Network architecture}

\begin{figure*}[htb] 
	\subcaptionbox{Sample neighborhood  \review{ (orange) for a depth $K=3$. The red node indicates the target node to be classified; orange nodes are nodes taken into account for the classification given a depth of three; blue nodes are not taken into account. Each depth $k$ represents a distance level from the target to the corresponding neighbor set.}}%
  [.3\linewidth]{\includegraphics[width=5.25cm]{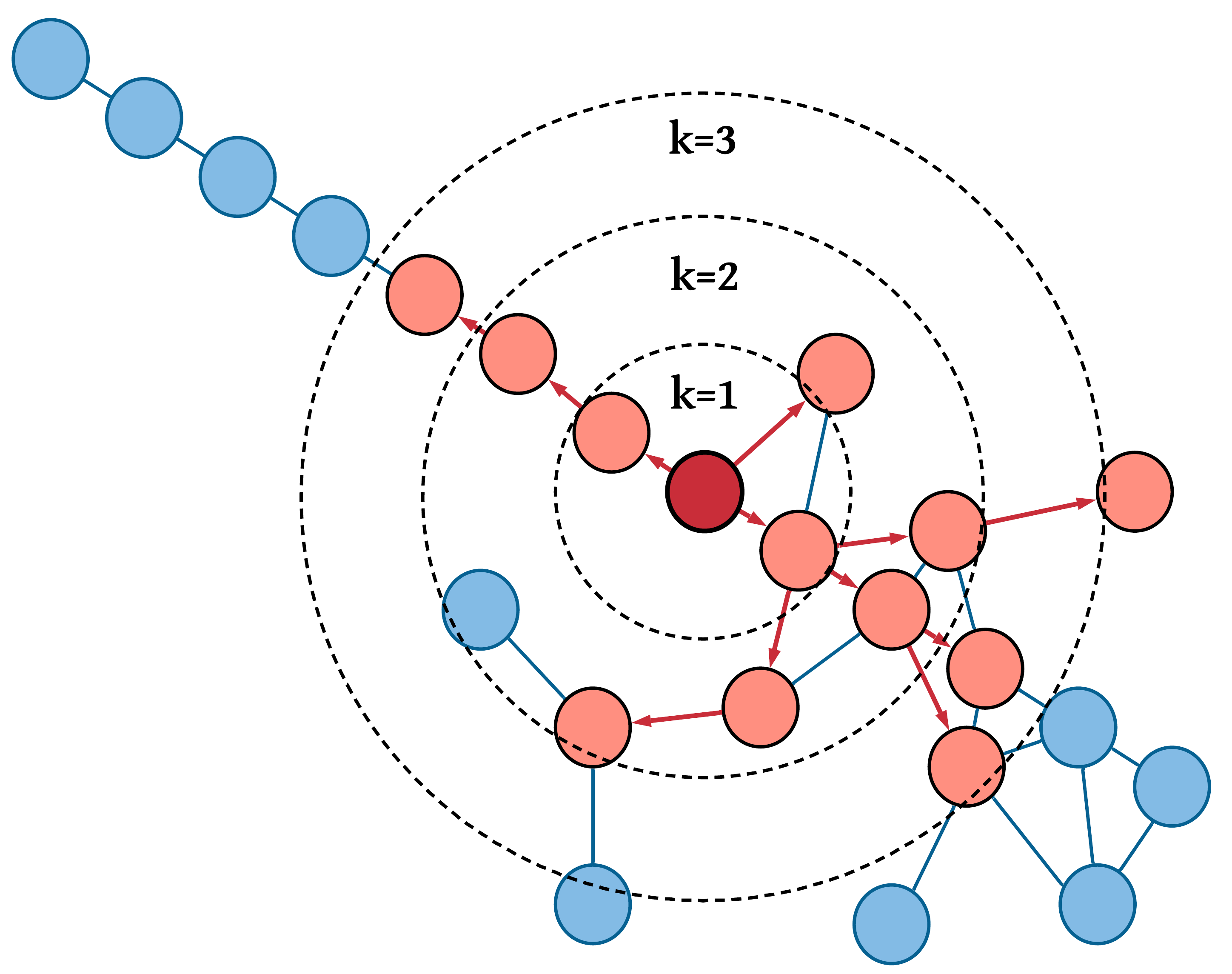}}
  \subcaptionbox{Aggregate feature information from neighbors. \review{Each node has features $f_1,\dots,f_N$, with purple, blue, and green nodes information being aggregated through aggregators 1, 2, and 3, respectively.}}%
  [.3\linewidth]{\includegraphics[width=5.25cm]{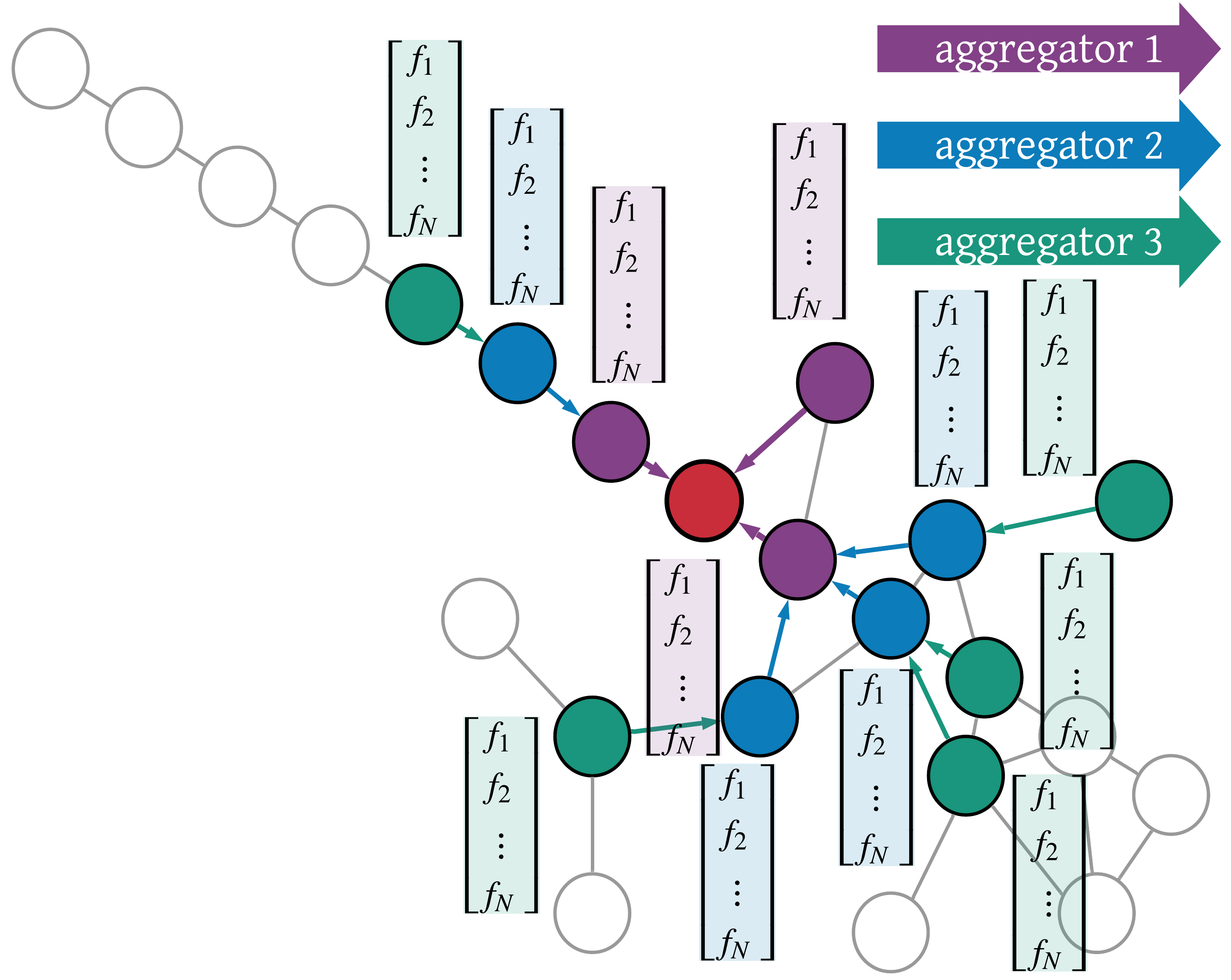}}
  \subcaptionbox{Use aggregated information as input for the fully connected layers and predict the label.}%
  [.3\linewidth]{\includegraphics[width=5.25cm]{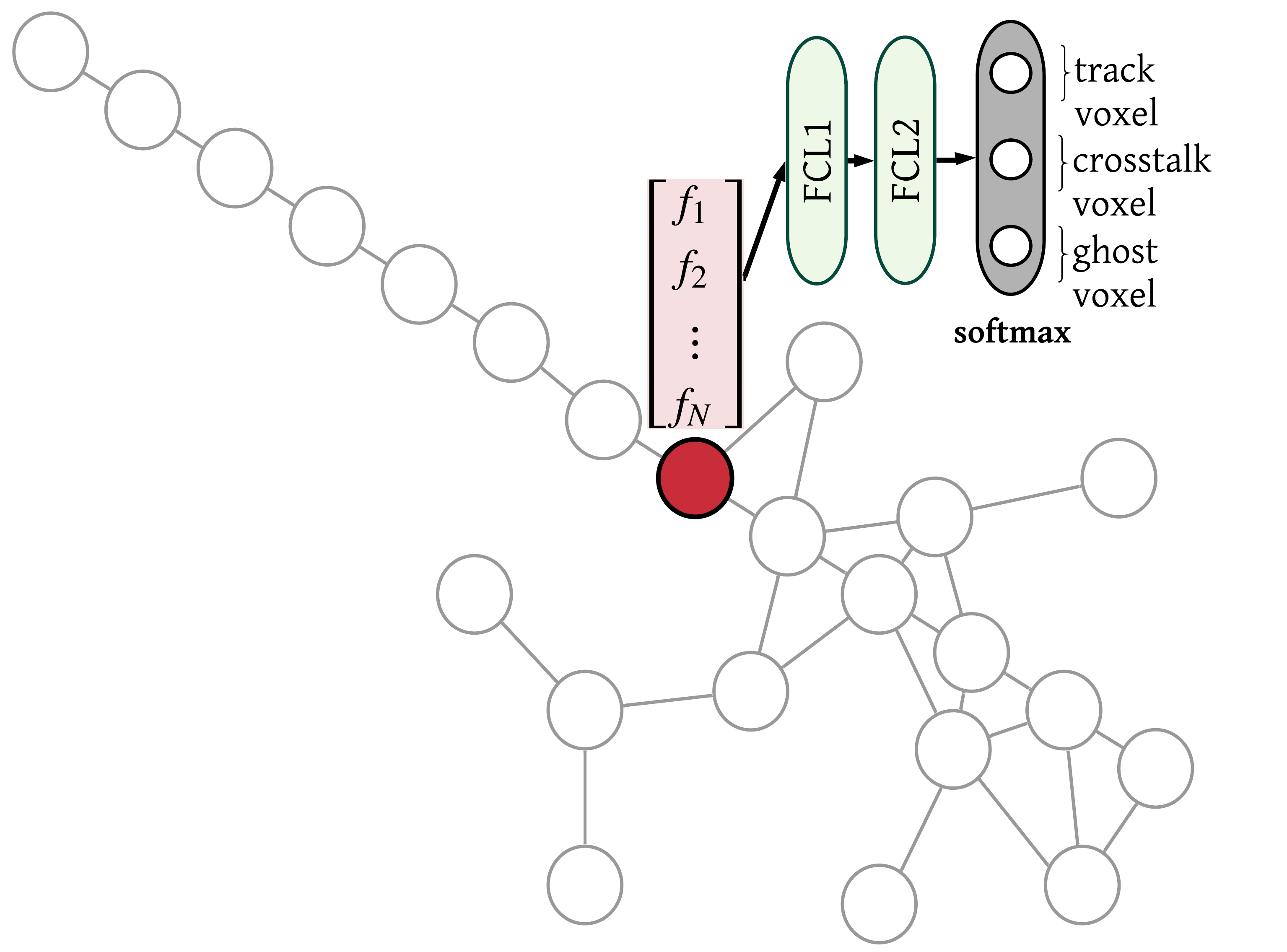}}
  \caption{Visual illustration of the GraphSAGE sample and aggregate approach with a depth of three \cite{hamilton2017inductive}.}
  \label{fig:graphsage}
\end{figure*}

Each graph in GraphSAGE is constructed using the proximity of two voxels in that graph. If both voxels are spatially located within a radius of 1.75\,cm\footnote{To link only those voxels within the 3$\times$3$\times$3 cube of voxels centred on the target voxel (the maximum diagonal distance from the center of this cube is $\sqrt{1^2+1^2+1^2}\approx 1.75$).}, then we consider them to be connected in the graph by an edge; we repeat the same procedure for each pair of voxels\footnote{If a voxel has no neighbors, it is discarded from the graph and cannot be classified; this happens for less than 0.6\% of the total number of voxels.}. Additionally, we consider a neighborhood depth of three, i.e., to produce the embedding of a voxel, we use the voxel features together with its first neighbors' features, the features of the neighbors of its neighbors, i.e, second neighbors' features, and the features of the neighbors of the neighbors of its neighbors, i.e., third neighbors' features. The aggregator used to combine the feature of the neighbors is the mean aggregator, which produces the average of the neighbors' values. This final embedding is then passed to an MLP consisting of two fully connected layers - each followed by a LeakyReLU activation function - and a final output layer followed by a softmax activation function. Figure~\ref{fig:graphsage} illustrates the GraphSAGE-based approach used, while Tab.~\ref{tab:architecture} shows the architectural parameters chosen. Categorical cross-entropy is chosen as the loss function to minimize during training, as it is considered the standard one for multi-class classification problems\review{, where each training example corresponds to a voxel}:
\begin{equation}
    J = - \frac{1} m \sum_{i=1}^{m} \sum_{j=1}^{c} y_{j}^{(i)} \log \hat{y}_{j}^{(i)},
\end{equation}
where:
    \begin{itemize}
        \item $\bm{y}^{(k)}$: true values corresponding to the \textit{k\textsuperscript{th}} training example. \review{$\bm{y}^{(k)}$ is a vector with all components equal to zero except for the class $j$, which is equal to one.}
        \item $\bm{\hat{y}}^{(k)}$: predicted values corresponding to the \textit{k\textsuperscript{th}} training example. \review{$\bm{\hat{y}}^{(k)}$ is a vector with each component $\hat{y}_{j}^{(k)}$ denoting the score (continuous value from 0 to 1) of being of class $j$.}
        \item $m$: number of training examples\review{, equal to the total number of voxels in the training sample.}
        \item $c$: number of classes/neurons corresponding to the output. In this case, the three classes are: track, crosstalk, and ghost.
    \end{itemize}

\begin{table}[htb]
    \centering
    \begin{tabular}{lc}
    \hline \hline
    \rule{0pt}{8pt}Parameter & value\\ \hline
    \rule{0pt}{8pt}Encoding size       & 128\\
    Depth                               & 3 \\
    Aggregator                          & mean \\
    Fully Connected Layer 1             & 128 neurons \\
    Fully Connected Layer 2             & 128 neurons \\
    Fully Connected Layer 3 (output)    & 3 neurons \\
    \hline \hline
    \end{tabular}
    \caption{Architectural parameters; for more information about the meaning of the parameters, see Sec.~\ref{sec:graphsage}.}
    \label{tab:architecture}
\end{table}

The output layer of the model consists of three neurons, one for each of the three classes, with values $v_i$ for $i=1, 2, 3$. The sum of neuron values is given by $\sum_{i=1}^3 v_i = 1$ such that each neuron value gives a fractional score that can be used to classify voxels. In other words, the model returns scores for each voxel to be one of the three desired outputs, which can be interpreted as the probability: track-like, crosstalk-like, or ghost-like.

\subsection{Training}
\label{sec:gnn-training}

\begin{figure}[htb] 
	\includegraphics[width=0.95\linewidth]{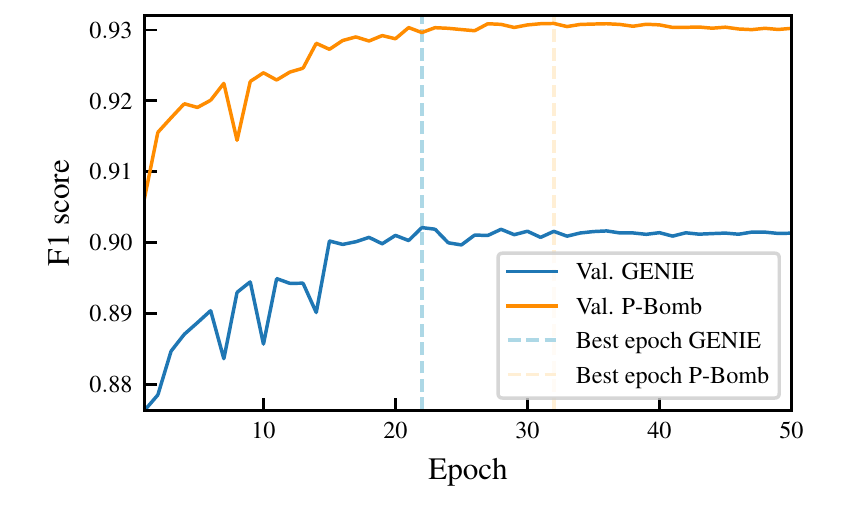}
	\caption{Validation F1 results on GENIE and P-Bomb samples.}
	\label{fig:validation}
\end{figure}

The network was trained for 50 epochs\footnote{Epoch: one forward pass and one backward pass of all the training examples. In other words, an epoch is one pass over the entire dataset.} using Python 3.6.9 and PyTorch 1.3.0~\cite{NEURIPS2019_9015} as the deep-learning framework, on an NVIDIA RTX 2080 Ti GPU. Adam~\cite{Kingma-2014-adam} is used as the optimizer, with a mini-batch size of 32, and an initial learning rate of 0.001 (divided by 10 when the error plateaus, as suggested in~\cite{He-et-al-2015-deep}). The model has a total of 105,347 parameters. \review{As is standard in machine learning, the dataset was split into three disjoint sets: the training set, to optimize the model's parameters; the validation set, to avoid overfitting and perform model selection; the test set, to verify the integrity of the model for new data.} Figure~\ref{fig:validation} shows the validation results during the training process, measured by the $F_1$-score metric:
\begin{align}
    F_1 = 2\;\frac{\mathrm{precision}\cdot \mathrm{recall}}{\mathrm{precision}+\mathrm{recall}}.
\end{align}
The precision and recall are defined as:
\begin{align}
    \mathrm{precision} &= \frac{\mathrm{true}_\mathrm{positives}}{\mathrm{true}_\mathrm{positives}+\mathrm{false}_\mathrm{positives}},\\
    \mathrm{recall} &= \frac{\mathrm{true}_\mathrm{positives}}{\mathrm{true}_\mathrm{positives}+\mathrm{true}_\mathrm{negatives}},
\end{align}
where the labels are compared as one class \review{(denoted as positive)} vs. all the others \review{(denoted as negative)}. The model used later for inference on new data is the one that maximizes the $F_1$-score for the validation set, as it has the best generalization for unseen data.

\subsection{Results}
\label{sec:results}

The GNN voxel-type predictions are compared against the true labels to evaluate the network performance and identify possible areas of improvement. Here, we choose the output class with the highest score as the predicted class of each voxel although, depending on the type of analysis, different selection criteria could be applied in the future.

The efficiencies and purities of these predictions are calculated by two methods: per voxel and per event, \review{the former are given by the following formulas for each type of voxels:}

\begin{align}
    \mathrm{efficiency}_\mathrm{i} &= \frac{\text{\# voxels with label}_\mathrm{true}\text{= label}_\mathrm{pred}\text{=i}}{\text{\# voxels with label}_\mathrm{true}\text{=i}},\\
    \mathrm{purity}_\mathrm{i} &= \frac{\text{\# voxels with label}_\mathrm{true}\text{= label}_\mathrm{pred}\text{=i}}{\text{\# voxels with label}_\mathrm{pred}\text{=i}}.
\end{align}

\review{The efficiencies and purities per event are defined as the mean of the efficiencies and purities of individual voxels for each event.} The results of both methods for four sets of training/testing samples are shown in Tab.~\ref{tab:eff&pur_table}, giving nearly identical performance that is independent of the dataset used to train and test the GNN.

\begin{figure*}[ht] 
    \subcaptionbox{
  Prediction: voxels are colored based on the GNN predictions. 
  \label{fig:prediction}}
  [.75\linewidth]{\includegraphics[width=13.25cm]{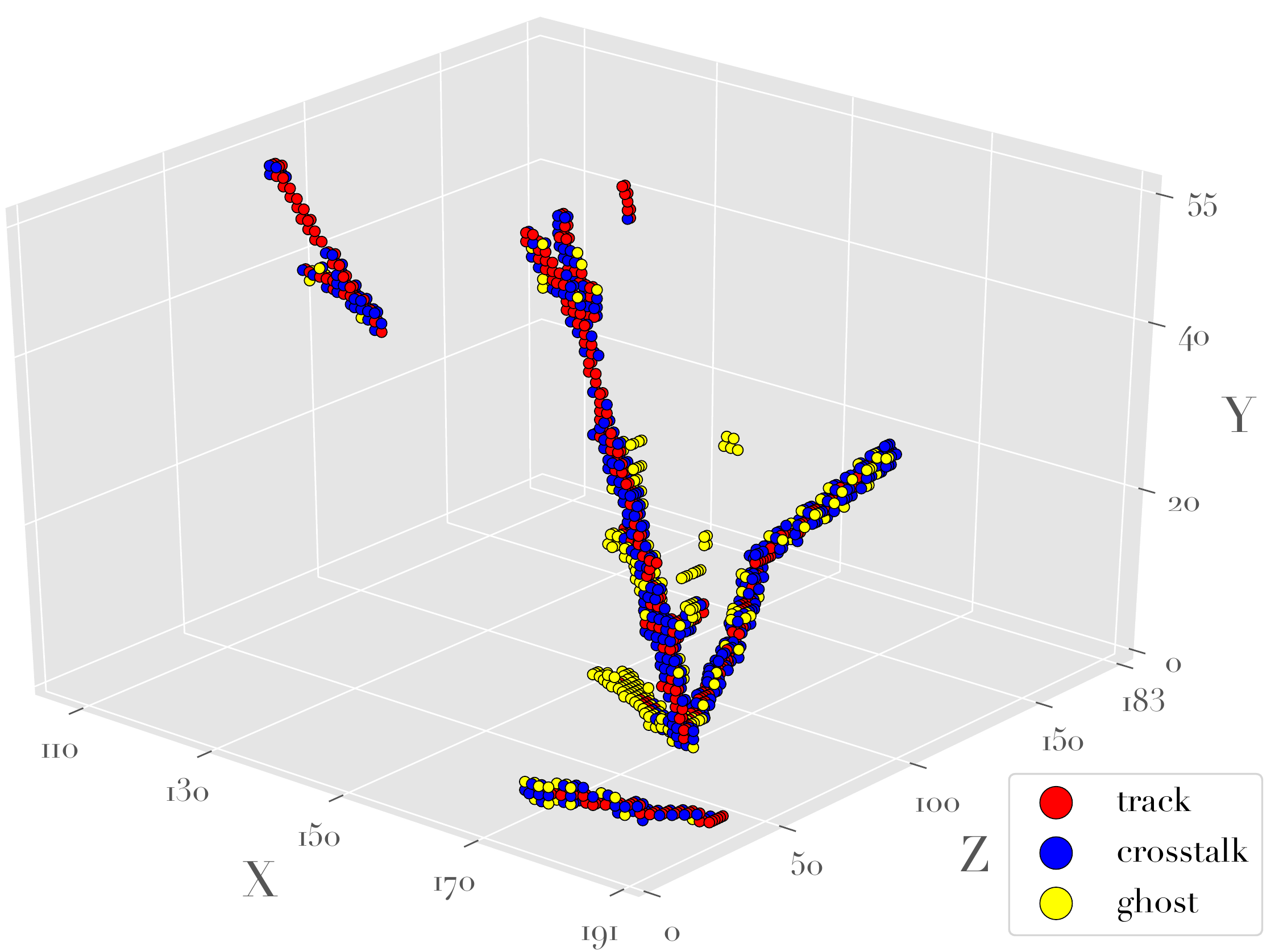}}
  \subcaptionbox{
  Accuracy: voxels correctly classified by the GNN are shown in green.
  \label{fig:accuracy}}
  [.75\linewidth]{\includegraphics[width=13.25cm]{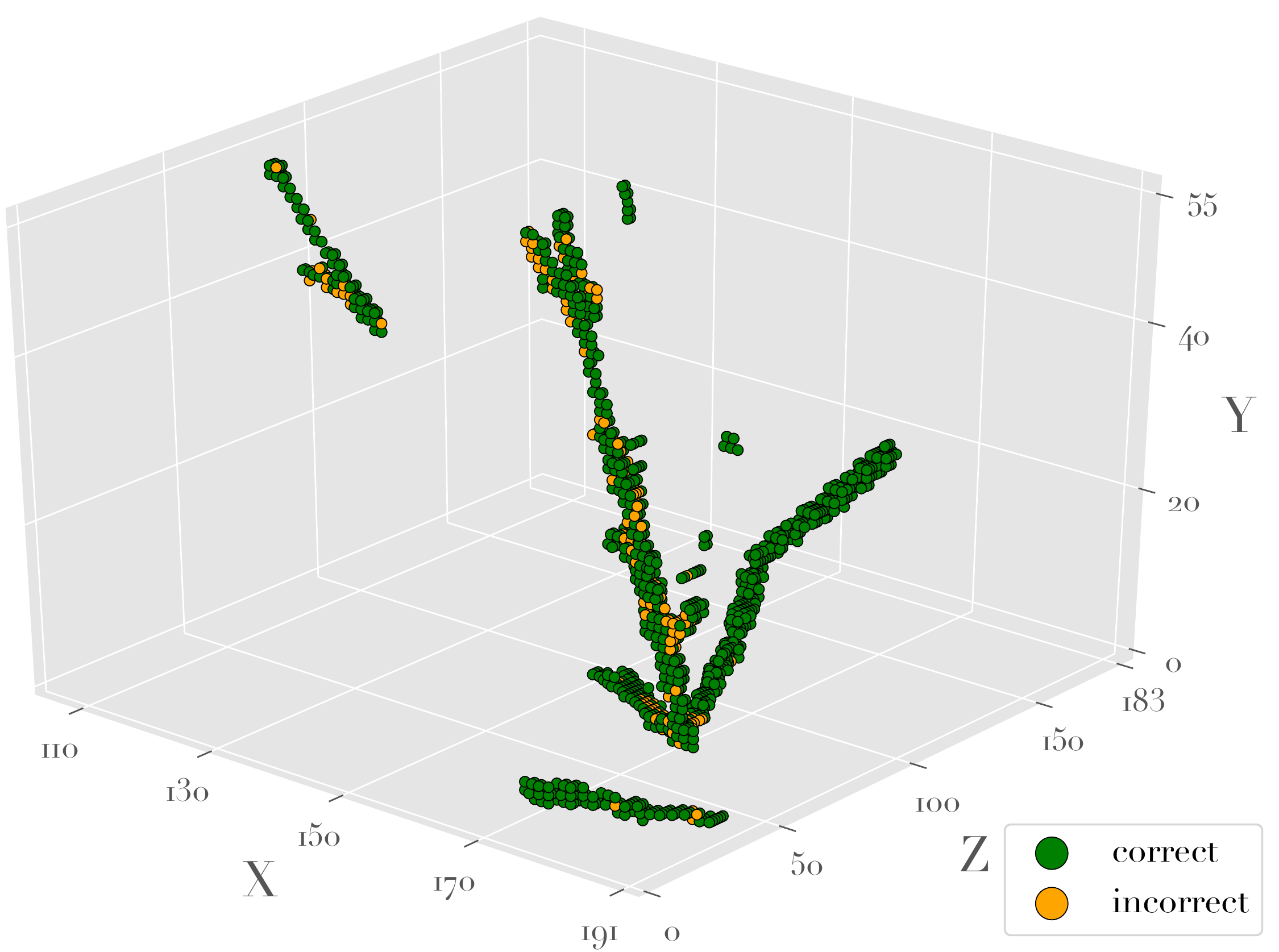}}
  \caption{Example GNN prediction results for the interaction shown in Fig.~\ref{fig:2Dto3D}. \review{The axes are in cm.}
 }
  \label{fig:results}
\end{figure*}

As an example, Fig.~\ref{fig:results} shows the voxel prediction results from the GNN when applied to the event shown in Fig.~\ref{fig:2Dto3D}, a GENIE event that features a track almost completely composed of ghost voxels. Figure~\ref{fig:prediction} shows the class predicted for each voxel, while Fig.~\ref{fig:accuracy} displays which voxels were correctly/incorrectly classified.

\begin{table*}[htb]
    \centering
    \begin{tabular}{l|l|lccc|lccc}
    \hline \hline
    \rule{0pt}{9pt} & & \multicolumn{4}{c|}{\textbf{GENIE Training}} & \multicolumn{4}{c}{\textbf{P-Bomb Training}}\\
    \hline
    \rule{0pt}{9pt}\multirowcell{6}{\textbf{GENIE}\\\textbf{Testing}} & \multirowcell{3}{\textbf{Per}\\\textbf{Voxel}} & & Track & Crosstalk & Ghost & & Track & Crosstalk & Ghost\\ 
    &&Efficiency & 93\% & 90\% & 84\% & Efficiency & 93\% & 89\% & 80\%\\
    &&Purity & 93\% & 87\% & 91\% & Purity & 91\% & 86\% & 89\%\\
    \cline{2-10}
    &\rule{0pt}{9pt}\multirowcell{3}{\textbf{Per}\\\textbf{Event}} & & Track & Crosstalk & Ghost & & Track & Crosstalk & Ghost\\ 
    &&Efficiency & 94\% & 94\% & 88\% & Efficiency & 94\% & 93\% & 88\%\\
    &&Purity & 96\% & 91\% & 92\% & Purity & 95\% & 91\% & 91\%\\
    \hline
    
    \rule{0pt}{9pt}\multirowcell{6}{\textbf{P-Bomb}\\\textbf{Testing}} & \multirowcell{3}{\textbf{Per}\\\textbf{Voxel}} & & Track & Crosstalk & Ghost & & Track & Crosstalk & Ghost\\ 
    &&Efficiency & 94\% & 93\% & 87\% & Efficiency & 95\% & 93\% & 88\%\\
    &&Purity & 95\% & 90\% & 92\% & Purity & 95\% & 91\% & 92\%\\
    \cline{2-10}
    &\rule{0pt}{9pt}\multirowcell{3}{\textbf{Per}\\\textbf{Event}} & & Track & Crosstalk & Ghost & & Track & Crosstalk & Ghost\\
    &&Efficiency & 94\% & 94\% & 87\% & Efficiency & 95\% & 93\% & 88\%\\
    &&Purity & 96\% & 90\% & 92\% & Purity & 96\% & 91\% & 92\%\\
    \hline \hline
    \end{tabular}
    \caption{Mean efficiencies and purities of voxel classification, calculated for the whole sample (per voxel) and as a mean of the event-by-event efficiencies and purities (per event).}
    \label{tab:eff&pur_table}
\end{table*}

A more in-depth analysis of the GNN performance can be carried out by studying the effects of different event properties on the efficiencies and purities of the predictions. For these studies, the results of the GNN trained and tested on the GENIE dataset are used. 

\begin{figure}[htb] 
	\begin{subfigure}{1.0\columnwidth}
		\includegraphics[width=0.95\linewidth]{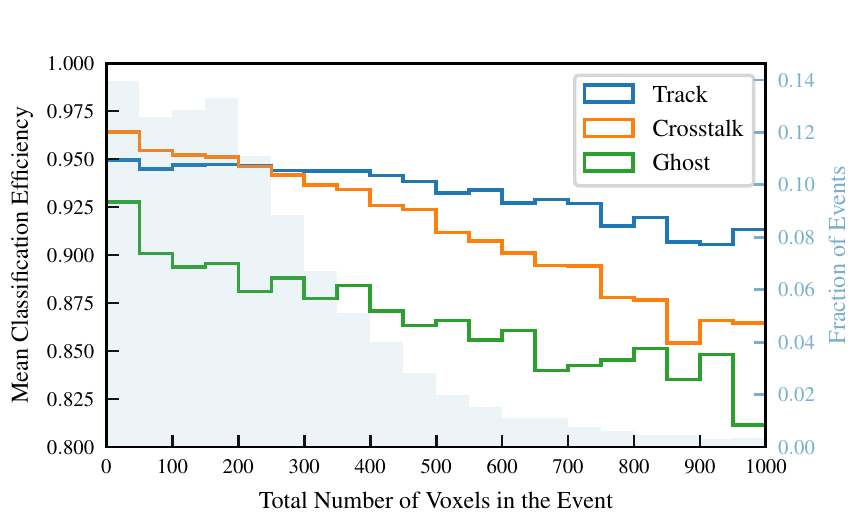}
		\caption{Efficiency.} 
		\label{fig:eff_NumOfVox}
	\end{subfigure}
	\begin{subfigure}{1.0\columnwidth}
		\includegraphics[width=0.95\linewidth]{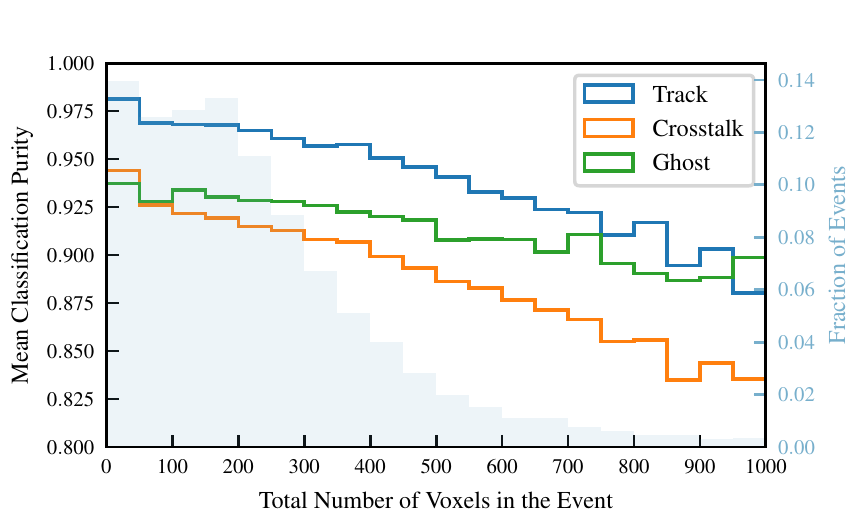}
		\caption{Purity.} 
		\label{fig:pur_NumOfVox}
	\end{subfigure}
	\begin{subfigure}{1.0\columnwidth}
		\includegraphics[width=0.95\linewidth]{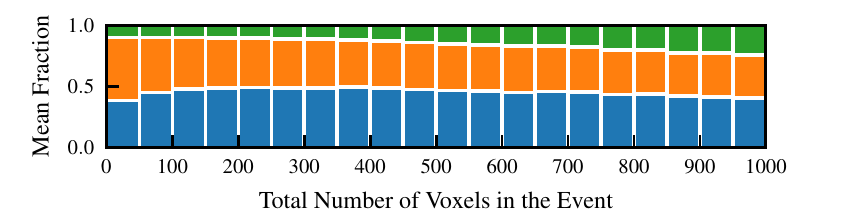}
		\caption{Mean fraction of each type of voxel as a function of the number of voxels in the event (blue~=~track, orange~=~crosstalk, green~=~ghost).} 
		\label{fig:type_NumOfVox}
	\end{subfigure}
	\caption{Efficiency and purity as a function of the number of voxels in the event for a sample trained and tested on GENIE simulated data.}
	\label{fig:eff/pur_NumOfVox}
\end{figure}

One of the factors expected to affect these predictions is the number of voxels in the event. Figure~\ref{fig:eff/pur_NumOfVox} shows the relationship between the mean efficiency and purity per event for each type of voxel as a function of the total number of voxels in the event. The figure also shows the mean number of events in each bin (in light blue). It is clear that both the efficiencies and purities of the three types of voxels decrease as the number of voxels in the event increases. This decrease is coupled with an increase of the fraction of ghost voxels as the total number of voxels increases, which are the hardest for the GNN to classify.

\begin{figure}[htb] 
	\begin{subfigure}{1.0\columnwidth}
		\includegraphics[width=0.95\linewidth]{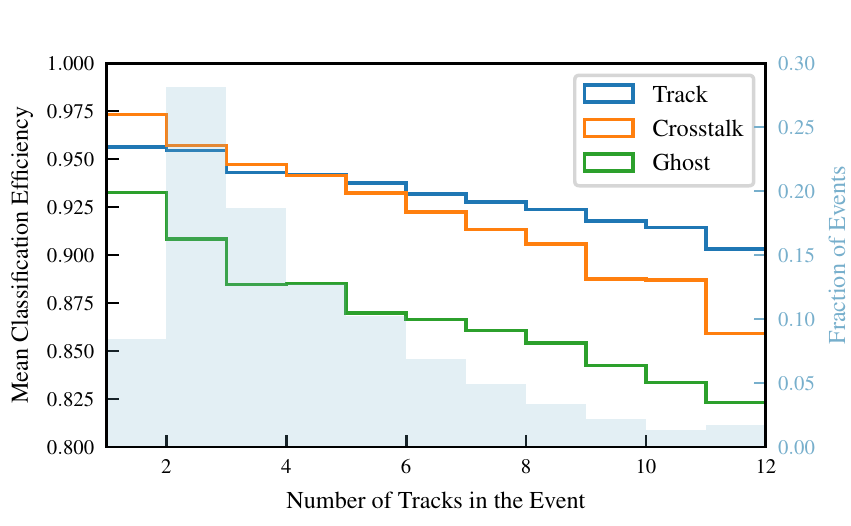}
		\caption{Efficiency.} 
		\label{fig:eff_NumOfTracks}
	\end{subfigure}
	\begin{subfigure}{1.0\columnwidth}
		\includegraphics[width=0.95\linewidth]{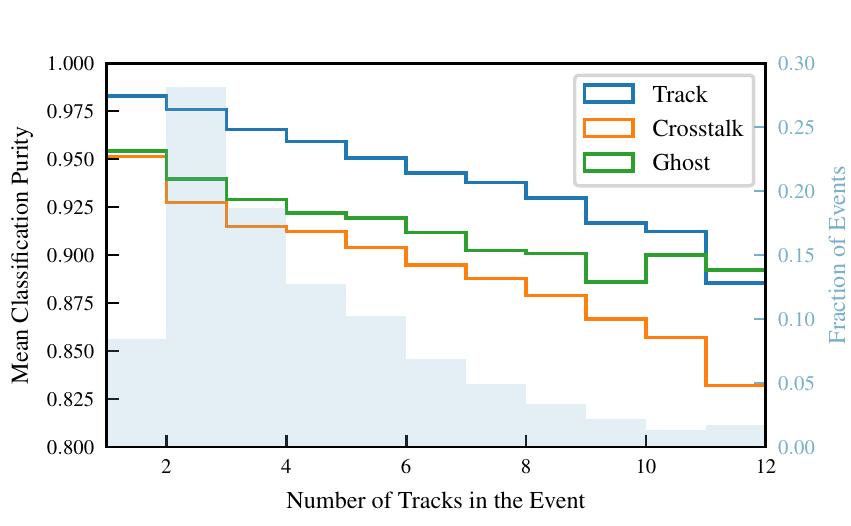}
		\caption{Purity.} 
		\label{fig:pur_NumOfTracks}
	\end{subfigure}
	\begin{subfigure}{1.0\columnwidth}
		\includegraphics[width=0.95\linewidth]{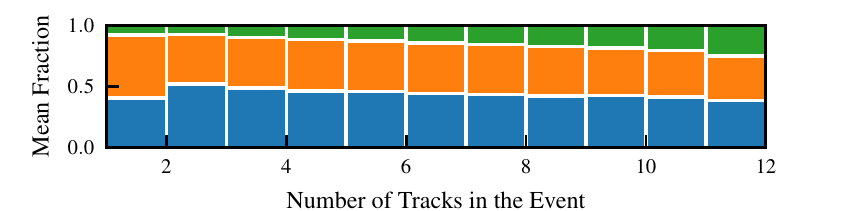}
		\caption{Mean fraction of each type of voxel as a function of the number of tracks in the event (blue~=~track, orange~=~crosstalk, green~=~ghost).} 
		\label{fig:type_NumOfTracks}
	\end{subfigure}
	\caption{Efficiency and purity as a function of the number of tracks in the event for a sample trained and tested on GENIE simulated data.}
	\label{fig:eff/pur_NumOfTracks}
\end{figure}

The number of tracks in the event is an estimate of the complexity of its topology. According to Fig.~\ref{fig:eff/pur_NumOfTracks}, the classification efficiencies and purities drop as the number of tracks increases. This behaviour is also correlated with the increasing fraction of ghost voxels in the events.

\begin{figure}[htb] 
	\begin{subfigure}{1.0\columnwidth}
		\includegraphics[width=0.95\linewidth]{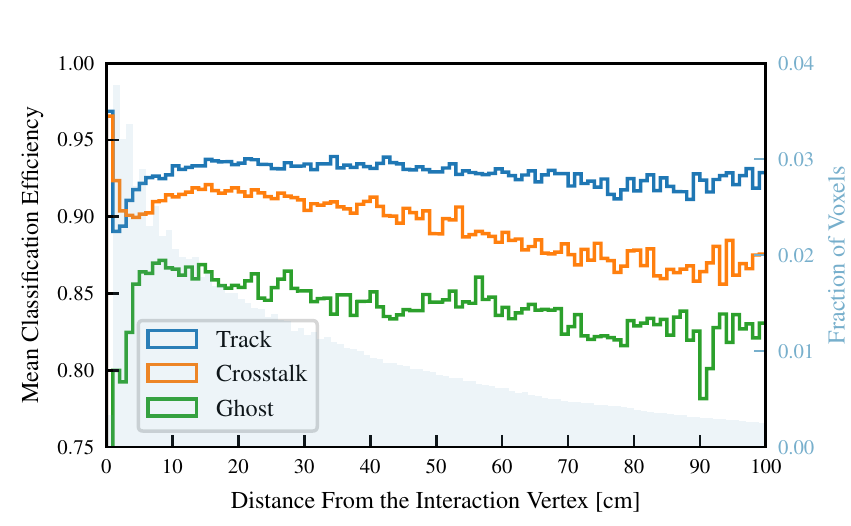}
		\caption{Efficiency.} 
		\label{fig:eff_DistToVtx}
	\end{subfigure}
	\begin{subfigure}{1.0\columnwidth}
		\includegraphics[width=0.95\linewidth]{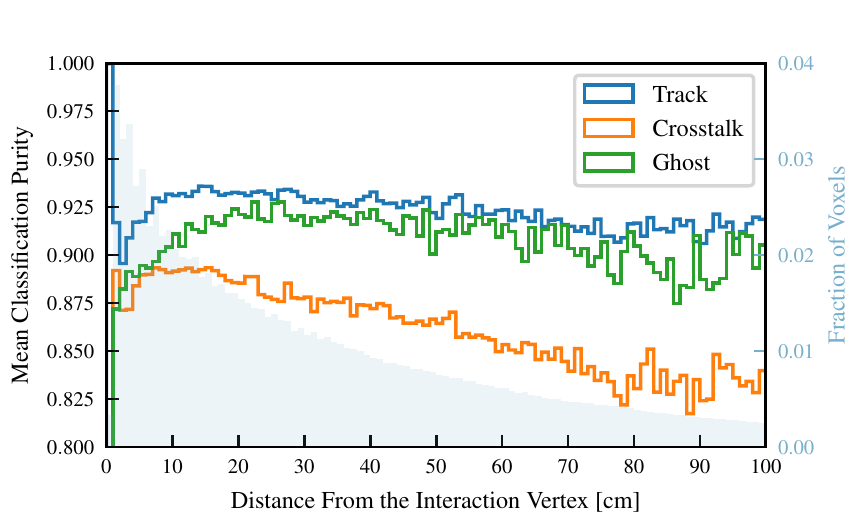}
		\caption{Purity.} 
		\label{fig:pur_DistToVtx}
	\end{subfigure}
	\begin{subfigure}{1.0\columnwidth}
		\includegraphics[width=0.95\linewidth]{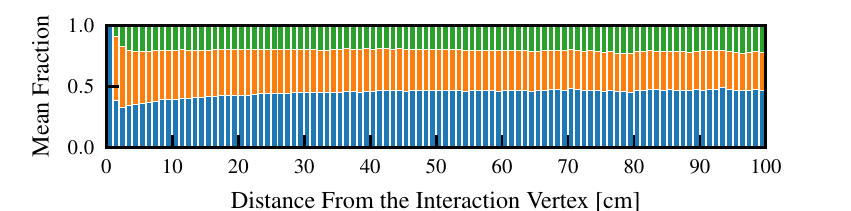}
		\caption{Mean fraction of each type of voxel as a function of the distance to the vertex (blue~=~track, orange~=~crosstalk, green~=~ghost).} 
		\label{fig:type_DistToVtx}
	\end{subfigure}
	\caption{Efficiency and purity as a function of the distance to the neutrino interaction vertex for a sample trained and tested on GENIE data.}
	\label{fig:eff/pur_DistToVtx}
\end{figure}

The region around the interaction vertex is of particular interest in the event. It is expected that a high spatial density of voxels within a certain volume of the detector may pose a challenge for the GNN to correctly identify the voxel type. This can be observed by studying the efficiencies and purities as a function of the distance to the interaction vertex, as shown in Fig.~\ref{fig:eff/pur_DistToVtx}. At the interaction vertex itself, it is clear that there are only track voxels and the GNN can identify them with over 96\% efficiency and 100\% purity. The following 2\,cm exhibit only a small fraction of ghost voxels, mainly due to the high spatial density of voxels with real signals in that volume, which is mainly occupied by track and crosstalk voxels. As we go further from the vertex, \review{the spatial density of voxels decreases and the tracks emerging from the vertex diverge allowing for easier voxel classification. However, this trend is reversed around 10\,cm from the vertex where the protons emerging from the CCQE (Charged-Current Quasi-Elastic) interaction vertex would have reached their range and only the low-ionizing muon tracks remain. The lower average voxel charge at these distances complicates the process of classification as most of the variables used as an input to the GNN are based on charge, which can be observed in the dropping efficiencies and purities.}

As the main goal of this GNN is to identify ghost voxels in order to eliminate them from the events, it is important to make sure that true track and crosstalk voxels are not lost in the process. According to the \review{GENIE sample results}, only 1.1\% of all true track voxels and 3.3\% of crosstalk voxels are incorrectly classified as ghost voxels by the GNN. In addition, it is important not to miss ghost voxels: the GNN correctly identified 84.5\% of all ghost voxels, where 72.1\% of those classified incorrectly were predicted as crosstalk. Therefore, although not ideal, this issue is not critical as crosstalk voxels have a smaller influence on future studies than track voxels.

\begin{figure}[htb] 
	\includegraphics[width=0.95\linewidth]{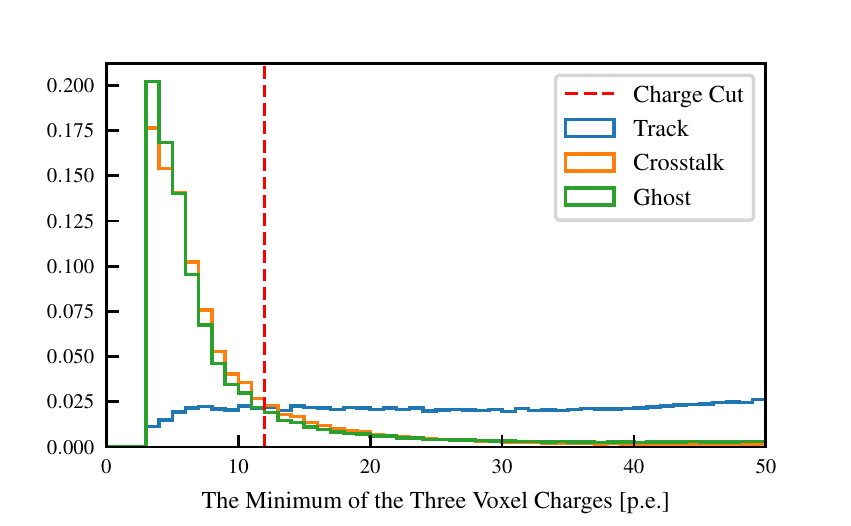}
	\caption{The distribution of the minimum charge among the three voxel charges for the GENIE sample.}
	\label{fig:chargecut_minQ}
\end{figure}

Lastly, we compare the results of the GNN against 
a conventional method of voxel classification which relies on a charge cut. As described in Appendix~\ref{sec:inputvariables}, each voxel has three charges that correspond to the signals from the three fibers passing through it. Since other voxels along the same fiber may have signals causing a larger amplitude to be recorded, we consider the smallest of these three charges to be the most accurate estimation of the true voxel charge. Hence, this minimum charge is used for the purposes of this charge cut.
Since, by definition, we expect higher energy deposition in track voxels compared to crosstalk and ghost voxels, we set a lower limit for the minimum charge in a voxel such that any voxels with a higher minimum charge than the threshold are classified as track voxels. Figure~\ref{fig:chargecut_minQ} shows the distribution of the minimum voxel charge for the three types of voxels. From this figure, it is clear that it is not possible to separate ghost from crosstalk voxels. Thus, this classification is only binary such that we have two categories: track or other. We decide to place this cut at 12 p.e., where the track and non-track voxel \review{curves} intersect.

To compare the results of this cut with those of our GNN, we combine the predictions of the crosstalk and ghost categories. Table~\ref{tab:chargecut} shows the efficiency and purity of the classifications for the two methods. It is evident that using only a charge cut can still yield a comparable track voxel classification efficiency to the GNN. However, it struggles to correctly classify non-track voxels which, in turn, reduces the purity of the predicted track voxels. 

Another \review{advantage of the GNN over the charge cut is the improved capability of reducing the number of ``fake'' tracks}, i.e. a cluster of ghost voxels that closely resembles the structure of a real particle track.
Since fake tracks are usually produced by the shadowing of real tracks, the corresponding number of p.e. measured in the three readout views is higher than 12 p.e., hence the charge cut cannot reject them easily.
The \review{superiority of the GNN in reducing ghost tracks} is shown in Appendix~\ref{sec:eventgallery} for a number of neutrino interactions and compared to the charge cut method.

\begin{table}[htb]
    \centering
    \begin{tabular}{lcc|lcc}
    \hline \hline 
    \multicolumn{3}{c|}{\textbf{GNN}} & \multicolumn{3}{c}{\textbf{Charge Cut}}\\
    \hline
    \rule{0pt}{9pt} & Track & Other & & Track & Other\\ 
    Efficiency & 94\% & 96\% & Efficiency & 93\% & 80\%\\
    Purity & 96\% & 95\% & Purity & 80\% & 91\%\\
    \hline \hline
    \end{tabular}
    \caption{Mean efficiencies and purities of voxel classification for the GNN and a simple charge cut.}
    \label{tab:chargecut}
\end{table}

\begin{figure}[htb] 
	\includegraphics[width=0.95\linewidth]{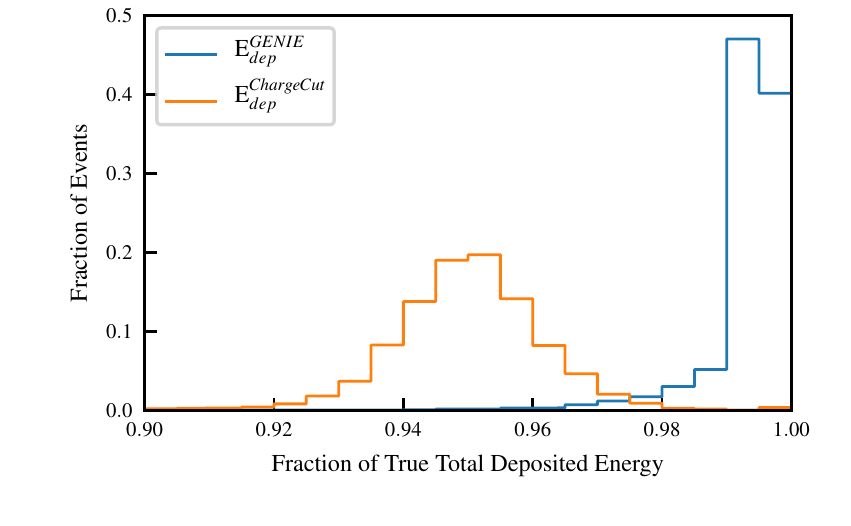}
	\caption{The fraction of the true total deposited energy obtained when using the GNN (trained on GENIE) or the charge cut as a classification method.}
	\label{fig:fracOfEdep}
\end{figure}

Figure~\ref{fig:fracOfEdep} shows the advantage of the three-fold classification of the GNN over the binary classification of the charge cut when comparing the fraction of true total deposited energy obtained using each method. In the case of the GNN, the total deposited energy in an event is the sum of the true energy deposited in all non-ghost voxels. For the charge cut, only the energy deposited in track voxels is used. This causes an average energy loss of 5\% per event when using a method that also excludes the crosstalk voxels, compared to less than 1\% when using the GNN that can isolate ghost voxels.

\section{Systematic uncertainty considerations}
\label{sec:systematics}

The results presented in Sec.~\ref{sec:results} show that the GNN is a very powerful technique for removing ghost voxels and identifying optical crosstalk in 3D-reconstructed neutrino interactions. \review{ In this section, we investigate potential sources of systematic uncertainty and test the robustness of this technique.}

One of the main limitations in the measurement of the neutrino oscillation parameters in long-baseline experiments comes from uncertainties in the modeling of neutrino interactions, not yet fully constrained by data and partially incomplete for describing all the details of the interaction final state. For example, the modeling of hadron multiplicity and kinematics may considerably change the image of the neutrino interaction, particularly near the neutrino vertex, or the total energy deposited by all the particles produced by the neutrino interaction.
Hence, it is hard to obtain a data-driven control sample to train a neural network without making any prior assumptions. Since the GNN is trained only on a subset of the parameter space, the results could be biased if the detected neutrino interactions belong to a region of the parameter space not well covered by the MC generator. 
\review{To account for a potentially incomplete sampling of the parameter space, different training samples (GENIE and P-Bomb) were generated, as described in Sec.~\ref{sec:results}. The difference in terms of neutrino interaction modeling between these two datasets, by construction, is expected to be much larger than the difference between 
GENIE and real neutrino interactions, see Appendix~\ref{sec:datasets_distributions}. As presented in Tab.~\ref{tab:eff&pur_table} the performance is still very good even when the samples used for training and testing were largely different in terms of modeling, supporting the safeness of training using MC events to classify real data.}

The robustness of the GNN against model dependencies can be verified by training different neural networks on different event samples and applying them to the same set of neutrino interactions. 
A difference in the observables used in the physics measurement, such as particle momenta, energy deposit, etc., obtained by the different training can be assigned as a systematic uncertainty introduced by the method.

\begin{figure}[h] 
	\includegraphics[width=0.95\linewidth]{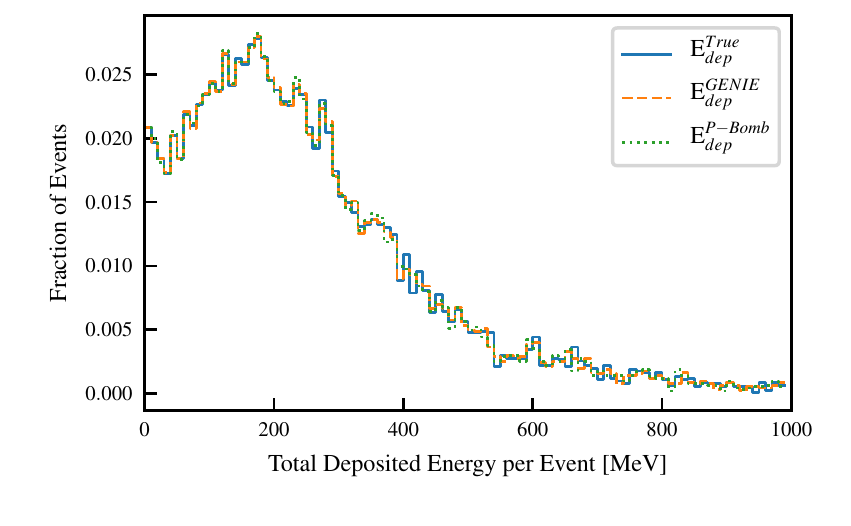}
	\caption{Distribution of the total true deposited energy after rejecting the ghost voxels classified either with GENIE- (dashed orange) or P-Bomb- (dotted green) trained GNNs and without any ghost rejection (solid blue). The mean total deposited energy per event is about 288\,MeV.}
	\label{fig:TotalEdep}
\end{figure}

\begin{figure}[h] 
	\includegraphics[width=0.95\linewidth]{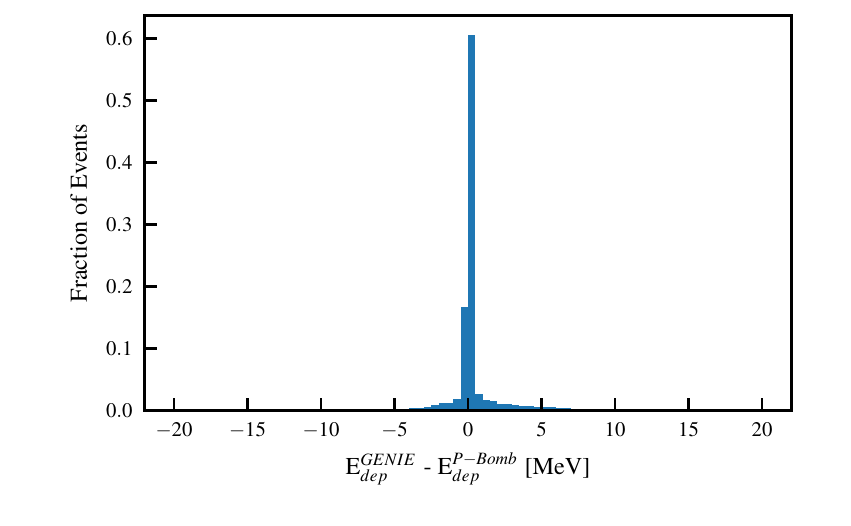}
	\caption{Difference between the total true deposited energy computed after rejecting the ghost voxels classified with GENIE- and P-Bomb- trained GNNs. The mean is 0.78\,MeV while the standard deviation is 5.5\,MeV. 
	About 40\% of events show no difference between P-Bomb and GENIE, 68\% have a difference within $\pm$0.192\,MeV, while only 5\% of the events have a difference outside the range $\pm$6.35\,MeV.	}
	\label{fig:Edep_diff}
\end{figure}

A study was performed to evaluate the impact of the method on the total true energy deposited in the detector. The difference between the total energy deposit computed after rejecting the voxels classified as ghosts for both network trainings was computed. Figure~\ref{fig:TotalEdep} shows the distribution of the total true deposited energy before and after discarding the voxels classified as ghosts. Both GENIE- and P-Bomb- trained GNNs give very similar results over the full range of total deposited energy. \review{The total true deposited energy computed with and without ghost rejection differ on average by less than 1\,MeV with a standard deviation of approximately 5.5\,MeV, mainly due to a few outlier entries, and 68\% of the events with a difference better than 0.192\,MeV, as shown in Fig.~\ref{fig:Edep_diff}. Hence, it is expected to be improved by increasing the statistics of the training samples.}

\begin{figure}[h] 
	\includegraphics[width=0.95\linewidth]{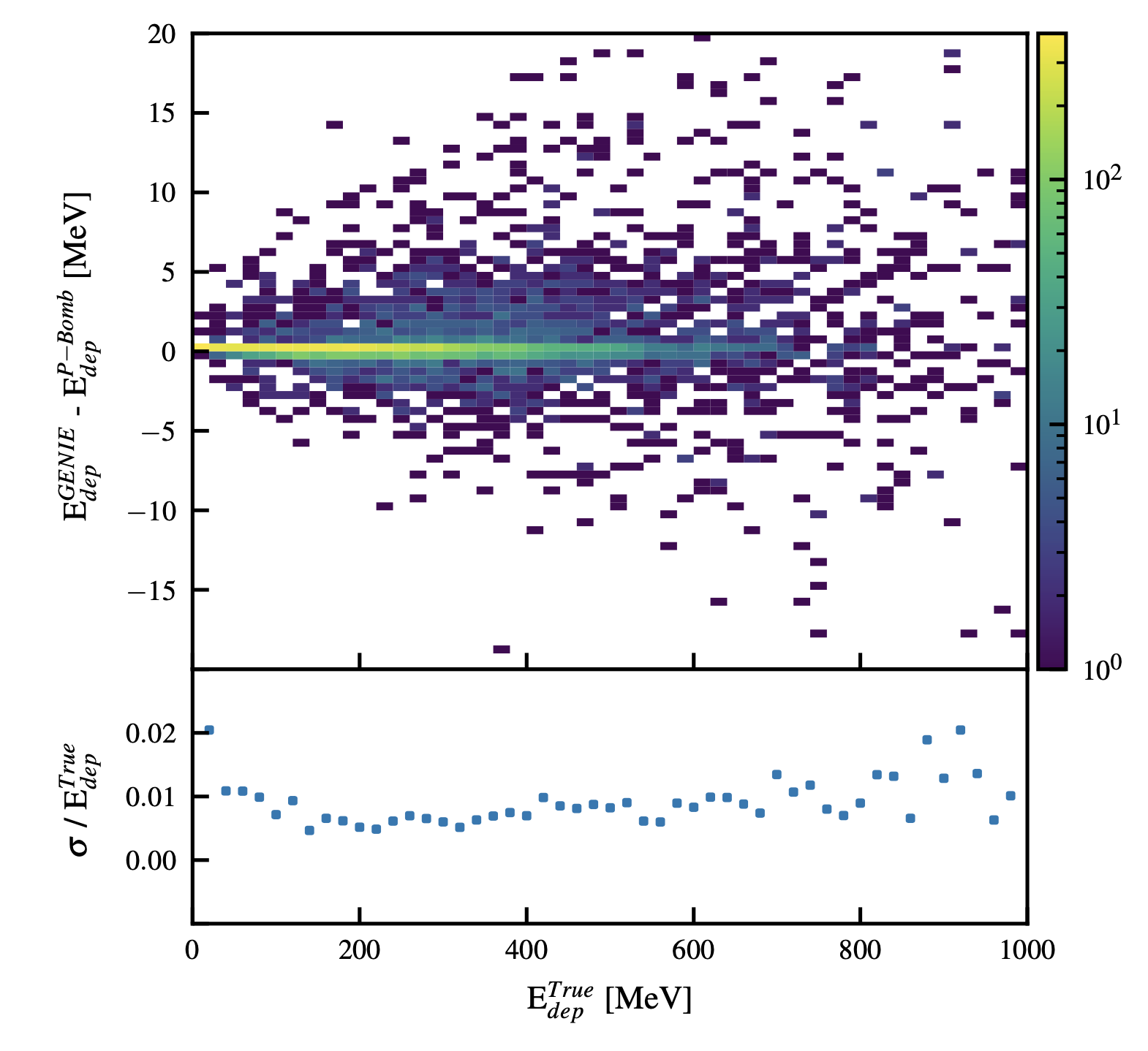}
	\caption{Top: difference between the total true deposited energy computed after rejecting the ghost voxels classified with GENIE- and P-Bomb- trained GNNs as a function of the total true deposited energy. Bottom: fractional standard deviation of the difference of the total true deposited energy computed after rejecting the ghost voxels classified with GENIE- and P-Bomb- trained GNNs as a function of the total true deposited energy.}
	\label{fig:Edep_diff_vs_True_rms}
\end{figure}

This corresponds to less than 2\% of the mean total deposited energy per event. In Fig.~\ref{fig:Edep_diff_vs_True_rms} the impact of the different training sample is shown as a function of the total deposited energy. The fractional standard deviation, defined as the standard deviation of the difference between deposited energy computed from different GNN trainings and divided by the true deposited energy, shown in the bottom panel, is less than 2\% and almost constant as a function of the deposited energy. This means that the performance of the method is about the same irrespective of the total deposited energy. This study confirms that GNN can be used for classifying 3D voxels potentially with limited systematic uncertainties in the deposited energy, while drastically improving the tracking capability.

Another potential issue could be given by a mismodeling of the amount of crosstalk. In addition to the nominal optical crosstalk (2.7\%), two further datasets were simulated using 2\% and 5\% crosstalk and the voxel classification was performed using the GNN trained with nominal crosstalk. As shown in Tab.~\ref{tab:eff&pur_evt_otherXtalk}, the efficiency and the purity is relatively stable even in the case where the crosstalk model is wrong, in particular for identifying track voxels.
\review{Whilst a drop in purity for track voxels is observed with 5\% crosstalk, such a large mismodeling is highly unlikely given that crosstalk can be measured even with small prototypes to sub-percent precision \cite{blondel2020superfgd}. Hence, crosstalk mismodeling is not considered to be a source of additional systematic uncertainty given that the GNN method is robust to small crosstalk variations.}

\begin{table}[htb]
    \centering
    \begin{tabular}{l|lccc}
    \hline \hline
    \rule{0pt}{9pt}\multirowcell{3}{\textbf{Nominal}\\\textbf{Crosstalk}\\\textbf{2.7\%}}& & Track & Crosstalk & Ghost\\
    & Efficiency & 93\% & 90\% & 84\% \\
    & Purity & 92\% & 87\% & 91\% \\
    \hline
    \rule{0pt}{9pt}\multirowcell{3}{\textbf{Crosstalk}\\\textbf{2\%}}& & Track & Crosstalk & Ghost\\
    & Efficiency & 92\% & 89\% & 81\% \\
    & Purity & 94\% & 83\% & 89\% \\
    \hline
    \rule{0pt}{9pt}\multirowcell{3}{\textbf{Crosstalk}\\\textbf{5\%}}& & Track & Crosstalk & Ghost\\
    & Efficiency & 94\% & 89\% & 88\%\\
    & Purity & 86\% & 91\% & 93\%\\
    \hline \hline
    \end{tabular}
    \caption{Mean efficiencies and purities of voxel classification, per voxel, for different crosstalk values, i.e. 2.7\% (nominal), 2\%, and 5\%. The GNN was trained with GENIE training samples with nominal crosstalk and tested on the same GENIE sample with different crosstalk values to study its robustness.}
    \label{tab:eff&pur_evt_otherXtalk}
\end{table}

\section{Conclusions}
\label{sec:conclusions}

A graph neural network inspired by GraphSAGE was developed and tested on simulated neutrino interactions in a 3D voxelized fine-granularity plastic-scintillator detector with three 2D readout views \review{ with the same geometry as SuperFGD, a detector that will be installed in the near detector (ND280) of T2K}.
The advantage of this neural network is that the graph data structures provide a natural representation of the neutrino interactions.

The neural network was able to identify ambiguities and scintillation light leakage between neighboring active scintillator detector volumes as well as real signatures left by particles with efficiencies and purities in the range of 94-96\% per event, with a clear improvement with respect to less sophisticated methods.
In particular, it can \review{reduce the number of} fake tracks produced by the shadowing of real tracks observed in the 2D readout views.
The performance was tested for neutrino events with different number of voxels, number of tracks and voxels at different distances from the vertex, variables that could hint to interaction model dependencies of the method. 
Efficiencies and purities were found to be relatively stable and the trends were consistent with the expectation.
The robustness of the neural network against possible systematic uncertainties introduced by the method was tested.
The results were obtained using neural networks trained on different samples, produced either with the GENIE event generator or by randomizing the number of final state particles and relative momentum to obtain a more generic sample that does not belong to any particular theoretical model. It was found that the bias introduced on the total deposited energy of the event by arbitrarily choosing a different training sample is, on average, less than 1\,MeV\review{, or less than 2\% for true deposited energies in the range 0.0-1.0\,GeV.} The impact of potential mismodeling of the light leakage between neighboring scintillator volumes was tested. Results show that the performance of the neural network is robust to expected changes in the crosstalk modeling.

To conclude, we showed that a graph neural network has great potential in assisting a 3D particle-\review{set} reconstruction of neutrino interactions.
Similar results may be expected for other types of detectors that aim to a 3D reconstruction of the neutrino event from 2D projections and that share analogous features like ambiguities and leakage of signal between detector voxels\review{, such as the very similar detector proposed as part of the DUNE near detector}.

\section{Acknowledgments}
\label{sec:acknowledgments}
D. Douqa and F. S\'anchez acknowledge the Swiss National Foundation Grant No. 200021\_85012. C. Jes\'us-Valls and T. Lux acknowledge funding from the Spanish Ministerio de Econom\'{i}a y Competitividad (SEIDI-MINECO) under Grants No.~FPA2016-77347-C2-2-P and SEV-2016-0588. S. Pina-Otey acknowledges the support of the Industrial Doctorates Plan of the Secretariat of Universities and Research of the Department of Business and Knowledge of the Generalitat of Catalonia. IFAE is partially funded by the CERCA program of the Generalitat de Catalunya. The authors also thank T. Jiang, T. Zhao, and D. Wang for their implementation of GraphSAGE\footnote{\url{https://github.com/twjiang/graphSAGE-pytorch}}, on which the software of this paper was based. 

This work was initiated in the framework of the T2K Near Detector upgrade project, fruitful discussions in this context with our colleagues are gratefully acknowledged. The authors acknowledge the T2K Collaboration for providing the neutrino interaction and detector simulation software. 

\appendix

\section{Comparison of GENIE and P-Bomb}\label{sec:datasets_distributions}
\begin{figure}[h!] 
	\begin{subfigure}{1.0\columnwidth}
		\includegraphics[width=0.95\linewidth]{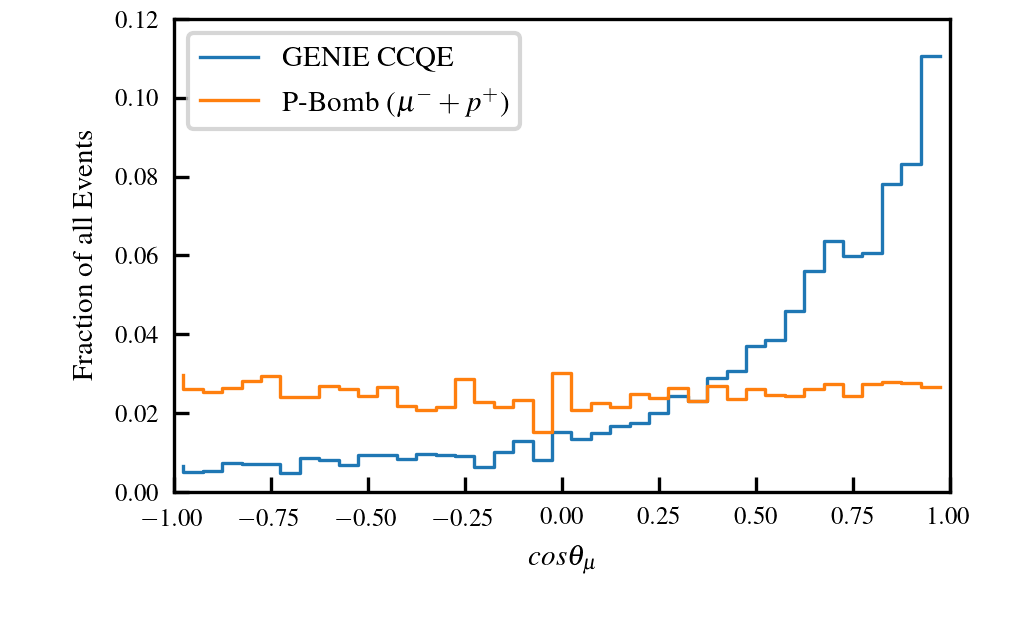}
		\caption{Angular distribution of muons.}
	\end{subfigure}
	\begin{subfigure}{1.0\columnwidth}
		\includegraphics[width=0.95\linewidth]{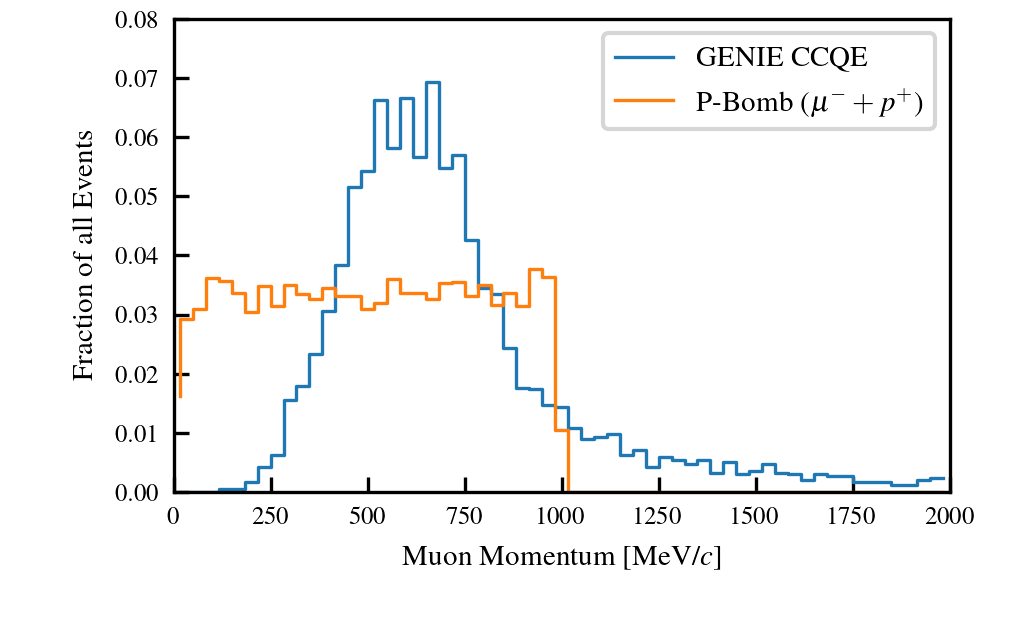}
		\caption{Momentum distribution of muons.}
	\end{subfigure}
	\begin{subfigure}{1.0\columnwidth}
		\includegraphics[width=0.95\linewidth]{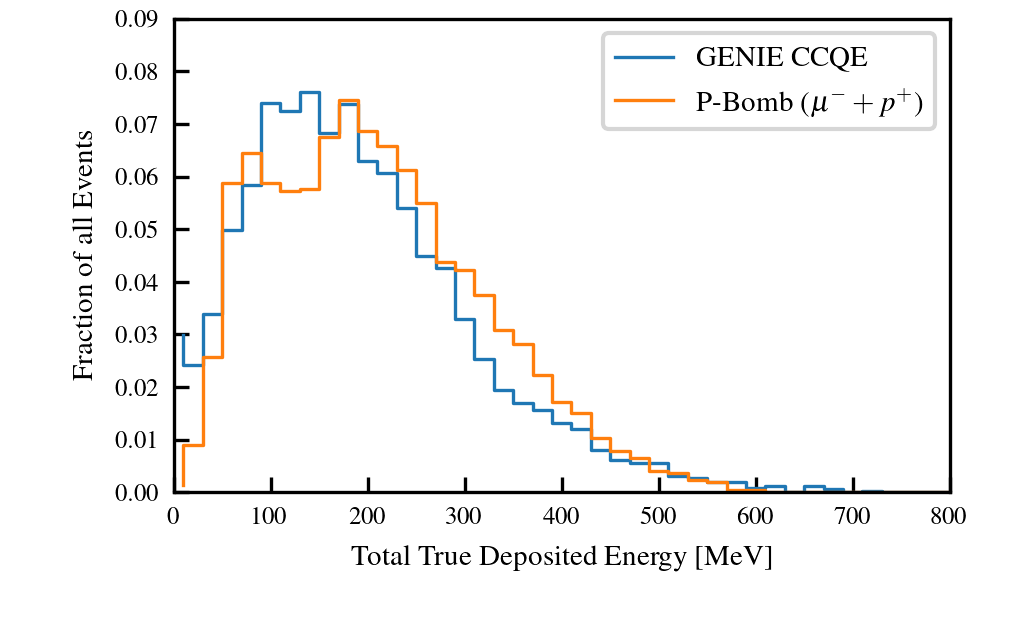}
		\caption{Total energy deposit.}
	\end{subfigure}
	\caption{Distributions of CCQE GENIE interactions compared to $\mu^{-}+p^{+}$ interactions in P-Bomb.}
	\label{fig:GENIEvsPbomb}
\end{figure}

\review{
Two different types of neutrino interactions have been studied, as described in Section~\ref{sec:generation}. The neutrino modelling differences can be easily visualized by comparing two of the simplest subsets of data from each dataset. GENIE charge current quasi elastic interactions (CCQE) typically produce an outgoing muon and proton in the final state. For illustration, we compare this sub-sample of the GENIE dataset with the $\mu^{-}+p^{+}$ sub-sample in the P-Bomb dataset in Fig\review{.}~\ref{fig:GENIEvsPbomb}. 
}

\section{Input variables}\label{sec:inputvariables}

The list of variables used as features for the graph nodes is given below. Each node is placed at XYZ coordinates matching the center of a cube, however, these center coordinates are not node variables by themselves since the detector response is isotropic. The numbers in front of each variable match those in Fig\review{.}~\ref{fig:correlationMatrices}. 

\begin{itemize}
    \item \texttt{0-2: peXY, peXZ, peYZ}\\
    \indent Number of photons detected in the XY, XZ or YZ-fiber intersecting the cube under consideration corrected by the expected attenuation.
    \item \texttt{3-5: mXY, mXZ, mYZ}\\
    \indent Number of active voxels intersected by the fiber associated to \texttt{peXY, peXZ} or \texttt{peYZ}
    \item \texttt{6: pewav}\\
    Average number of detected photons \texttt{peXY, peXZ, peYZ}, each weighted by the fiber multiplicity \texttt{mXY, mXZ, mYZ}.
    \begin{align*}
        \texttt{pewav} = \frac{\frac{\texttt{peXY}}{\texttt{mXY}}+\frac{\texttt{peXZ}}{\texttt{mXZ}}+\frac{\texttt{peYZ}}{\texttt{mYZ}}}{3}
    \end{align*}
    
    \item\texttt{7-9: pullX, pullY, pullZ}\\
    Relative difference between the light measured in two different 2D planes.
    \begin{align*}
        \texttt{pullX} = \frac{\texttt{peXY}-\texttt{peXZ}}{\texttt{peXY}+\texttt{peXZ}}
    \end{align*}
    \begin{align*}
        \texttt{pullY} = \frac{\texttt{peXY}-\texttt{peYZ}}{\texttt{peXY}+\texttt{peYZ}}
    \end{align*}
    \begin{align*}
        \texttt{pullZ} = \frac{\texttt{peXZ}-\texttt{peYZ}}{\texttt{peXZ}+\texttt{peYZ}}
    \end{align*}

    \item \texttt{10: residual}\\
    Similarity of the light yield measured in the three 2D planes, measured as the squared distance from each \texttt{peXY, peXZ, peYZ} to the average, weighted by the squared average.
    \begin{align*}
        \mu =  \frac{\texttt{peXY}+\texttt{peXZ}+\texttt{peYZ}}{3}
    \end{align*}
    \begin{align*}
        \texttt{residual} = \frac{(\texttt{peXY}-\mu)^2+(\texttt{peXZ}-\mu)^2+(\texttt{peYZ}-\mu)^2}{\mu^2}
    \end{align*}

    \item \texttt{11: pullXYZ}\\
    Similarity of the light yield measured in the three 2D planes, measured as a combination of 2D pulls ($a_1$,$a_2$,$a_3$) weighted by \texttt{pewav}.
    \begin{align*}
        a_1 = \frac{\frac{\texttt{peXY}}{\texttt{mXY}}-\frac{\texttt{peXZ}}{\texttt{mXZ}}}{\frac{\texttt{peXY}}{\texttt{mXY}}+\frac{\texttt{peXZ}}{\texttt{mXZ}}}
    \end{align*}
    \begin{align*}
        a_2 = \frac{\frac{\texttt{peXY}}{\texttt{mXY}}-\frac{\texttt{peYZ}}{\texttt{mYZ}}}{\frac{\texttt{peXY}}{\texttt{mXY}}+\frac{\texttt{peYZ}}{\texttt{mYZ}}}
    \end{align*}
    \begin{align*}
        a_3 = \frac{\frac{\texttt{peXZ}}{\texttt{mXZ}}-\frac{\texttt{peYZ}}{\texttt{mYZ}}}{\frac{\texttt{peXZ}}{\texttt{mXZ}}+\frac{\texttt{peYZ}}{\texttt{mYZ}}}
    \end{align*}
    \begin{align*}
        \texttt{pullXYZ} = \frac{a_1a_2+a_1a_3+a_2a_3}{\texttt{pewav}}
    \end{align*}

    \item \texttt{12: ratioMQ}\\
    Ratio between the average voxel multiplicity in the three fibers and \texttt{pewav}.
    \begin{align*}
        \texttt{ratioMQ} = \frac{\frac{\texttt{mXY}+\texttt{mXZ}+\texttt{mYZ}}{3}}{\texttt{pewav}}
    \end{align*}
    
    \item \texttt{13-14: R1, R3}\\
    Number of active neighbor voxels in a sphere of certain radius. \\
    \indent $\hookrightarrow$ \texttt{R1}, r=1\,cm. \\
    \indent $\hookrightarrow$ \texttt{R2}, r=2\,cm. \\
    \indent $\hookrightarrow$ \texttt{R3}, r=5\,cm.
    \texttt{R2} was not used as a variable due to the high correlation with \texttt{R1}, but is used to compute \texttt{RR}.
    \item \texttt{15-20: x+, x-, y+, y-, z+, z-}\\
    Boolean variables representing the existence of immediate neighbors in each of the 6 surrounding cubes
    \item \texttt{21: orthogonal\_neighbor}\\
    \indent It is 1 if any of \texttt{x+, x-, y+, y-, z+, z-} is 1. 
    \item \texttt{22: RR}\\
    Ratio between the number of close and far voxels. The $\epsilon=10^{-7}$ prevents numerical problems when \texttt{R3}=0.
    \begin{align*}
        \texttt{RR} = \frac{\texttt{R2}}{\texttt{R3}+\epsilon}
    \end{align*}
    
    \item \texttt{23: ratioDQ}\\
    Ratio between the average voxel distance \texttt{aveDist} around the voxel and the weighted average light yield \texttt{pewav}.
    \begin{align*}
    \texttt{ratioDQ} =\frac{\texttt{aveDist}}{\texttt{pewav}}
    \end{align*}

    \item \texttt{24: aveDist}\\
    Average distance from the voxel center $C$ to all fired voxel centers ($C_i$) within a sphere of radius 2.5\,cm.
    \begin{align*}
        \texttt{aveDist} =\frac{\texttt{1}}{N}\sum_{i}^{N}\textrm{EuclidianDist}(C,C_i)
    \end{align*}
\end{itemize}

A number of these variables are calculated from the same underlying properties of the energy deposits. In theory, an infinitely deep GNN trained on an infinite amount of training data would be able to extract all of the information required for classification from the few underlying properties. In practice, we use a larger number of derived variables to guide the GNN to allow it to more easily extract information from the data and to converge quickly in the training process.
Global position was intentionally not used as a variable to avoid the GNN to learn neutrino modelling specific behaviours.

\section{Comparison of GENIE and P-Bomb simulated data samples}\label{sec:correlations}

Figure~\ref{fig:correlationMatrices} shows the correlations of the input variables defined in Appendix~\ref{sec:inputvariables} for the GENIE and P-Bomb data samples. Differences between the two matrices arise from the different topologies of interactions produced by the two generator methods.

\begin{figure}[h!] 
	\begin{subfigure}{1.0\columnwidth}
		\includegraphics[width=0.95\linewidth]{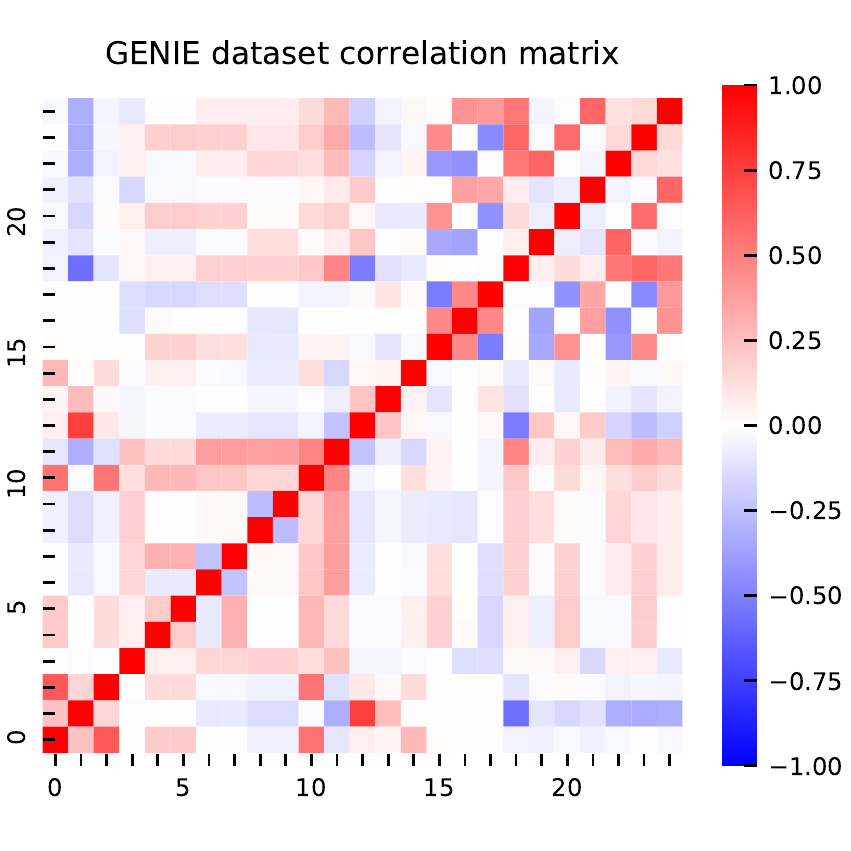}
		\caption{GENIE dataset correlation matrix.}
		\label{fig:GENIEcorrelation}
	\end{subfigure}
	\begin{subfigure}{1.0\columnwidth}
		\includegraphics[width=0.95\linewidth]{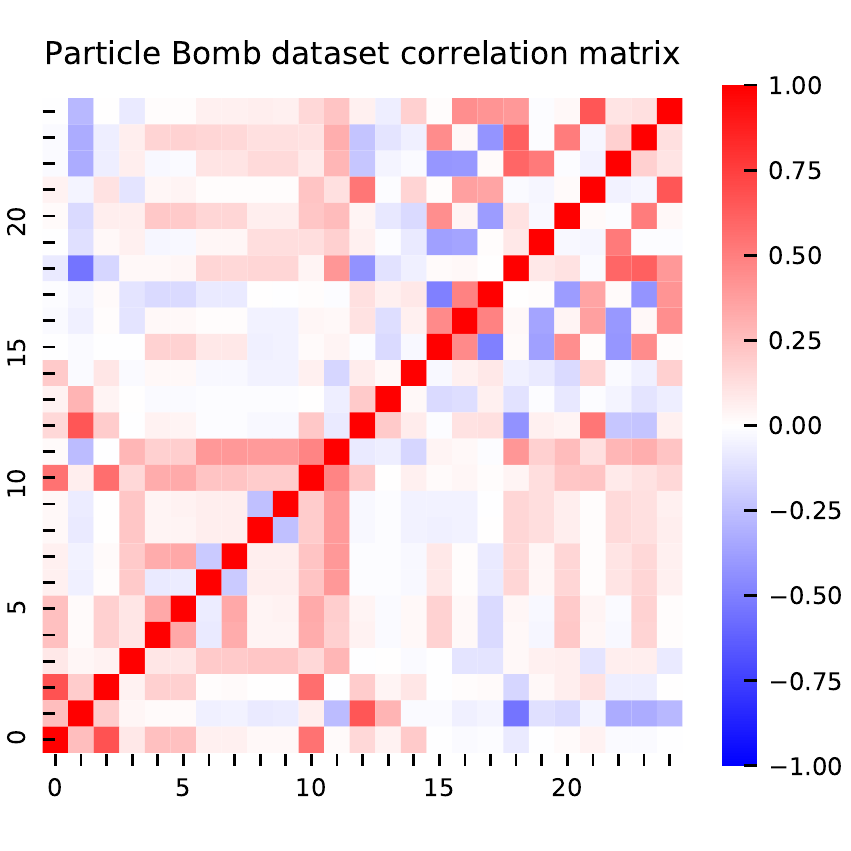}
		\caption{P-Bomb dataset correlation matrix.}
		\label{fig:pbombcorrelation}
	\end{subfigure}
	\caption{Correlation matrices for the input variables of the GENIE and P-Bomb datasets used. Appendix~\ref{sec:inputvariables} gives the mapping between the numbers on the axes and the variable names.}
	\label{fig:correlationMatrices}
\end{figure}

\section{Event Gallery}
\label{sec:eventgallery}
\begin{figure*}[ht!] 
  \subcaptionbox{The 3D voxels labelled as track (red), crosstalk (blue) and ghost (yellow) according to the truth information from the simulation are shown. \label{fig:event_truth_0}}%
  [.495\linewidth]{\includegraphics[width=8.8cm]{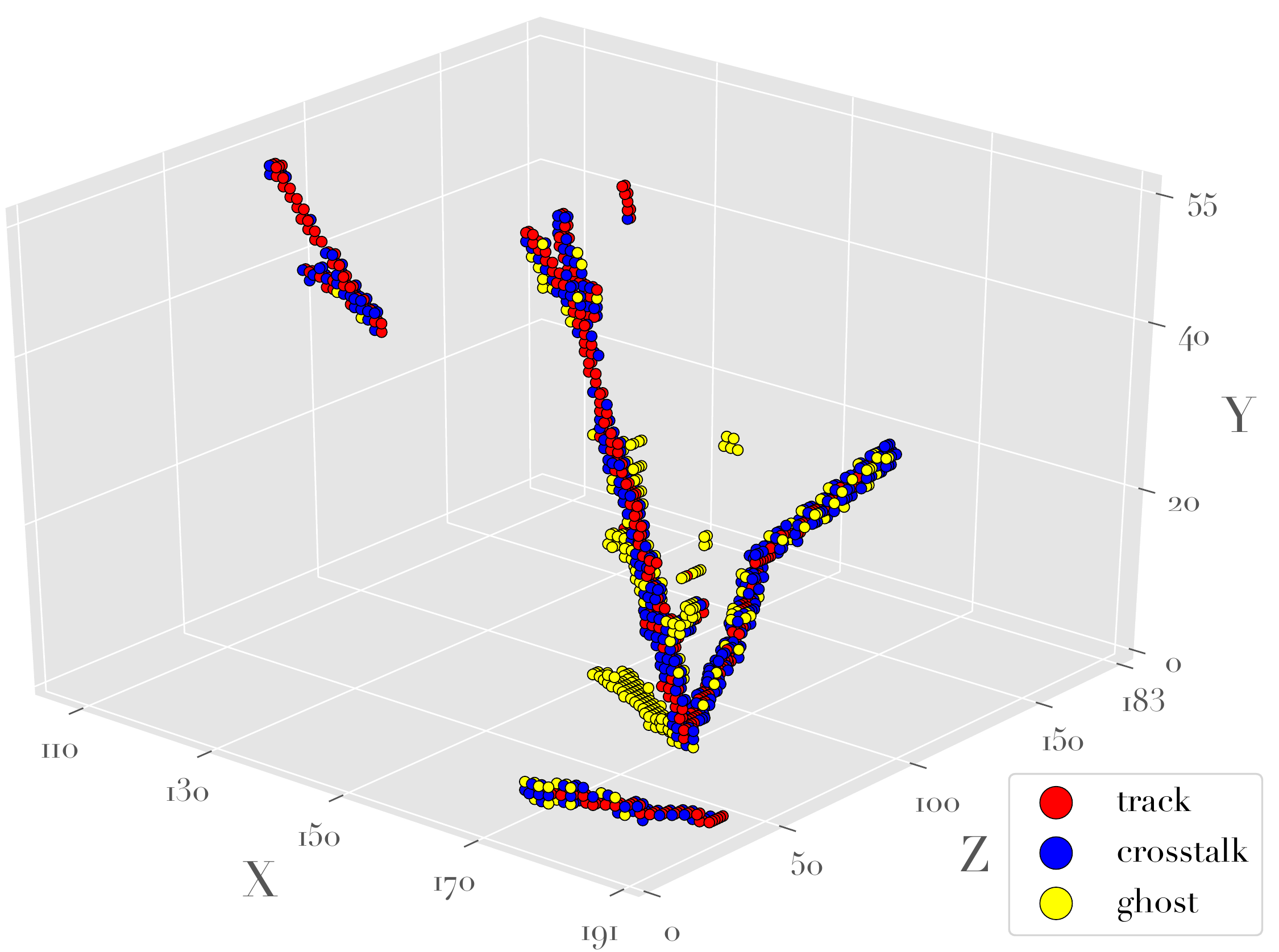}}
  \subcaptionbox{Only the 3D voxels labelled as track according to the truth information from the simulation are shown. \label{fig:event_track_truth_0}}%
  [.495\linewidth]{\includegraphics[width=8.8cm]{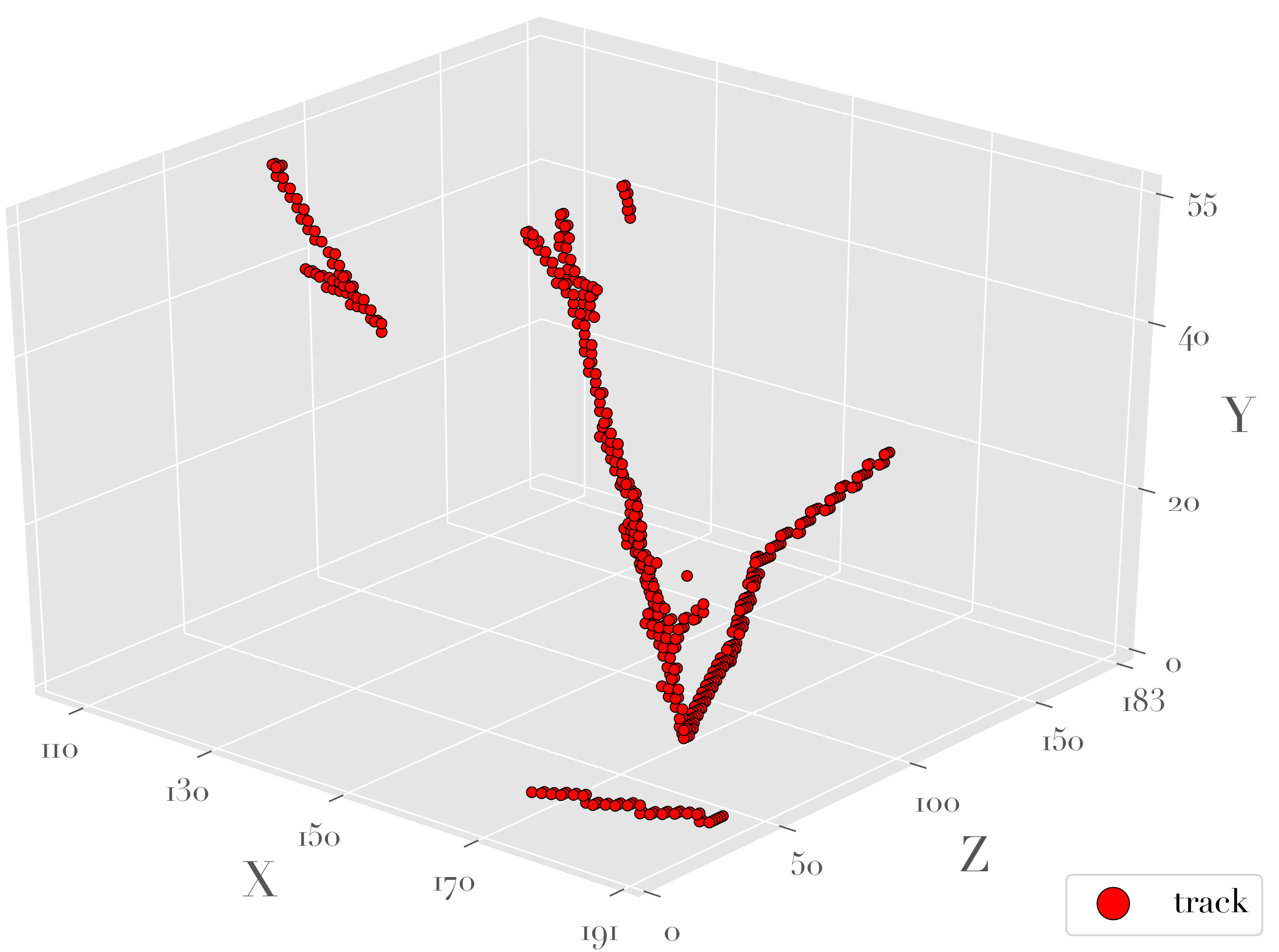}}
  \subcaptionbox{The 3D voxels labelled as track according to the charge cut classification are shown. \label{fig:event_track_chargecut_0}}%
  [.495\linewidth]{\includegraphics[width=8.8cm]{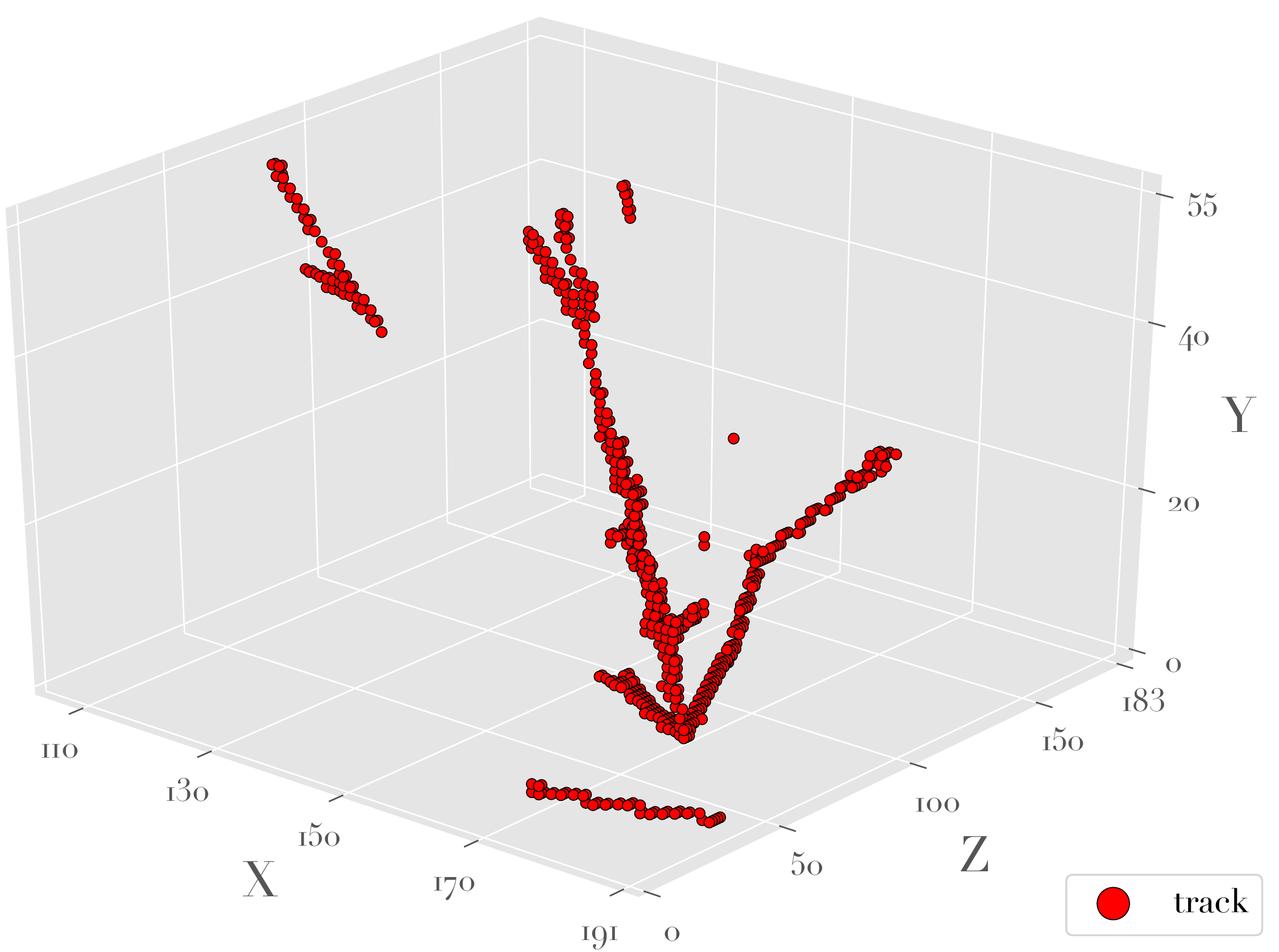}}
  \subcaptionbox{The 3D voxels labelled as track according to the GNN classification are shown.  \label{fig:event_track_gnn_0}}%
  [.495\linewidth]{\includegraphics[width=8.8cm]{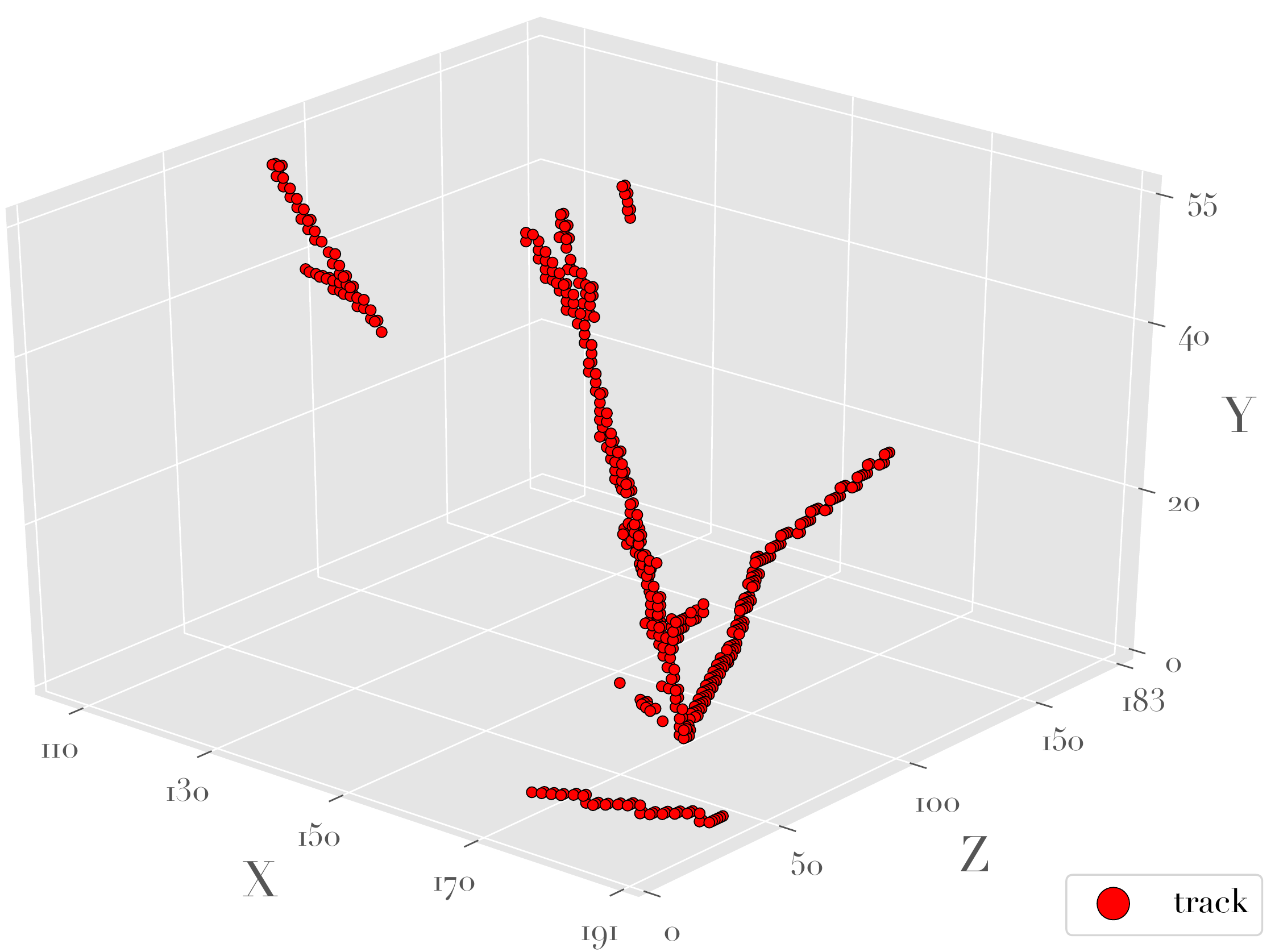}}
  \caption{3D visualization of a neutrino interaction in a finely segmented 3D scintillator detector after the 3D matching of the three 2D views.
  The GNN cut is able to almost entirely reject the fake track traveling on the XZ plane and stopping near to the vertex at X$\sim$160~\review{cm} and Z$\sim$70~\review{cm}, while the charge cut cannot. \review{The energy of the incoming neutrino is 4.754~GeV. The axes are in cm.}
  }
  \label{fig:event_1}
\end{figure*}

\begin{figure*}[ht!] 
  \subcaptionbox{The 3D voxels labelled as track (red), crosstalk (blue) and ghost (yellow) according to the truth information from the simulation are shown. \label{fig:event_truth_1}}%
  [.495\linewidth]{\includegraphics[width=8.8cm]{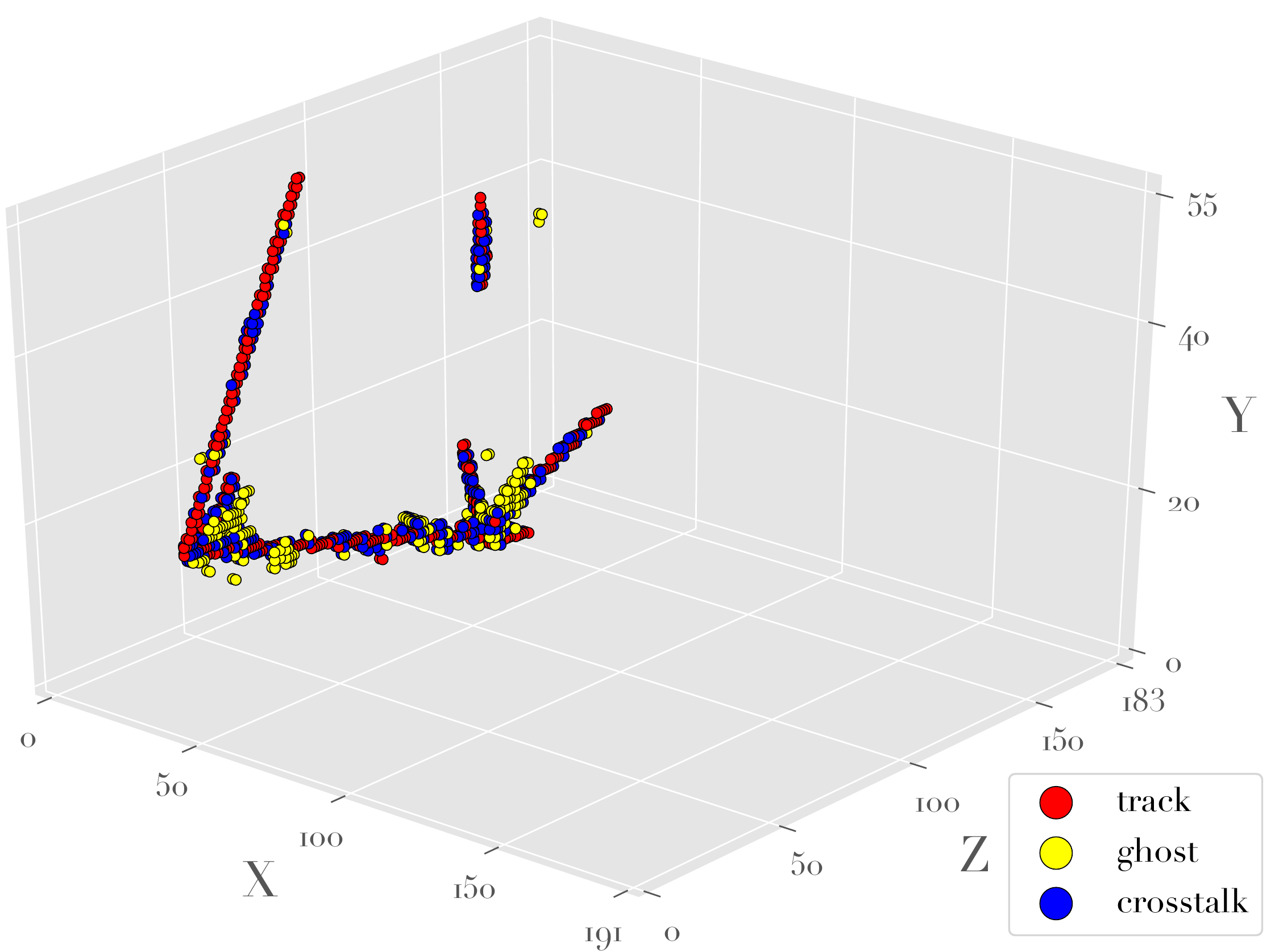}}
  \subcaptionbox{Only the 3D voxels labelled as track according to the truth information from the simulation are shown. \label{fig:event_track_truth_1}}%
  [.495\linewidth]{\includegraphics[width=8.8cm]{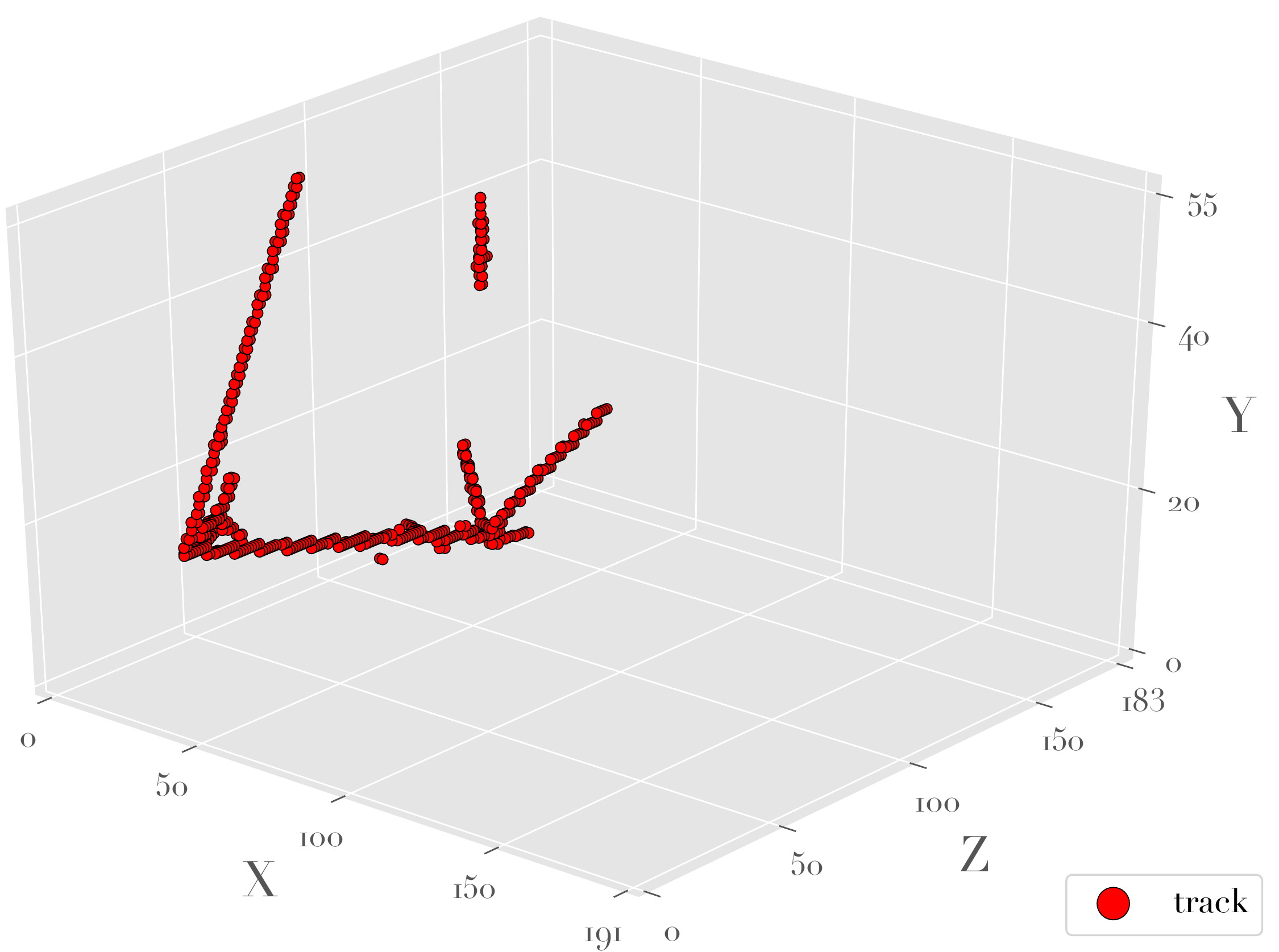}}
  \subcaptionbox{The 3D voxels labelled as track according to the charge cut classification are shown. \label{fig:event_track_chargecut_1}}%
  [.495\linewidth]{\includegraphics[width=8.8cm]{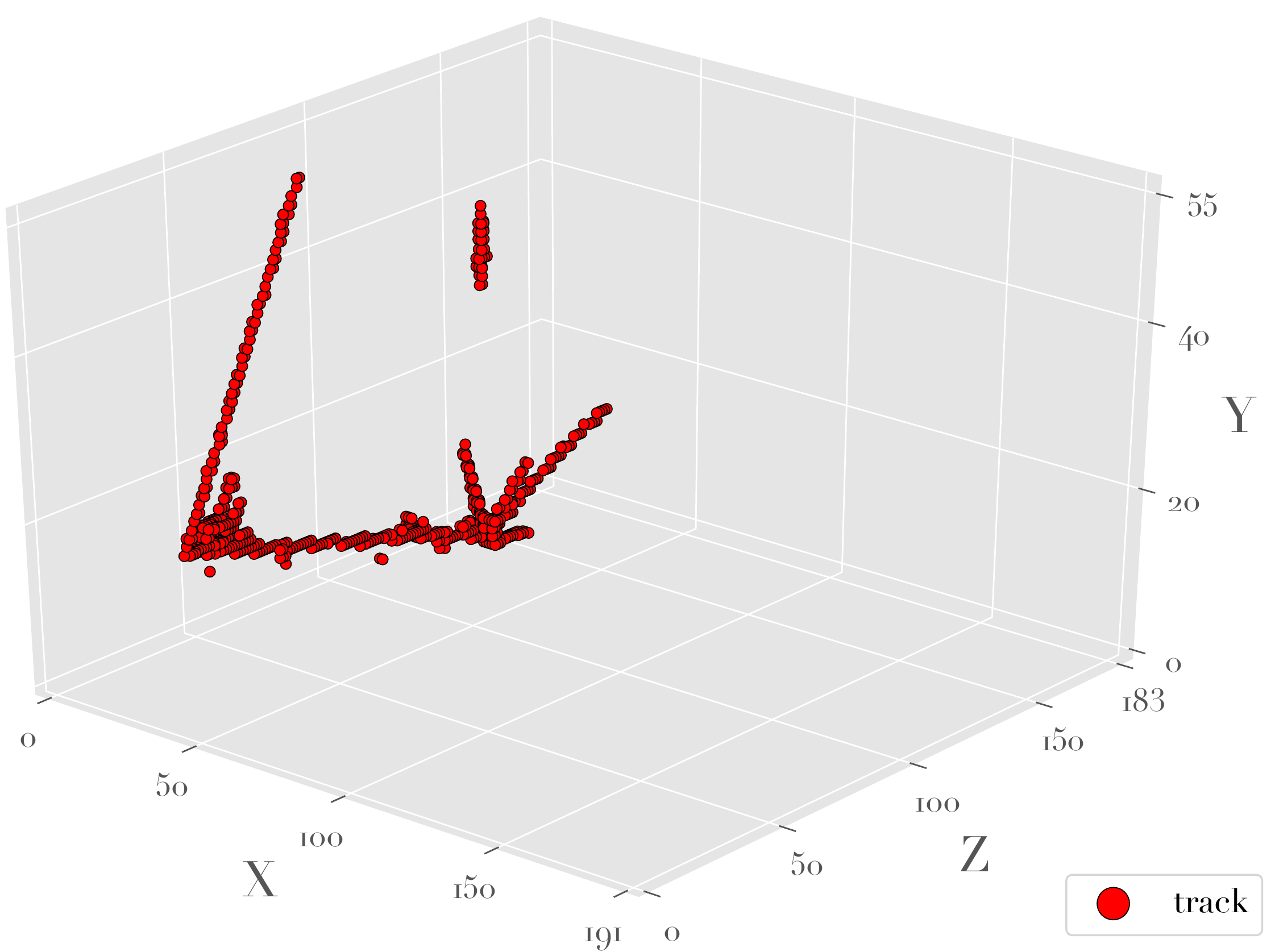}}
  \subcaptionbox{The 3D voxels labelled as track according to the GNN classification are shown.  \label{fig:event_track_gnn_1}}%
  [.495\linewidth]{\includegraphics[width=8.8cm]{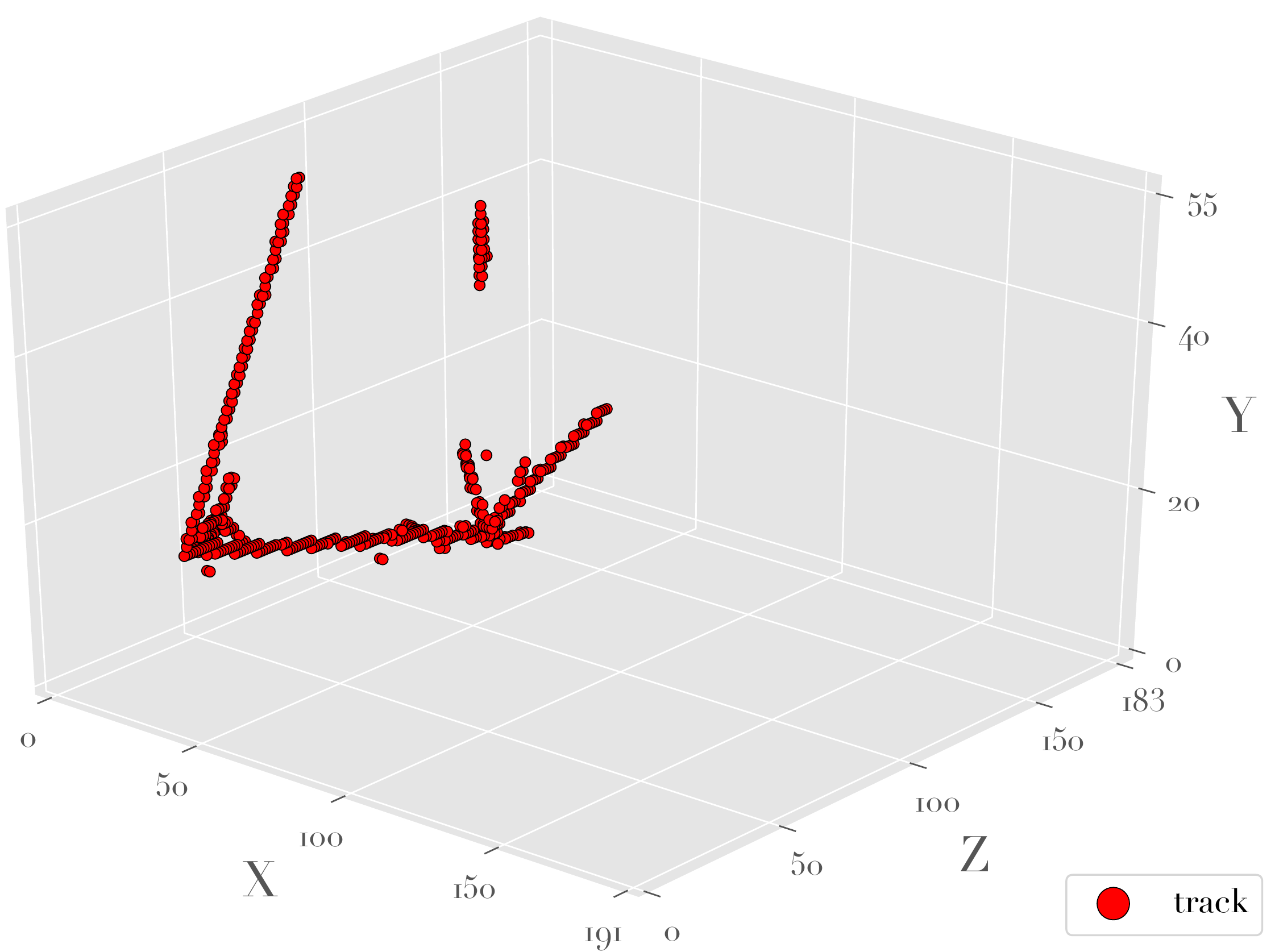}}
  \caption{3D visualization of a neutrino interaction in a finely segmented 3D scintillator detector after the 3D matching of the three 2D views.
  The charge cut is not able to reject two fake tracks, one coming from a vertex a X$<$50~\review{cm} Z$<$50~\review{cm} traveling on the XZ plane and stopping near to the vertex at X$\sim$160~\review{cm} and Z$\sim$70~\review{cm},. Moreover, the charge cut leave a bump of ghost voxels around the vertex that could mimic the interaction of a few low-energy protons, an effect that could bias the reconstruction of the neutrino energy. \review{The energy of the incoming neutrino is 760~MeV. The axes are in cm.}
  }
  \label{fig:event_2}
\end{figure*}

\begin{figure*}[ht!] 
  \subcaptionbox{The 3D voxels labelled as track (red), crosstalk (blue) and ghost (yellow) according to the truth information from the simulation are shown. \label{fig:event_truth_2}}%
  [.495\linewidth]{\includegraphics[width=8.8cm]{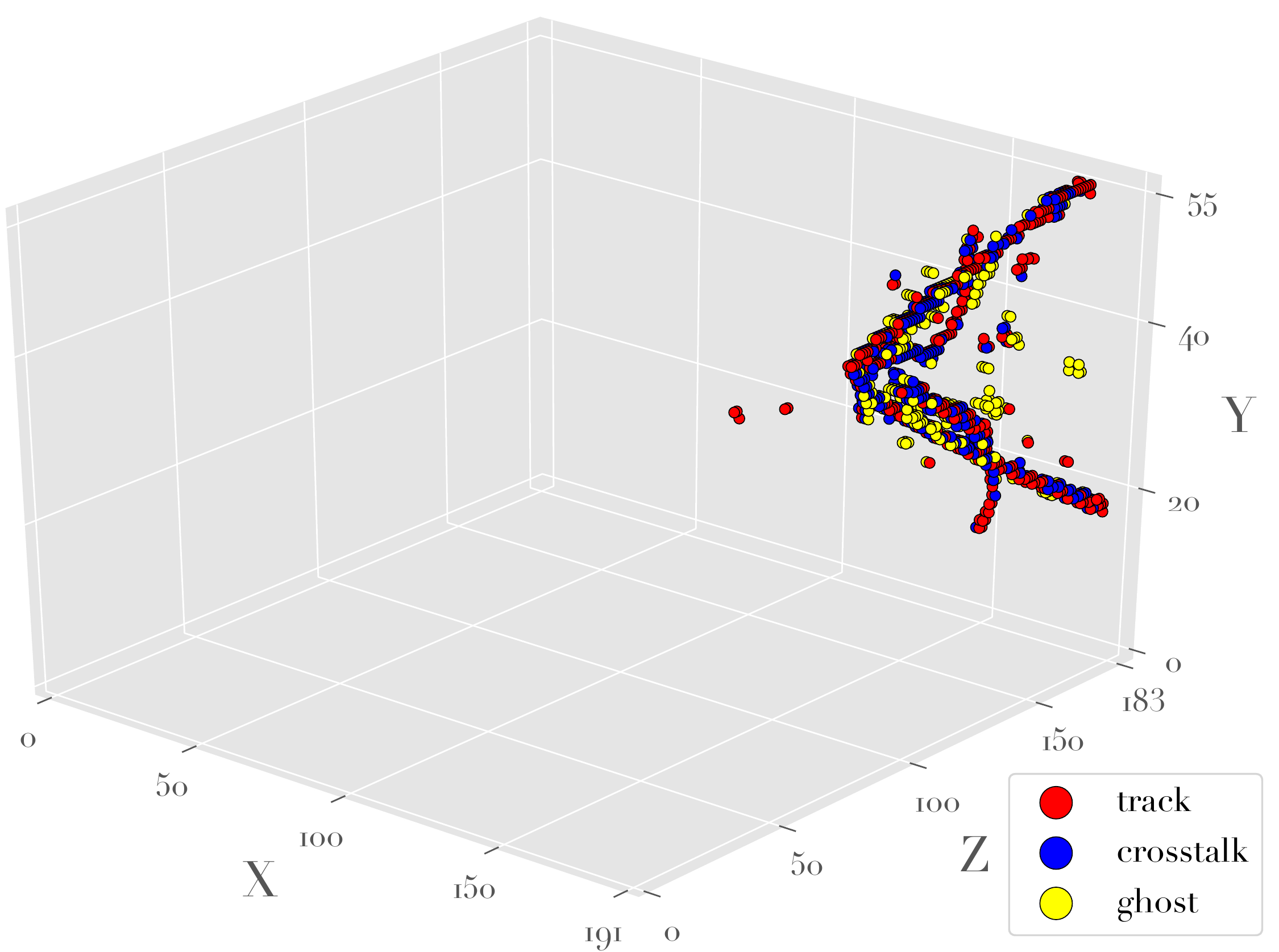}}
  \subcaptionbox{Only the 3D voxels labelled as track according to the truth information from the simulation are shown. \label{fig:event_track_truth_2}}%
  [.495\linewidth]{\includegraphics[width=8.8cm]{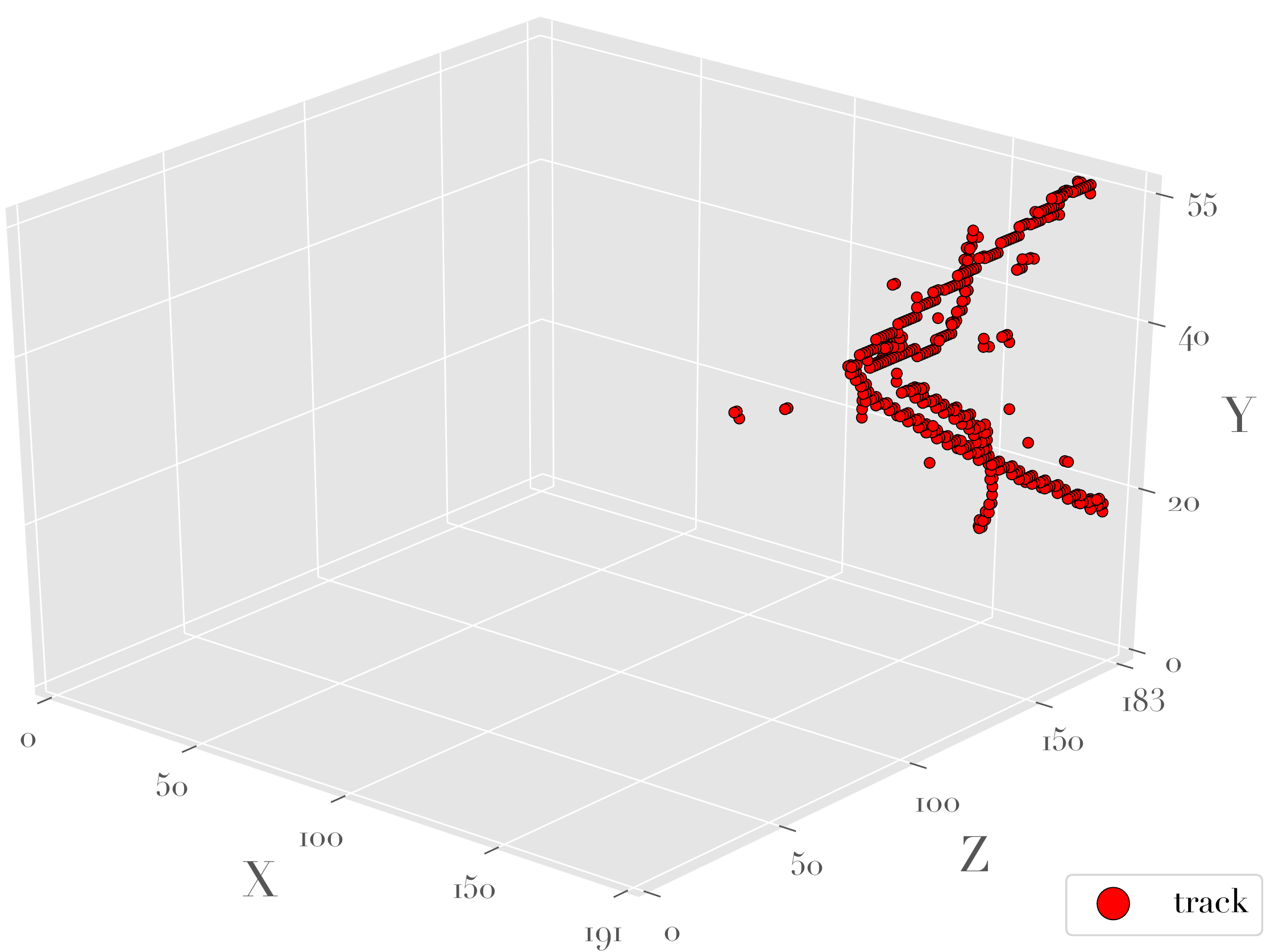}}
  \subcaptionbox{The 3D voxels labelled as track according to the charge cut classification are shown. \label{fig:event_track_chargecut_2}}%
  [.495\linewidth]{\includegraphics[width=8.8cm]{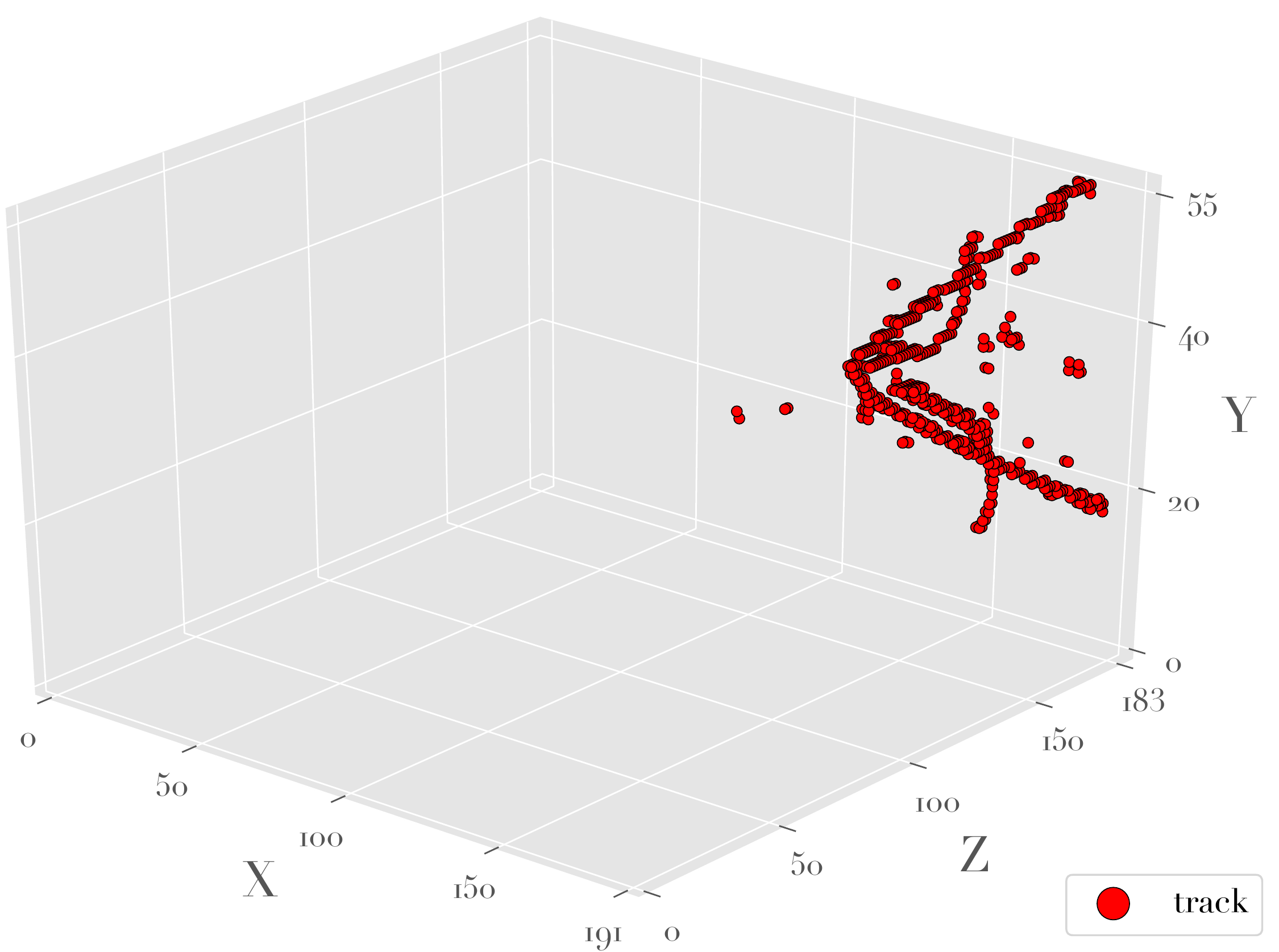}}
  \subcaptionbox{The 3D voxels labelled as track according to the GNN classification are shown.  \label{fig:event_track_gnn_2}}%
  [.495\linewidth]{\includegraphics[width=8.8cm]{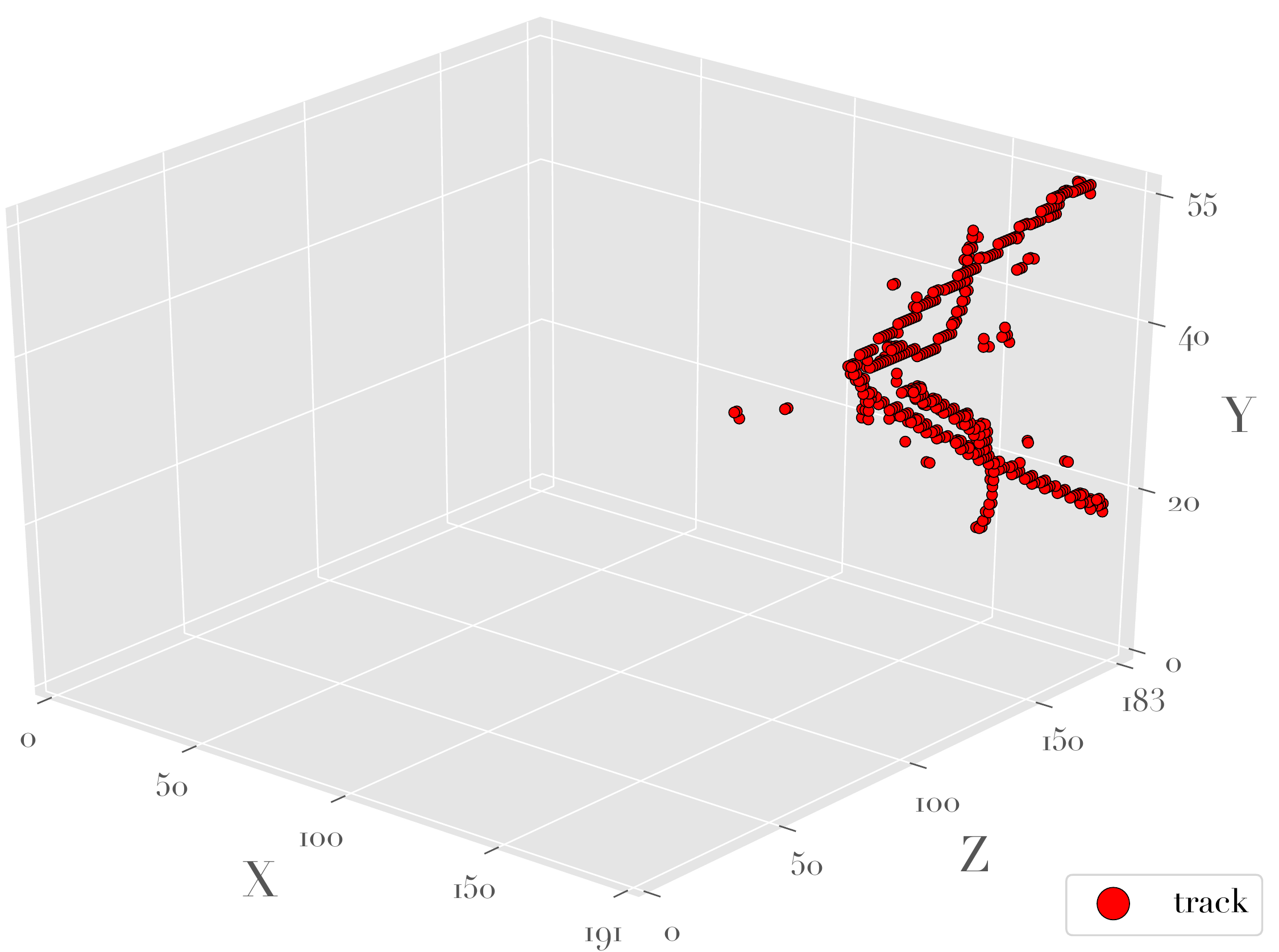}}
  \caption{3D visualization of a neutrino interaction in a finely segmented 3D scintillator detector after the 3D matching of the three 2D views.
  In this even the performance of GNN and the charge cut is quite similar because the ghost voxels are mainly given by the overlap of crosstalk hits in the 2D readout views. \review{The energy of the incoming neutrino is 5.076~GeV. The axes are in cm.}
  }
  \label{fig:event_3}
\end{figure*}

\begin{figure*}[ht!] 
  \subcaptionbox{The 3D voxels labelled as track (red), crosstalk (blue) and ghost (yellow) according to the truth information from the simulation are shown. \label{fig:event_truth_3}}%
  [.495\linewidth]{\includegraphics[width=8.8cm]{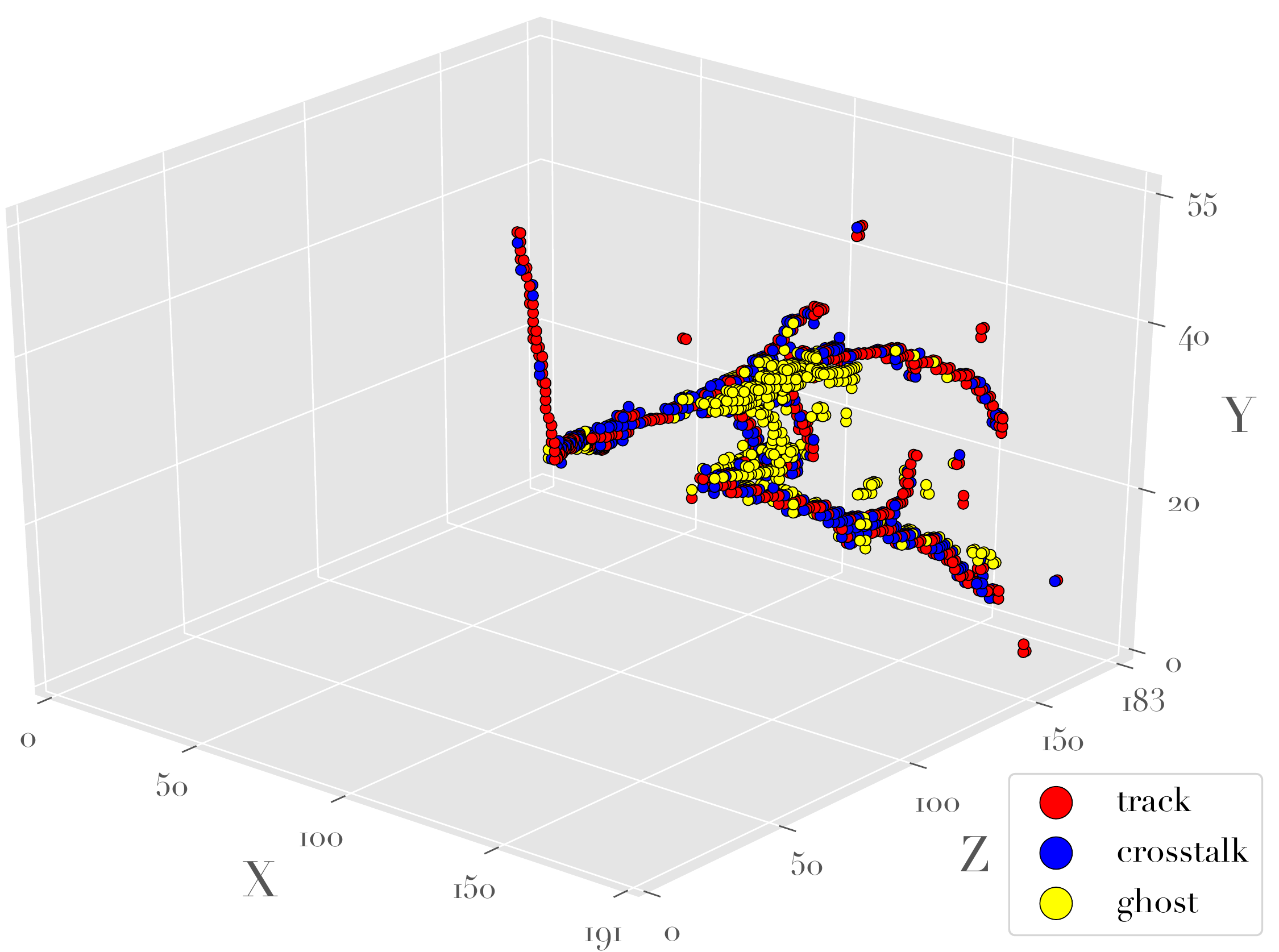}}
  \subcaptionbox{Only the 3D voxels labelled as track according to the truth information from the simulation are shown. \label{fig:event_track_truth_3}}%
  [.495\linewidth]{\includegraphics[width=8.8cm]{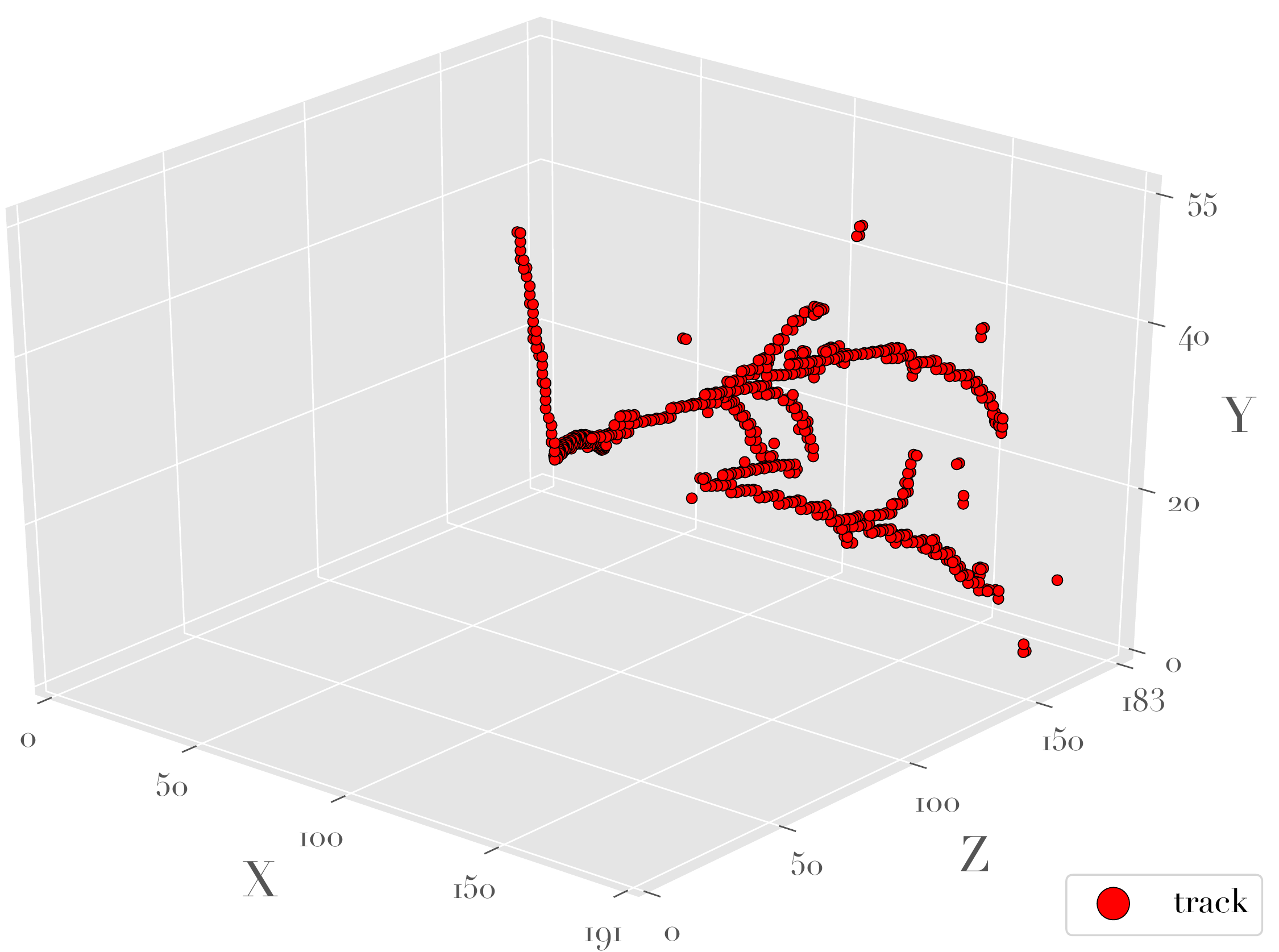}}
  \subcaptionbox{The 3D voxels labelled as track according to the charge cut classification are shown. \label{fig:event_track_chargecut_3}}%
  [.495\linewidth]{\includegraphics[width=8.8cm]{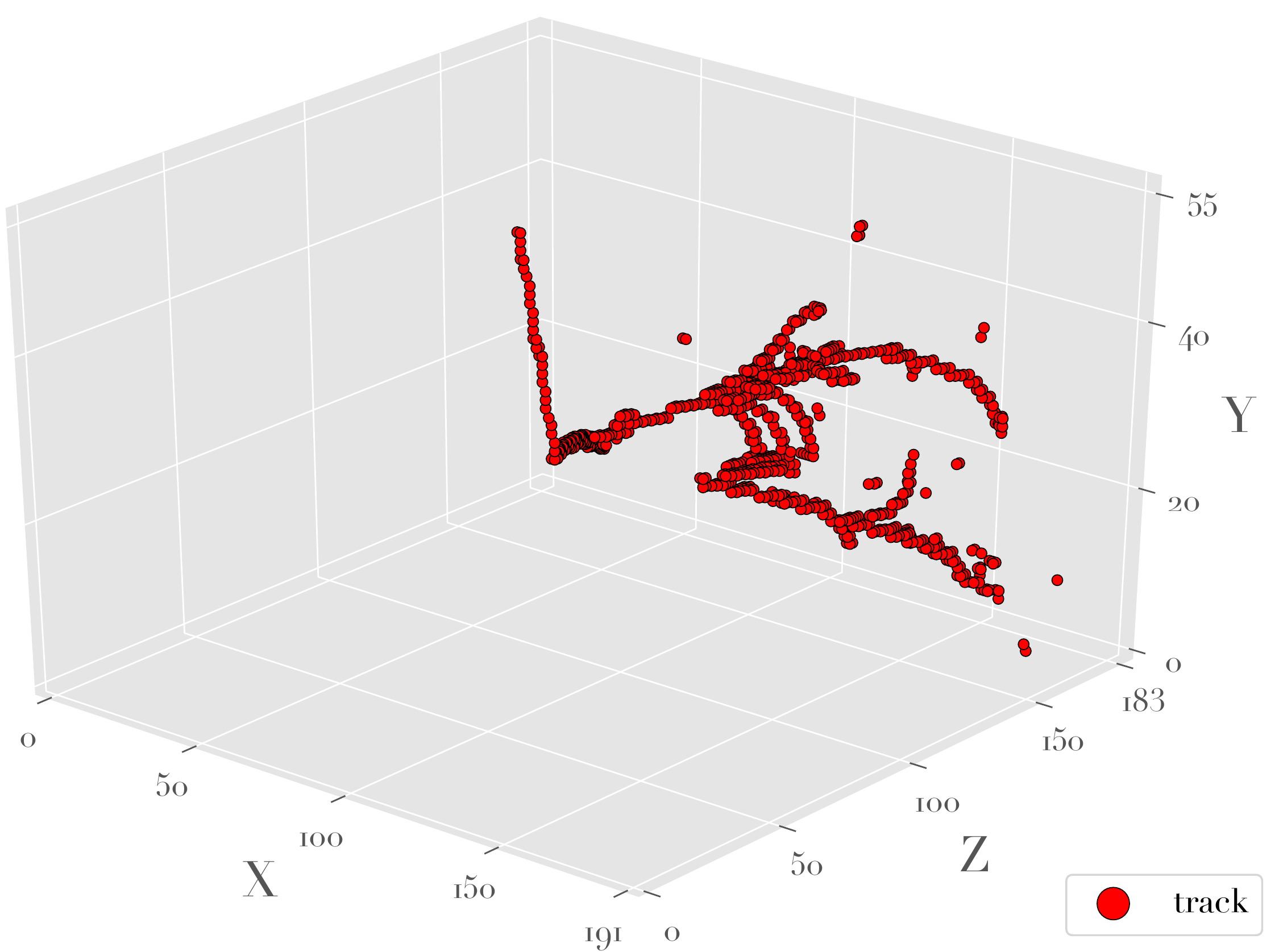}}
  \subcaptionbox{The 3D voxels labelled as track according to the GNN classification are shown.  \label{fig:event_track_gnn_3}}%
  [.495\linewidth]{\includegraphics[width=8.8cm]{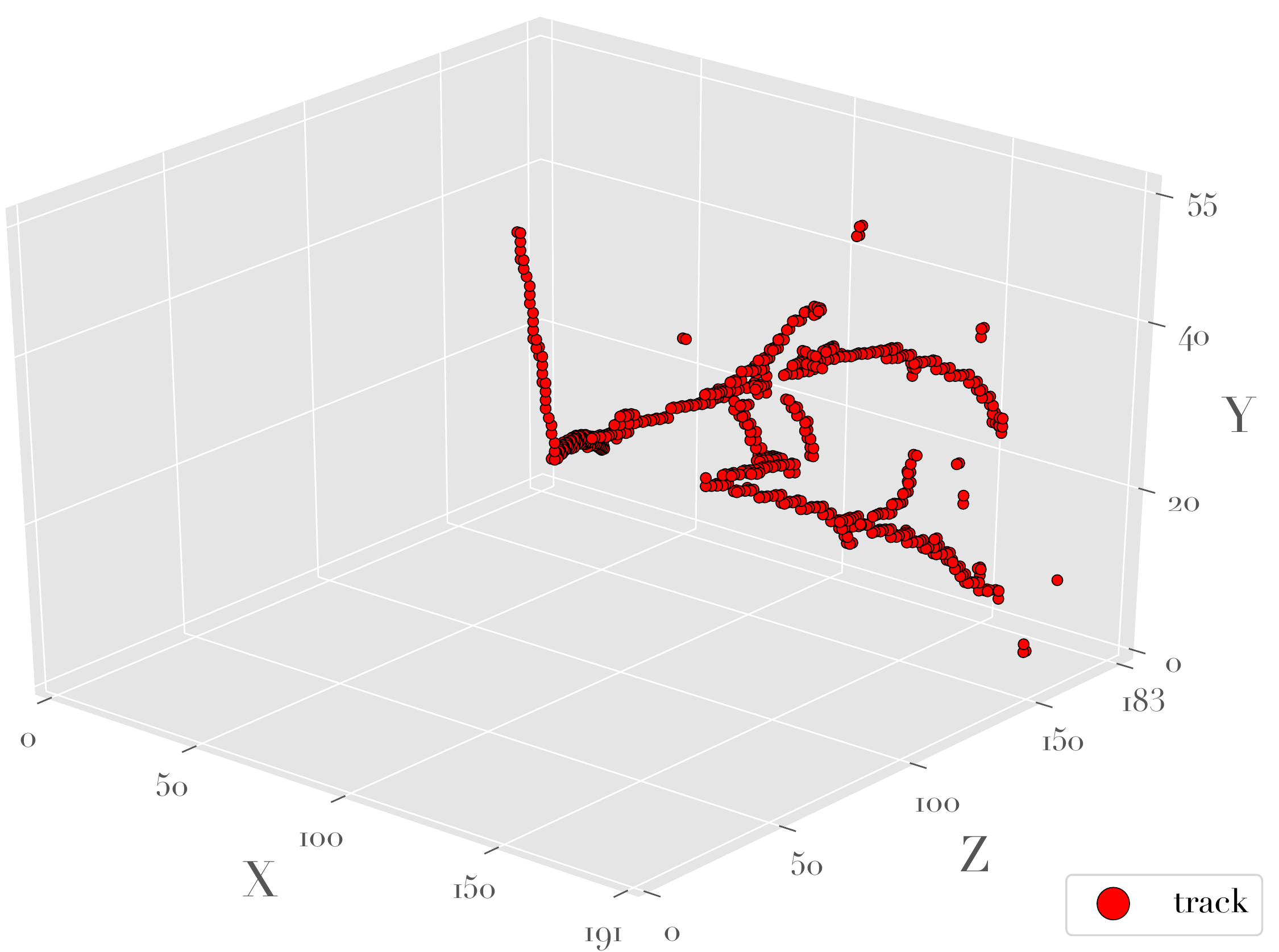}}
  \caption{3D visualization of a neutrino interaction in a finely segmented 3D scintillator detector after the 3D matching of the three 2D views.
  This neutrino event has a quite high multiplicity and tracks are quite close each other. This produce relatively big clusters of ghost voxels that produce at least two fake tracks even after the charge cut. Instead GNN allows to classify ghosts more precisely and correctly visualize the correct number of tracks. Moreover, the charge cut makes true tracks more fat making their separation harder and, potentially, less precise the particle momentum reconstruction. \review{The energy of the incoming neutrino is 1.064~GeV. The axes are in cm.}
  }
  \label{fig:event_4}
\end{figure*}

\begin{figure*}[ht!] 
  \subcaptionbox{The 3D voxels labelled as track (red), crosstalk (blue) and ghost (yellow) according to the truth information from the simulation are shown. \label{fig:event_truth_4}}%
  [.495\linewidth]{\includegraphics[width=8.8cm]{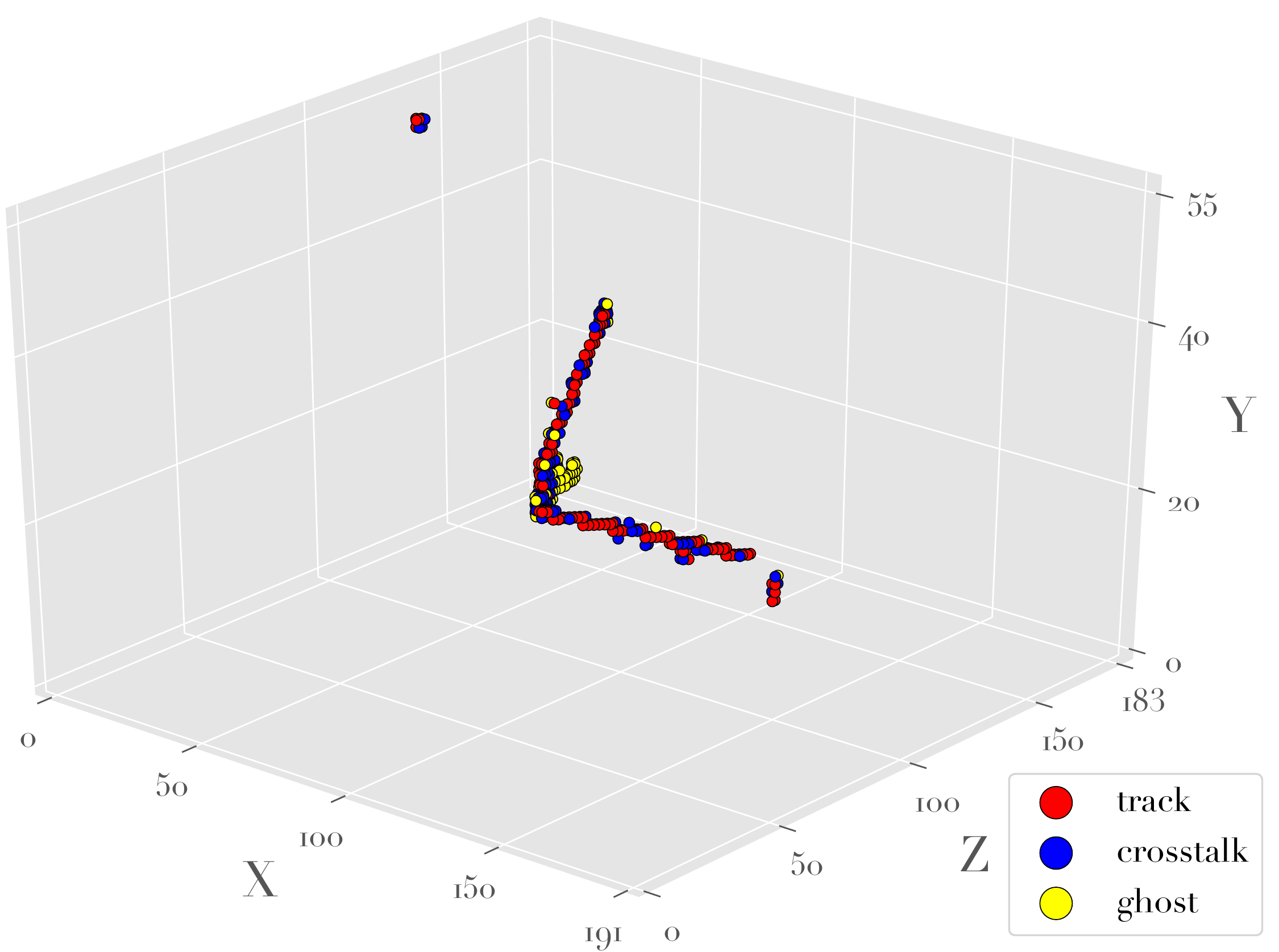}}
  \subcaptionbox{Only the 3D voxels labelled as track according to the truth information from the simulation are shown. \label{fig:event_track_truth_4}}%
  [.495\linewidth]{\includegraphics[width=8.8cm]{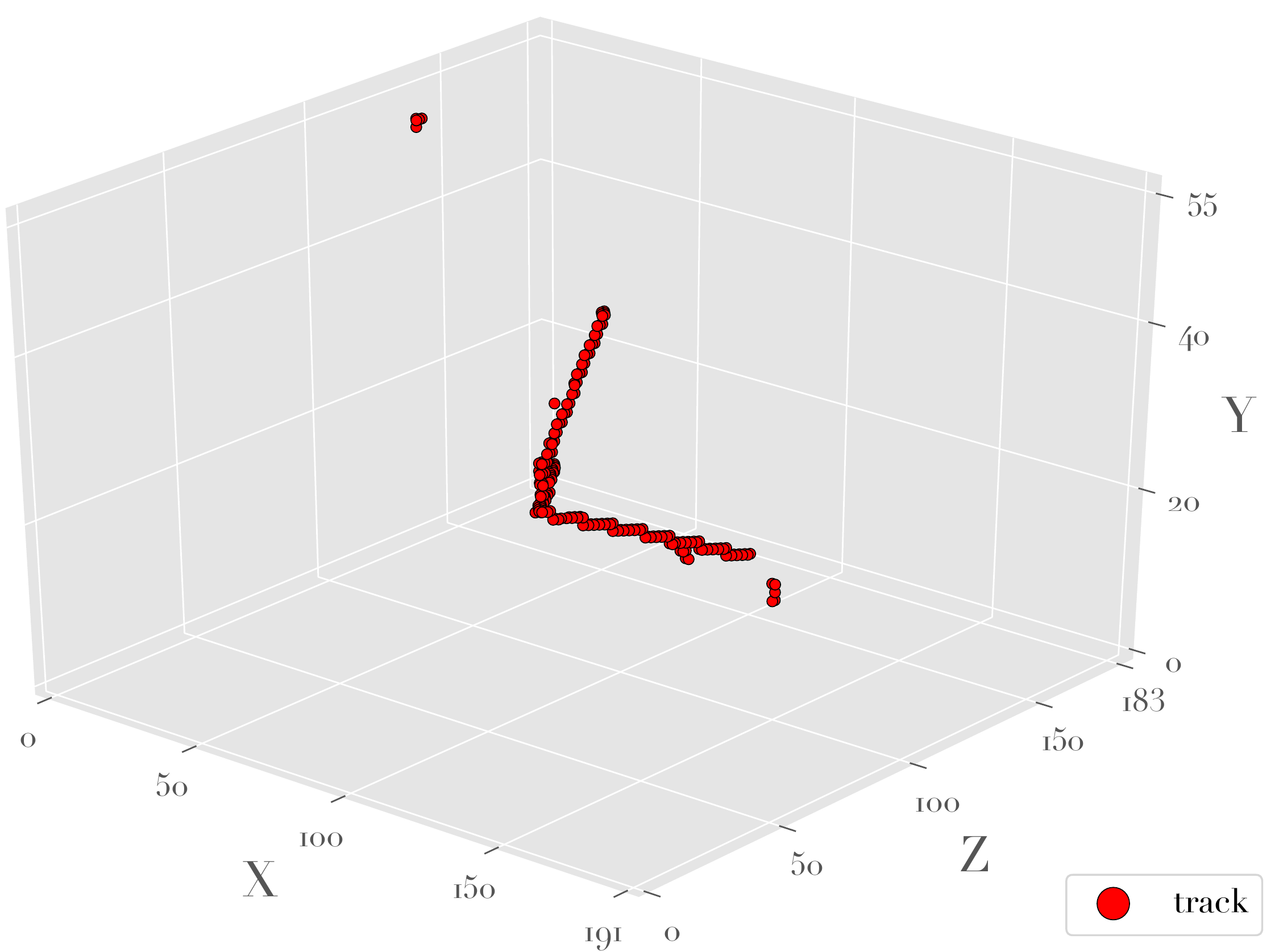}}
  \subcaptionbox{The 3D voxels labelled as track according to the charge cut classification are shown. \label{fig:event_track_chargecut_4}}%
  [.495\linewidth]{\includegraphics[width=8.8cm]{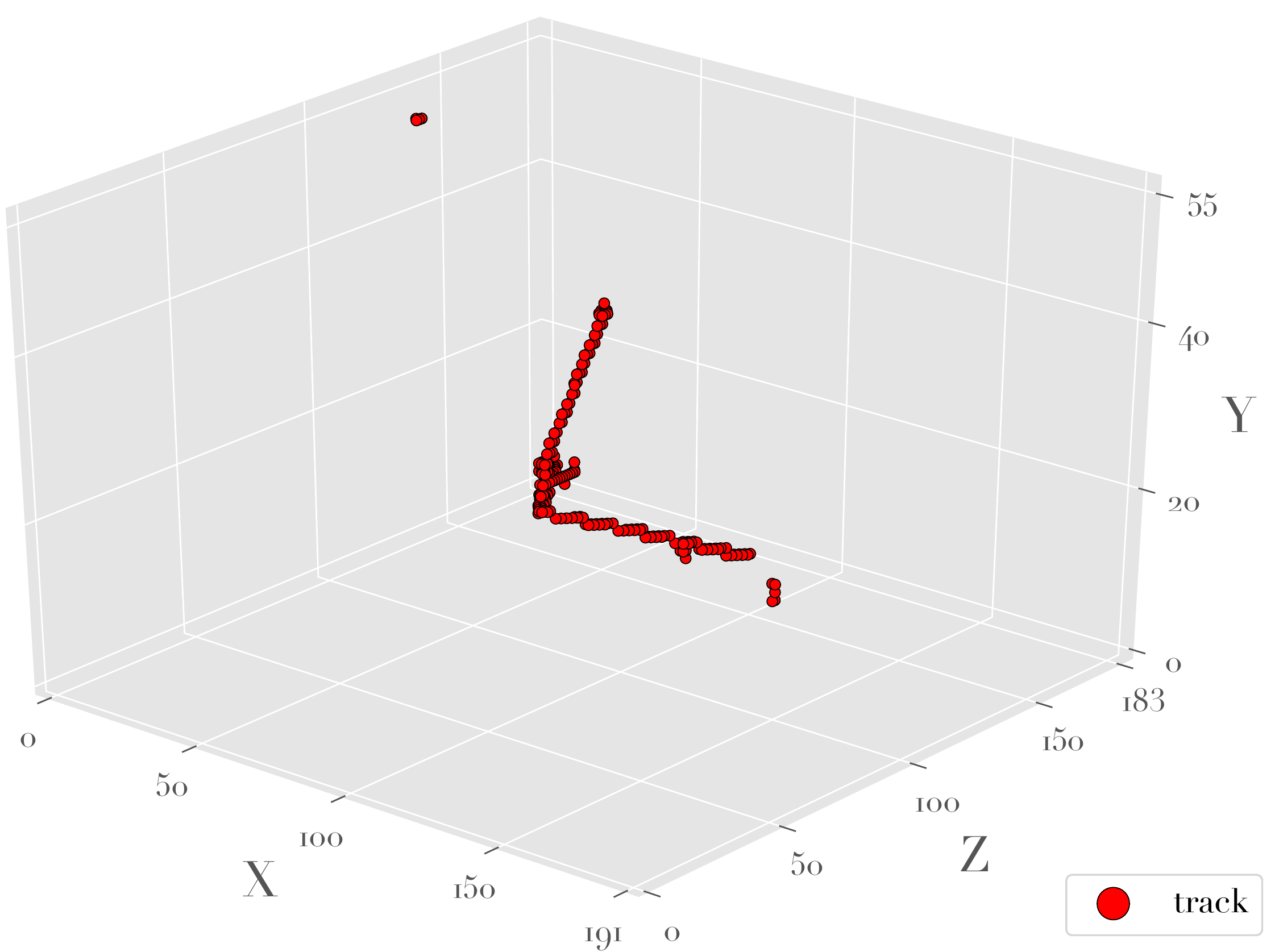}}
  \subcaptionbox{The 3D voxels labelled as track according to the GNN classification are shown.  \label{fig:event_track_gnn_4}}%
  [.495\linewidth]{\includegraphics[width=8.8cm]{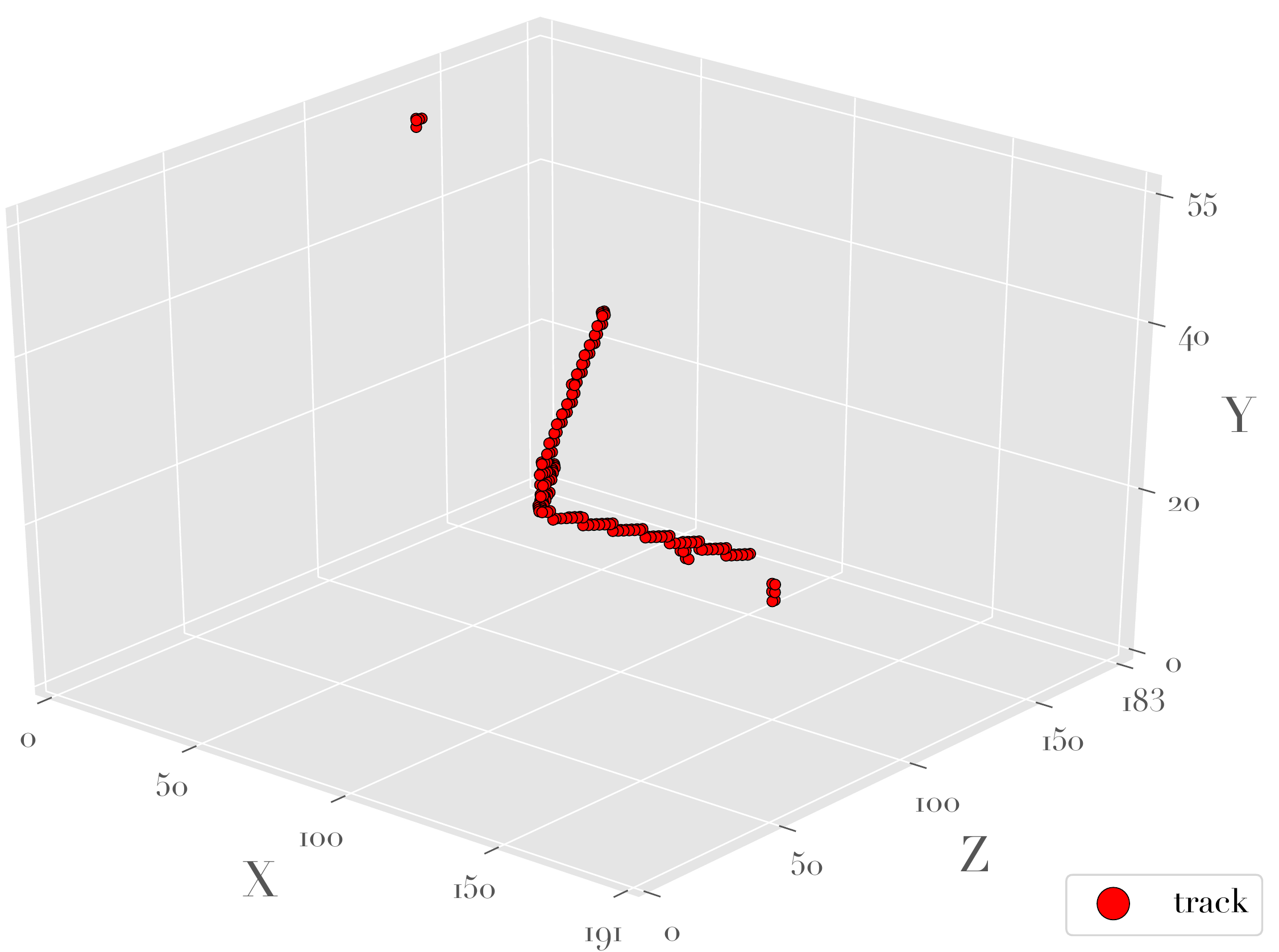}}
  \caption{3D visualization of a neutrino interaction in a finely segmented 3D scintillator detector after the 3D matching of the three 2D views.
  Although this is a relatively simple neutrino event, the charge cut is not able to reject a fake track stopping near the neutrino interaction vertex while GNN can provide a much cleaner reconstruction. \review{The energy of the incoming neutrino is 1.132~GeV. The axes are in cm.}
  }
  \label{fig:event_5}
\end{figure*}

\begin{figure*}[ht!] 
  \subcaptionbox{The 3D voxels labelled as track (red), crosstalk (blue) and ghost (yellow) according to the truth information from the simulation are shown. \label{fig:event_truth_5}}%
  [.495\linewidth]{\includegraphics[width=8.8cm]{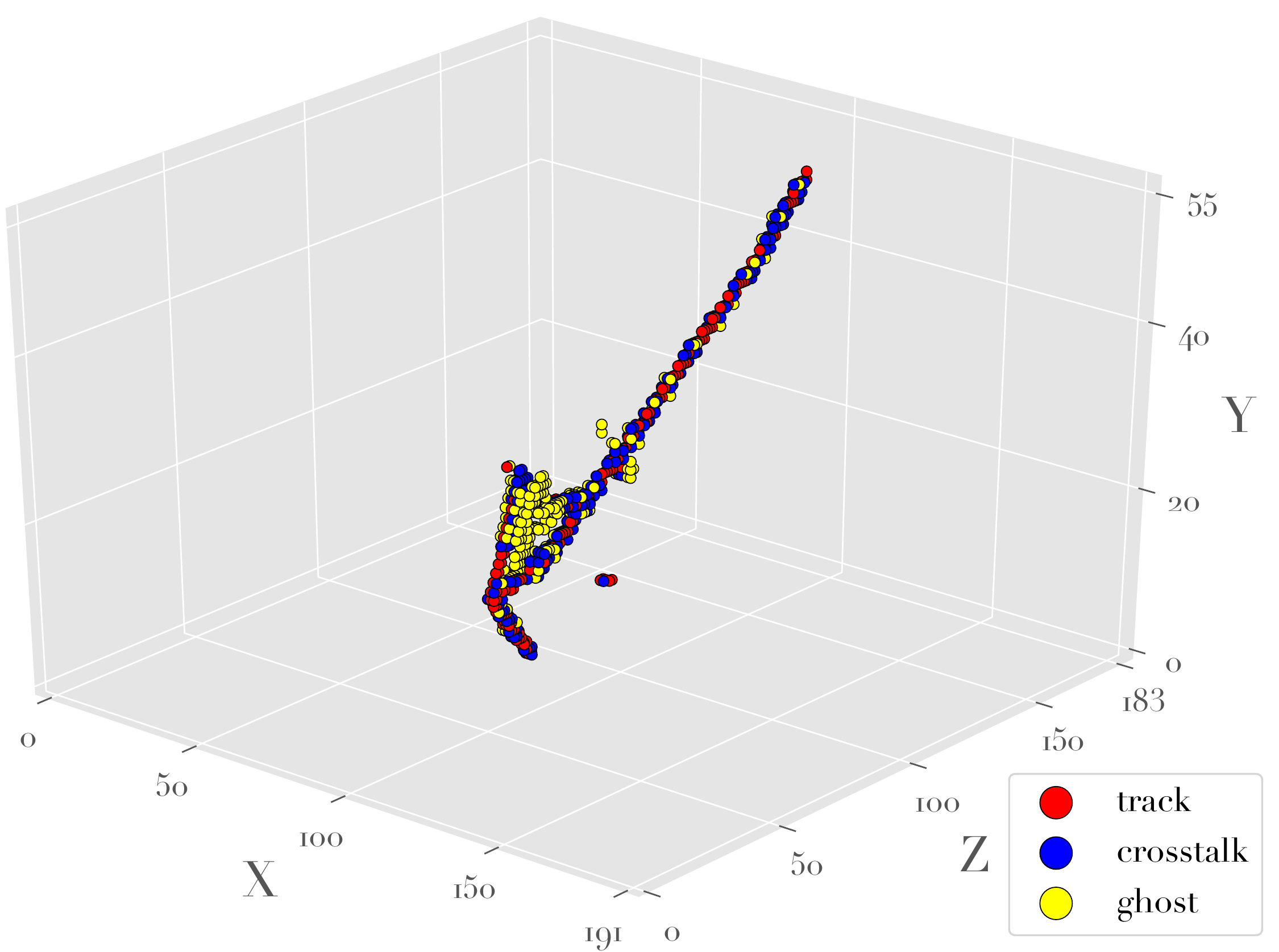}}
  \subcaptionbox{Only the 3D voxels labelled as track according to the truth information from the simulation are shown. \label{fig:event_track_truth_5}}%
  [.495\linewidth]{\includegraphics[width=8.8cm]{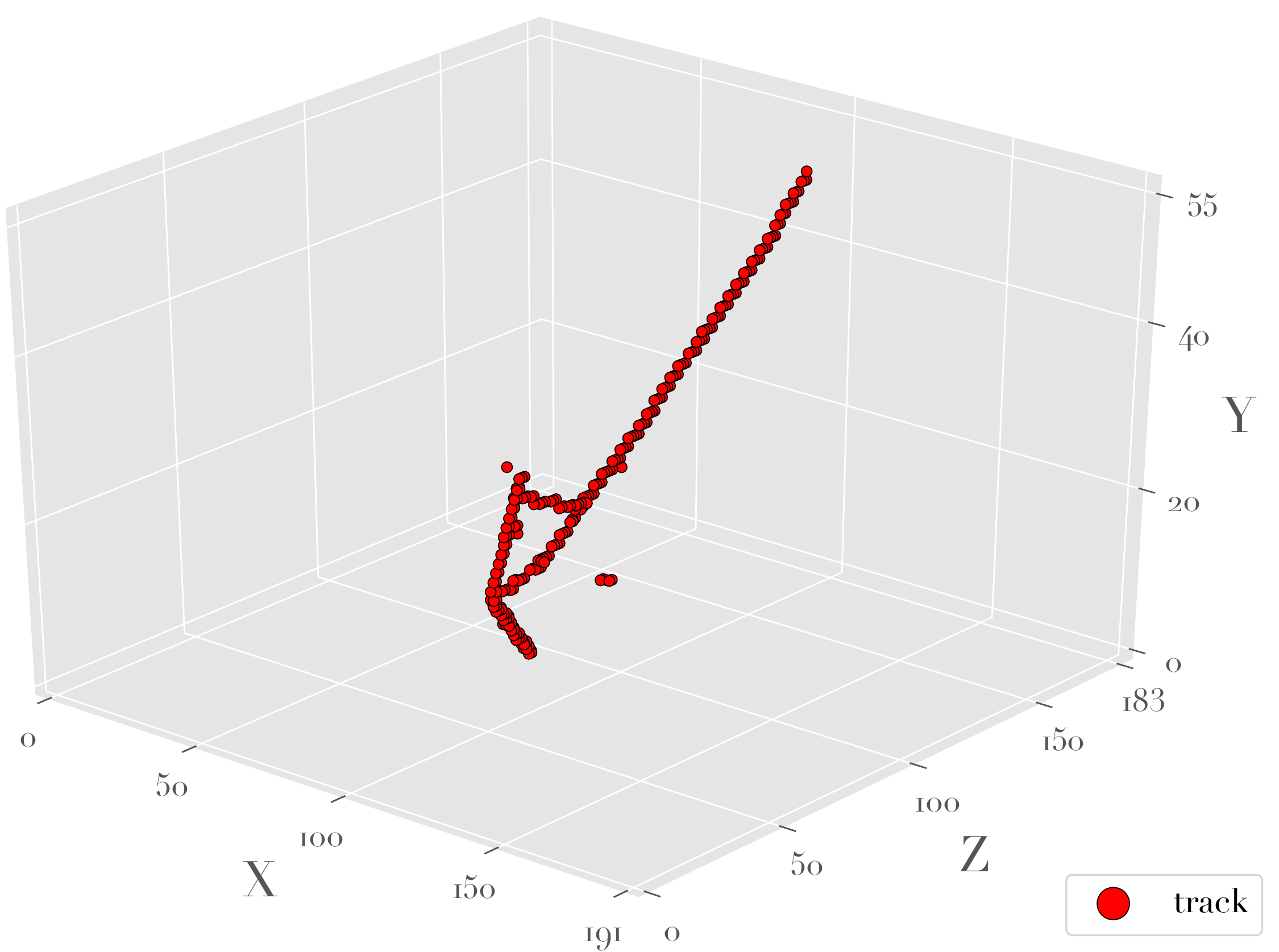}}
  \subcaptionbox{The 3D voxels labelled as track according to the charge cut classification are shown. \label{fig:event_track_chargecut_5}}%
  [.495\linewidth]{\includegraphics[width=8.8cm]{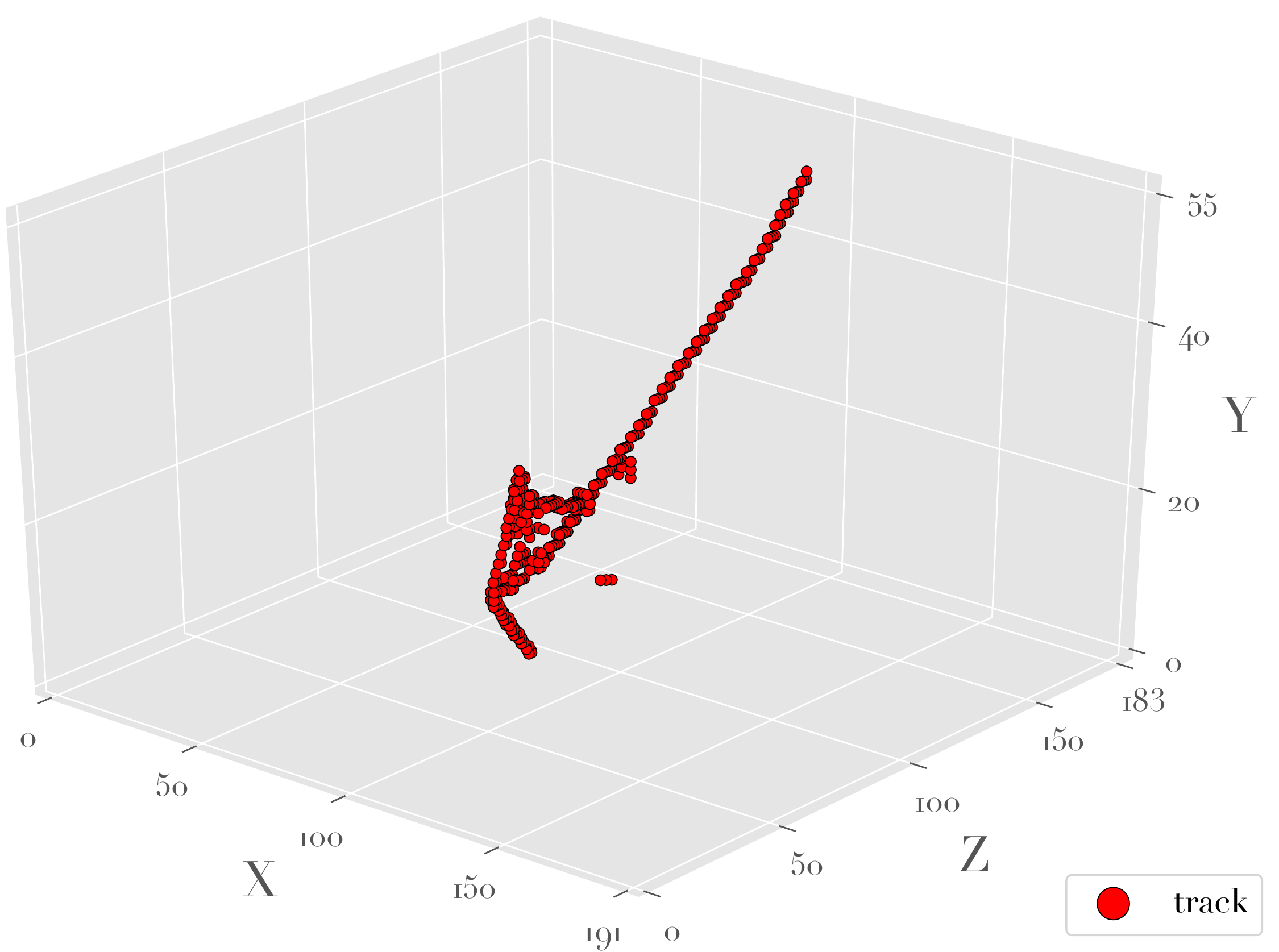}}
  \subcaptionbox{The 3D voxels labelled as track according to the GNN classification are shown.  \label{fig:event_track_gnn_5}}%
  [.495\linewidth]{\includegraphics[width=8.8cm]{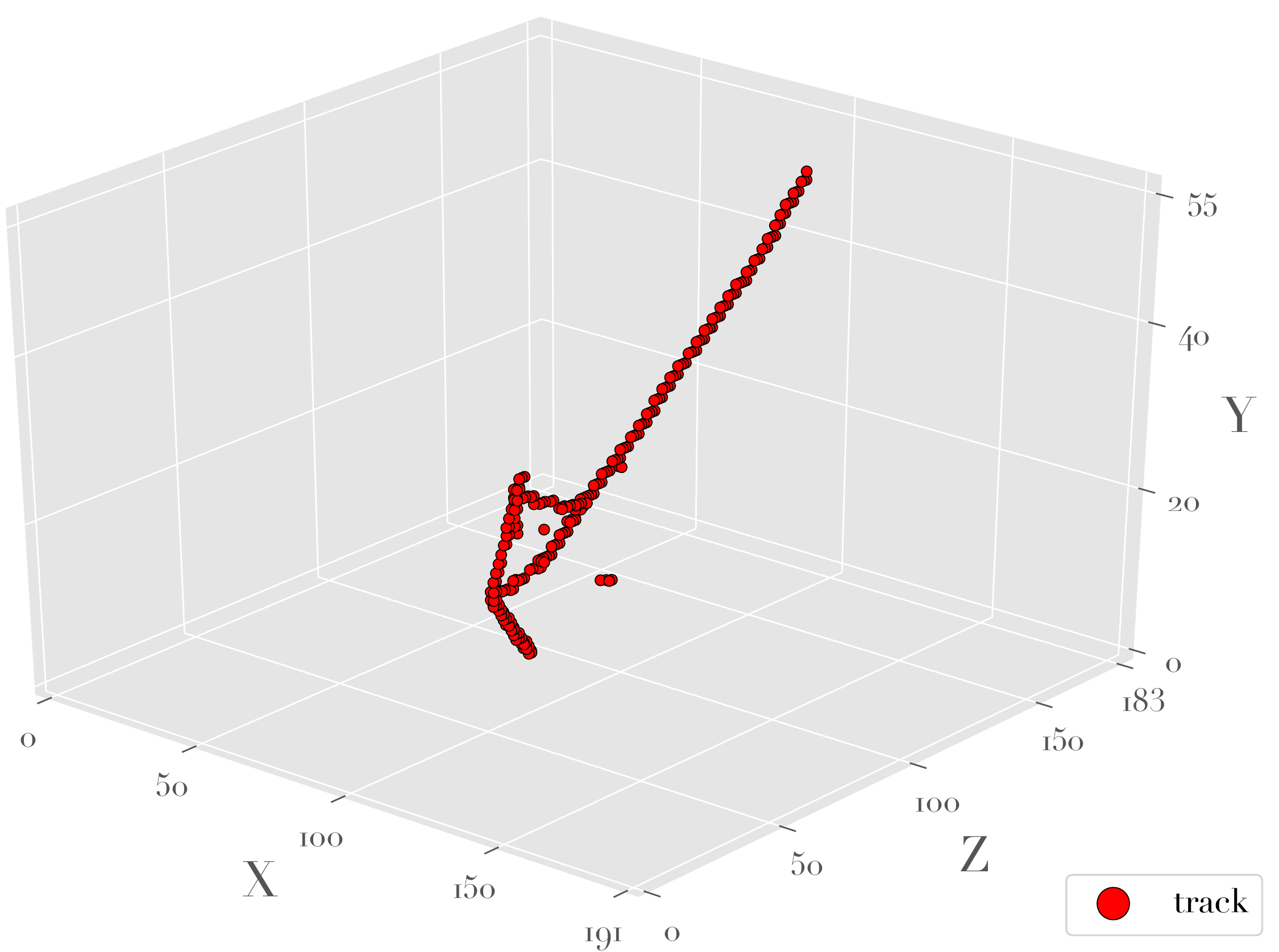}}
  \caption{3D visualization of a neutrino interaction in a finely segmented 3D scintillator detector after the 3D matching of the three 2D views.
  In the neutrino event GNN can easily reject the relatively big cluster of ghost voxels that would make difficult a proper reconstruction of the number of tracks and corresponding energy, in particular near to the interaction vertex. \review{The energy of the incoming neutrino is 1.897~GeV. The axes are in cm.}
  }
  \label{fig:event_6}
\end{figure*}

This section contains a number of visualizations to show the classification performance of the GNN for a number of neutrino interactions with different complexity and topology. Displays are shown for different events in Figs~\ref{fig:event_1} -~\ref{fig:event_6}: all voxels with their true classification, only the true track voxels, the classified track voxels using the charge cut method, and the classified track voxels using the GNN. The interactions shown here are examples of interactions containing many ghost voxels in order to showcase the GNN performance. 

The track voxel classification ability of the charge cut and GNN methods can be seen by comparing subfigures (c) and (d) with (b), respectively, for each interaction. The GNN is able to reject ghost voxels very well, as shown in Figs~\ref{fig:event_2},~\ref{fig:event_4},~\ref{fig:event_5} and~\ref{fig:event_6} where ghost tracks remain using the charge cut method. In general, the performance improvement from the GNN increases with the complexity of the interactions. For simple interactions with only a single muon in the final state both methods perform similarly.

\clearpage
\newpage

\bibliographystyle{apsrev4-2}
\bibliography{biblio}

\end{document}